\documentclass[pdflatex,sn-mathphys-num]{sn-jnl}

\usepackage{graphicx}%
\usepackage{multirow}%
\usepackage{amsmath,amssymb,amsfonts}%
\usepackage{amsthm}%
\usepackage{mathrsfs}%
\usepackage[title]{appendix}%
\usepackage{xcolor}%
\usepackage{textcomp}%
\usepackage{manyfoot}%
\usepackage{booktabs}%
\usepackage{algorithmic}%
\usepackage{listings}%
\usepackage{color}
\usepackage{comment}
\usepackage[ruled,vlined]{algorithm2e}
\usepackage{amsmath}
\usepackage{tabularx}
\usepackage[export]{adjustbox}
\usepackage{subfigure}
\usepackage{array}
\usepackage{multirow}
\usepackage{longtable}
\usepackage{hhline}
\usepackage{caption}
\usepackage[breakable]{tcolorbox}
\definecolor{custompink}{HTML}{DD99BB} 
\definecolor{customred}{HTML}{F66B2E} 
\definecolor{customyellow}{HTML}{F3AE36} 
\definecolor{customgreen}{HTML}{B9D5B1} 
\definecolor{customblue}{HTML}{4D9DE0} 
\definecolor{custompurple}{HTML}{AEADF0} 
\definecolor{customwhite}{HTML}{fbf9f4}

\newcommand{\deepretro}{DeepRetro\xspace}

\begin{document}

\title[DeepRetro: Retrosynthetic Pathway Discovery using Iterative LLM Reasoning]{DeepRetro: Retrosynthetic Pathway Discovery using Iterative LLM Reasoning}


\author*[1]{\fnm{Shreyas} \sur{Vinaya Sathyanarayana}}\email{shreyas@deepforestsci.com}
\equalconttwo{Co-first Author}

\author[1,6]{\fnm{Sharanabasava} \sur{D. Hiremath}}
\equalconttwo{Co-first Author}

\author[1]{\fnm{Rahil} \sur{Shah}}
\equalcont{These authors contributed equally to this work.}

\author[1,3,5]{\fnm{Rishikesh} \sur{Panda}}
\equalcont{These authors contributed equally to this work.}

\author[1]{\fnm{Rahul} \sur{Jana}}
\workdoneat{Deep Forest Sciences}
\author[1]{\fnm{Riya} \sur{Singh}} 
\author[1]{\fnm{Rida} \sur{Irfan}}
\author[1]{\fnm{Ashwin} \sur{Murali}}
\workdoneat{Deep Forest Sciences}

\author*[1]{\fnm{Bharath} \sur{Ramsundar}}\email{bharath@deepforestsci.com}

\affil[1]{\orgname{Deep Forest Sciences}, \orgaddress{\state{California}, \country{USA}}}


\affil[3]{\orgdiv{Department of Biology}, \orgname{Birla Institute of Technology \& Science, Pilani}, \orgaddress{\state{Goa}, \country{India}}}


\affil[5]{\orgdiv{Department of Electrical \& Electronics Engineering}, \orgname{Birla Institute of Technology \& Science, Pilani}, \orgaddress{\state{Goa}, \country{India}}}

\affil[6]{\orgdiv{Department of Chemical Sciences}, \orgname{IISER Kolkata}, \orgaddress{\state{West Bengal}, \country{India}}}

\abstract{

The synthesis of complex natural products remains one of the grand challenges of organic chemistry. We present DeepRetro, a major advancement in computational retrosynthesis that enables the discovery of viable synthetic routes for complex molecules typically considered beyond the reach of existing retrosynthetic methods. DeepRetro is a novel, open-source framework that tightly integrates large language models (LLMs), traditional retrosynthetic engines, and expert human feedback in an iterative design loop. Prior approaches rely solely on template-based methods or unconstrained LLM outputs. In contrast, DeepRetro combines the precision of template-based methods with the generative flexibility of LLMs, controlled by rigorous chemical validity checks and enhanced by recursive refinement. This hybrid system dynamically explores and revises synthetic pathways, guided by both algorithmic checks and expert chemist feedback through an interactive user interface.
While DeepRetro achieves strong performance on standard retrosynthesis benchmarks, its true strength lies in its ability to propose novel, viable pathways to highly complex natural products—targets that have historically eluded automated planning. Through detailed case studies, we illustrate how this approach enables new routes for total synthesis and facilitates human-machine collaboration in organic chemistry.
Beyond retrosynthesis, DeepRetro represents a working model for how to leverage LLMs in scientific discovery. We provide a transparent account of the system’s design, algorithms, and human-feedback loop, enabling broad adaptation across scientific domains. By releasing DeepRetro as an open-source tool, we aim to empower chemists to tackle increasingly ambitious synthetic targets, accelerating progress in drug discovery, materials design, and beyond.
}

\keywords{DeepRetro, LLMs, retrosynthesis, CASP, AI, machine learning, chemistry, drug discovery}



\maketitle

\section{Introduction}
\label{sec:intro}

The ability to design and execute efficient, predictable synthetic routes for organic compounds is foundational to progress across the chemical sciences. 
Challenging syntheses are not only critical for the developent of small molecule therapeutics but also central to advances in semiconductors, energy, and agrochemicals. Yet, devising viable synthetic pathways, particularly for structurally complex molecules, remains a major bottleneck \citep{blakemore.2018.organic}, with iterative `design-make-test' cycles frequently constrained by the synthesis (``make'') step \citep{schneider.2018.automating}.

At the core of this challenge lies retrosynthesis, a strategy pioneered in the early-20th century \citep{corey_computer-assisted_1969,corey_computer-assisted_1985,corey_general_1967}, in which a target molecule is deconstructed stepwise into simpler precursors through a series of hypothetical ``disconnections" corresponding to known chemical reactions. 
This backwards reasoning enables chemists to identify viable synthetic pathways that can then guide forward synthesis in the laboratory. While conceptually elegant, navigating the vast space of possible reactions and intermediates while considering reaction yields, selectivity, cost, and safety remains a daunting task, making retrosynthesis a long-standing grand challenge for both chemistry and artificial intelligence. 

Early efforts to automate retrosynthesis via computer-aided synthesis planning (CASP) such as LHASA (Logic and Heuristics Applied to Synthetic Analysis) \cite{LHASA}, used expert systems to attempt to codify chemical knowledge. These early rule-based systems showed promise but were limited by the need for laborious encoding of chemical rules and the difficulty of maintaining up-to-date reaction databases, often leaving pathway identification as a manual, chemist-driven process.


In recent years, the convergence of machine learning (ML), large reaction databases, and increased computational power has yielded significant progress in CASP \citep{casp_recent}. Tools, such as ASKCOS \citep{coley.2019.askcos}, AiZynthFinder \citep{genheden.2020.aizynthfinder}, Synthia\citep{szymkuc_synthia_2016}, and IBM RXN have incorporated diverse ML techniques, including template-based approaches that apply curated reaction templates from databases \citep{segler_neural-symbolic_2017,fortunato_data_2020,coley_computer-assisted_2017,seidl_improving_2022,ishida_prediction_2019,dai_retrosynthesis_2019,chen_deep_2021}, template-free methods based on graph neural networks or sequence-to-sequence architectures \citep{zheng_predicting_2019,chen_learning_2019,yang_molecular_2019,lin_automatic_2020,tetko_state---art_2020,seo_gta_2021,kim_valid_2021,irwin_chemformer_2022,zhong_root-aligned_2022,sacha_molecule_2021,mao_molecular_2021,mann_retrosynthesis_2021,ucak_substructure-based_2021,ucak_retrosynthetic_2022,liu_retrosynthetic_2017}, and powerful search algorithms such as Monte Carlo Tree Search (MCTS) \citep{genheden_aizynthfinder_2020,coley.2019.askcos}.

Despite these advances, conventional CASP tools struggle with complex or unconventional synthetic targets. Template-based methods are constrained by their underlying reaction databases. The exponential growth of possible pathways necessitates heuristic searches, risking the pruning of optimal solutions. Capturing the nuanced intuition of expert chemists remains difficult, and data scarcity for specific reaction classes can impede model performance for both template-based and template-free methods.

Large Language Models (LLMs), typically transformer-based architectures \citep{vaswani_attention_2017} trained on vast text and code datasets, are emerging as a powerful new tool in computational chemistry. These models can directly operate on string-based molecular representations (e.g., SMILES) and have shown promise in molecular property prediction \citep{chemberta,chemberta2}, reaction outcome forecasting \citep{llm_reaction}, novel molecule generation \citep{llm_molgen,molt5}, and literature mining \citep{zhang_fine-tuning_2024}. ChemCrow \cite{chemcrow} demonstrates the ability of LLMs to autonomously plan and execute complex chemical tasks by orchestrating a suite of expert-designed computational tools.
The ability to extract implicit chemical knowledge and generalize across diverse chemical contexts positions LLMs as a promising foundation for next-generation synthesis planning. Recent work has begun to explore LLMs for retrosynthesis, using them as route generators or as a guide for traditional search algorithms \citep{wang_llm-augmented_2025}. 

\begin{figure}[!ht]
\centering
	\includegraphics[width=0.90\textwidth]{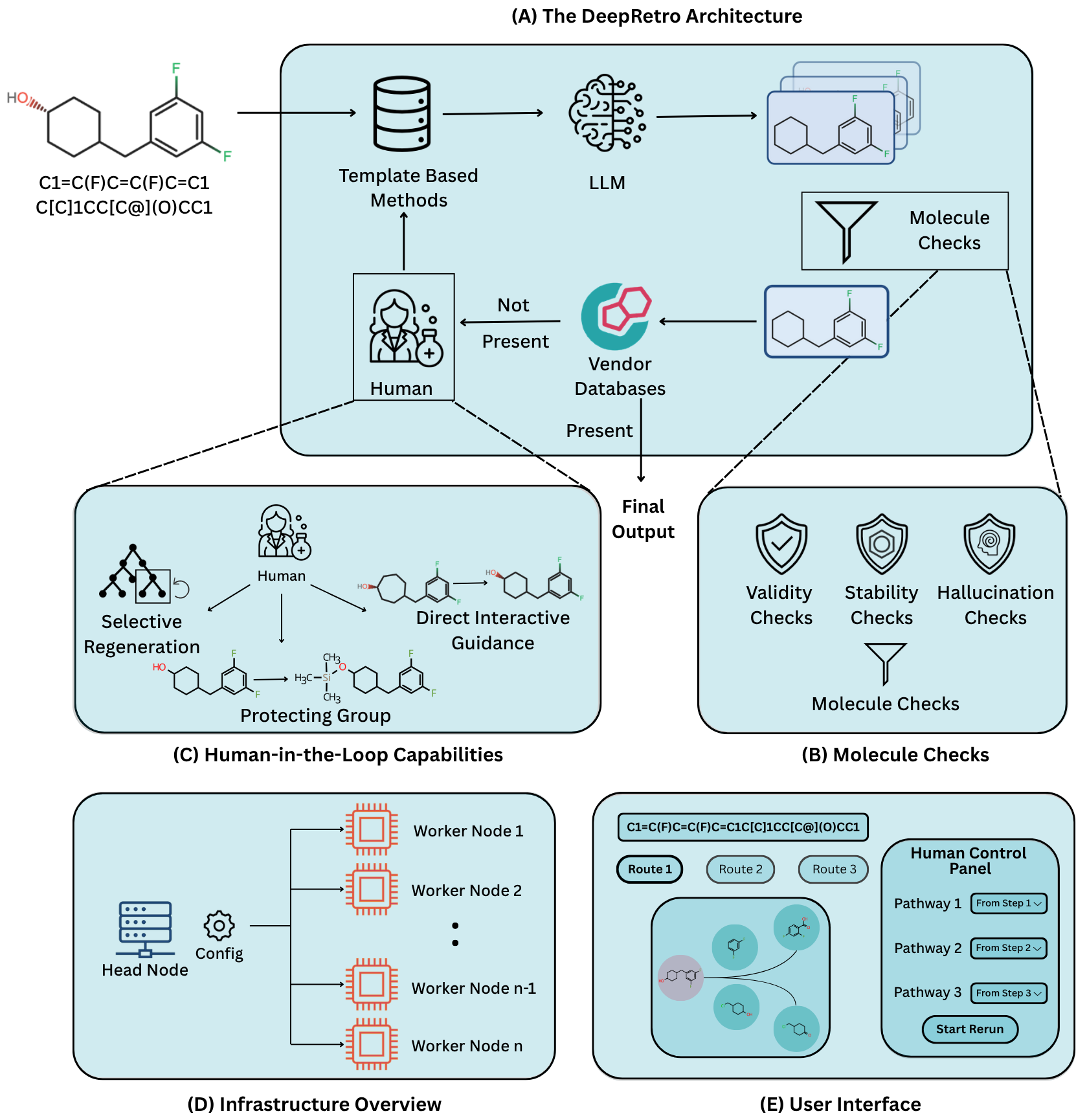}
	\caption{(a) The DeepRetro framework. Retrosynthesis starts with a template based tool invocation. If this fails, an LLM proposes single steps, which undergo validation checks. If proposed molecules are not available in a vendor database, the molecule continues in the pipeline. The pipeline then moves into an optional human intervention   before recursive evaluation. (b) The different molecule checks that are incorporated into DeepRetro. This includes Validity checks (Valency, allowed atoms), Stability Checks (discussed in detail in \ref{fig:checks}) and Hallucination Checks (verification that LLM provides sensible outputs). (c) Describes the different types of human interventions compatible with DeepRetro. ``Selective Regeneration" enables regeneration of erroneous parts of the pathway. ``Direct Interactive Guidance" enables chemists to make small changes to fix hallucinations. ``Protecting Group" allows for addition of protecting groups. (d) DeepRetro operates a head node which controls several worker nodes. The number of worker nodes can be scaled for complex syntheses. (e) The GUI that allows chemists to visualize pathways, select nodes for regeneration, and directly edit molecular structures.}
	\label{fig:deepretro_hero}
\end{figure}

In this work, we introduce DeepRetro, an iterative hybrid framework that significantly expands the reach of retrosynthetic planning by combining LLMs with traditional CASP tools and human-in-the-loop feedback.  Rather than attempting to generate full synthetic routes in a single pass, DeepRetro employs an iterative control loop wherein an LLM proposes single-step disconnections, which are then subjected to strict chemical validity, stability, and hallucination checks. Validated precursors are recursively fed back into the planning loop, allowing for step-wise refinement and dynamic course correction. This approach preserves the flexibility of LLM reasoning while enforcing chemical rigor at each step, resulting in more interpretable, reliable, and creative synthetic strategies.
This framework is detailed in figure \ref{fig:deepretro_hero}.

Crucially, DeepRetro enables the synthesis planning of complex natural products that have historically remained out of reach for retrosynthetic systems. Through detailed case studies, we show that DeepRetro can, with expert human guidance, discover novel synthetic routes to such molecules. Beyond benchmark performance, these examples illustrate DeepRetro’s ability to generalize across reaction families, propose unconventional pathways, and complement human chemists in solving hard synthetic problems.


To support chemist collaboration and ensure practical usability, we develop an interactive graphical user interface (GUI) that allows domain experts to visualize and intervene in retrosynthetic reasoning in real time. This human-in-the-loop component mitigates failure modes such as hallucinations and enables creative co-design between expert chemists and the LLM. We open-source DeepRetro and its interface, providing detailed descriptions of our algorithms, workflows, and validation pipelines to facilitate broad reuse and extension.

We believe DeepRetro represents a compelling working model for how LLMs can be integrated into scientific discovery pipelines. By combining generative reasoning, chemical validation, and expert interaction, DeepRetro enables more powerful, interpretable, and creative retrosynthetic planning—laying a foundation for the synthesis of previously inaccessible compounds in drug discovery, materials design, and beyond.

\section{Results}
\label{sec:Results}
To evaluate the performance and capabilities of DeepRetro, we conducted experiments on standard benchmark datasets and illustrative case studies. We have chosen five molecules for our case studies, namely Ohauamine C \cite{chen_ohauamines_2025}, a Tetracyclic Azepine derivative\cite{gala_development_2004},  Erythromycin\cite{erythromycin}, Reserpine\cite{chen_reserpine_2005,reserpine_total_synthesis} and Discodermolide\cite{discodermolide}. These molecules were chosen to test DeepRetro's ability to solve retrosynthesis for interesting and complex natural products. These case studies required human-in-the-loop guidance at certain critical steps.

\subsection{Datasets}
\label{sec:exp:datasets}
Our experimental evaluations and model development primarily utilized reaction datasets derived from the United States Patent and Trademark Office (USPTO) collection, a widely recognized benchmark in retrosynthesis research \citep{uspto}. Specifically, the USPTO-190 test dataset, comprising approximately 190 reactions, was employed for benchmarking multi-step retrosynthesis predictions.
A subset of the USPTO-50k test set containing 250 reactions was used for single step benchmarking. (An evaluation on the full USPTO-50k test set would have been prohibitively expensive due to the need for external LLM calls and is left to future work as LLM pricing falls.) These 250 reactions were selected by clustering circular fingerprints of reactant compounds to select a maximally diverse set of reactions.

For broader evaluations and as a baseline for our template/MCTS tool $T$ (an AiZynthFinder adaptation), we leveraged the model and reaction policies provided by the original AiZynthFinder developers, which are trained on the larger USPTO dataset. This ensured comparability with established benchmarks. In addition to USPTO data, we also utilized the Pistachio dataset (2024Q4 version) from NextMove Software \citep{Pistachio_2,lowe_2012_pistachio,nextmovesoftware2024pistachio}, a comprehensive reaction database primarily extracted from chemical patents and containing several million reactions. For specific developmental aspects of our hybrid pipeline and for experiments requiring an independently trained template-based component, we trained our instance of the AiZynthFinder tool on this Pistachio dataset (2024Q4 version). This allowed us to explore the system's performance with a distinct and extensive reaction knowledge base.

\subsection{Evaluation Metrics}
\label{sec:exp:metrics}
Evaluating retrosynthesis pathways is complex, as multiple valid routes can exist, and computational metrics may not fully capture chemical feasibility or elegance. We used a combination of quantitative and qualitative metrics:

\subsubsection{Single-Step Prediction Accuracy}
\label{sec:exp:metrics:single_step_accuracy}
To evaluate the performance of single-step retrosynthetic predictions, we adapted the conventional top-$k$ accuracy metric to provide a more nuanced understanding of model performance. We defined two primary measures:

\paragraph{All Correct Accuracy} This stringent metric quantifies instances where the complete set of reactants predicted by the model precisely matches the full set of ground truth reactants documented in the reference dataset for a given target molecule. A prediction is only considered correct under this measure if all proposed reactant molecules are identical to those in the ground truth.

\paragraph{Any Correct Accuracy} Recognizing that a retrosynthetic model might propose a chemically valid transformation involving the correct key precursor(s) alongside differing co-reactants or reagents, or might identify an alternative valid disconnection leading to one or more of the same key precursors, we also employ an ``Any Correct Accuracy". This metric considers a prediction successful if at least one of the reactant molecules proposed by the model matches any of the ground truth reactant molecules for the target. This measure is particularly useful as it acknowledges predictions where the core transformation leading to a critical precursor is correctly identified, even if the full set of associated molecules (e.g., minor reagents, byproducts considered as reactants in the reverse reaction) differs from the dataset's specific annotation, or if the model proposes a legitimate alternative synthetic approach to a key intermediate.

\subsubsection{Multi-Step Predictions}
\paragraph{Pathway Success Rate} For the end-to-end multi-step evaluation, we measured the percentage of target molecules in the multi-step test set for which DeepRetro successfully found any complete pathway terminating in the defined stock materials within a given computational budget (time limits, API \& Compute Cost requirements).

\subsubsection{Nuances in Evaluating Performance with Standard Metrics} 
\label{sec:exp:metrics:limit}
While we report standard metrics like Top-K accuracy to ensure comparability with established benchmarks, it's crucial to recognize their inherent limitations when evaluating a generative framework like DeepRetro. These metrics often fail to capture the full scope of chemical plausibility and synthetic utility.

A primary issue is that a single ``correct" answer, as defined by a ground truth entry in a dataset, does not account for multiple viable synthetic strategies. For instance, a Top-K accuracy metric would penalize DeepRetro for proposing a chemically sound and perhaps even more elegant disconnection if it doesn't precisely match the specific reactants listed in the test set. The model's ability to explore a wider chemical space and identify valid alternative pathways—a key strength of LLM-driven approaches—can paradoxically result in a lower score under this rigid evaluation.

This is why we also employ the ``Any Correct Accuracy" metric, which partially mitigates this issue by rewarding the identification of at least one key ground-truth precursor. Similarly, the ``Pathway Success Rate" is a useful indicator of the system's ability to find a complete route within set constraints, but it does not measure the quality, efficiency, or novelty of the discovered pathway.

Ultimately, these quantitative measures are insufficient on their own. They must be contextualized with qualitative assessments, such as the detailed case study analyses (Section \ref{sec:casestudies}), which are better suited to evaluate the creativity and practical value of the routes generated by our hybrid approach.

\subsubsection{Case Study Analysis} 
To overcome the limitations of automated metrics, we performed a detailed case-study analysis on selected complex targets to qualitatively assess the value of our hybrid approach. This involved a direct comparison of pathways generated by our pipeline against those from the baseline MCTS-based tool ($T$), allowing us to isolate the LLM's contribution. Each LLM-proposed step was manually evaluated for chemical plausibility and novelty, specifically identifying instances where it successfully bypassed the constraints of the baseline's reaction templates. Furthermore, by evaluating pathways for targets with no established literature precedence, we assessed the framework's potential to facilitate novel chemical discovery.

\subsection{Single-Step Benchmarks}
When evaluated on the USPTO-50k test subset of 250 molecules selected based on the circular fingerprints of the reactant molecules, a DeepRetro model trained on Pistachio achieved an ``Any Correct" accuracy of 54\% (135/250) 
in predicting the ground truth reactants compared with an accuracy of 46\% (115/250)for ASKCOS. 
The choice of 250 test compounds may affect this comparison, so these results should be taken as qualitative comparisons until larger more rigorous benchmarks can be completed. 
These results are shown in Table \ref{tab:single_step_results}. DeepRetro tends to perform worse in ``All Correct" as DeepRetro often provides alternate retrosynthetic routes (including reagents) for most of the molecules (as mentioned in section \ref{sec:exp:metrics:limit}). However, DeepRetro using Claude 4 Opus and Pistachio obtains 44\% (110/250) and still slightly outperforms ASKCOS 42\% (105/250) in ``All Correct".

\begin{table}
\centering
\caption{
This table showcases the Single-Step Retrosynthesis Prediction Accuracy (Top-1) on a 250 subset of USPTO-50k. The numbers reported are out of 250 tested molecules. DeepRetro's performance depends on the choice of underlying LLM and training dataset for the template-based algorithm. With Claude 4 Opus and Pistachio, DeepRetro outperforms strong baselines like ASKCOS.  DeepRetro was run in automatic mode with no human intervention.}
\label{tab:single_step_results}
\begin{tabular}{|lllcc|}
\hline
Model & LLM & Dataset & All Correct Accuracy (/250) & Any Correct Accuracy (/250) \\
\hline
ASKCOS 			 & - & Reaxys & 105 & 115 \\
\hline
Aizynthfinder 	 & - & Pistachio 					& 73 & 83 \\
DeepRetro        & Claude 3 Opus  & Pistachio       & 95 & 110\\
DeepRetro        & Claude 3.5 Sonnet & Pistachio    & 90 & 102 \\
DeepRetro        & Claude 3.7 Sonnet  & Pistachio   & 95 & 113 \\
DeepRetro        & Claude 4 Opus & Pistachio        & 110 & 135 \\
DeepRetro        & Claude 4 Sonnet & Pistachio      & 107 & 129 \\
DeepRetro        & DeepSeek R1 & Pistachio          & 95 & 110 \\
\hline
Aizynthfinder    & - & USPTO                    & 63 & 70 \\
DeepRetro        & Claude 3 Opus  & USPTO       & 80 & 90 \\
DeepRetro        & Claude 3.5 Sonnet & USPTO    & 82 & 89 \\
DeepRetro        & Claude 3.7 Sonnet  & USPTO   & 85 & 95 \\
DeepRetro        & Claude 4 Opus & USPTO        & 95 & 107 \\
DeepRetro        & Claude 4 Sonnet & USPTO      & 93 & 103 \\
DeepRetro        & DeepSeek R1 & USPTO          & 83 & 92 \\
\hline
\end{tabular}
\end{table}

\subsection{Multi-Step Benchmarks}
The primary evaluation multi-step benchmark focused on the end-to-end pathway finding capability on the USPTO-190 test set.

\begin{table}[!ht]
\centering
\caption{This table showcases the number of solved molecules of Different Retrosynthesis Models. The numbers reported are out of 190.  DeepRetro was run in automatic mode with no human intervention.}
\label{tab:model_success_rates}
\begin{tabular}{|cccc|}
\hline
\textbf{Model} & \textbf{LLM} & \textbf{Dataset} & \textbf{Number of solved molecules (/190)} \\
\hline
	DeepRetro        & Claude 3 Opus  & Pistachio       & 170 \\
	DeepRetro        & Claude 3.5 Sonnet & Pistachio    & 166 \\
	DeepRetro        & Claude 3.7 Sonnet  & Pistachio   & 175 \\
	DeepRetro        & Claude 4 Opus & Pistachio        & 183 \\
	DeepRetro        & Claude 4 Sonnet & Pistachio      & 180 \\
	DeepRetro        & DeepSeek R1 & Pistachio          & 165 \\
	\hline
	DeepRetro        & Claude 3.7 Sonnet & USPTO        & 174 \\
	DeepRetro        & DeepSeek R1 & USPTO              & 168 \\
	Retro*           & NA & USPTO                       & 139 \\
	PDVN             & NA & USPTO                       & 177 \\
\hline
\end{tabular}
\end{table}

Table \ref{tab:model_success_rates} presents a comparison of number of solved molecules (out of 190) for several retrosynthesis models. The baseline model PVDN demonstrates a high success rate of 93.15\% (177/190). Our evaluations of the DeepRetro model show that specific configurations can achieve comparable top-tier performance. The DeepRetro Claude 4 Opus configuration when utilized with the Pistachio dataset gave the best performance of 96.31\% (183/190).

\subsection{Case Studies}
\label{sec:casestudies}

To illustrate the practical application and capabilities of DeepRetro, we present case studies for five distinct target molecules. These molecules were chosen to represent varying levels of complexity and to test different aspects of our methodology. Case study molecules were intended to be challenging for an expert human chemist to solve, but do not include any molecules that have not already been solved by human chemists. For 2 of the 5 case study molecules, DeepRetro succeeds in identifying novel pathways not reported in the literature to the best of our knowledge.

It is important to note that the case studies results below depend on both human and machine contributions. We have separated the contributions of the human and LLM in Table \ref{tab:human_llm_contri}. Full details of pathways with explicit annotation of human interventions is provided in Appendix~\ref{app:mol_paths}. As an important note, case study molecules required between 6-14 runs of DeepRetro, with human-in-the-loop guidance, in order to generate viable pathways. All case studies were run three times to check reproducibility of pathways.

    

\begin{table}[htbp]
  \centering
  \caption{This table showcases the specific individual contributions of the both the LLM and Human in obtaining the output shared in this paper. It also gives an overview of the number of regenerations DeepRetro requires to reproduce a pathway comparable to the pathway shared in this section. The ``*" for Erythromycin (molecule 3) is added to indicate that the pathway could not be generated without one key human intervention. All pathways were regenerated 3 times to verify reproducibility. The number of regenerations are obtained with DeepRetro with Claude 3.7 Sonnet}
    \begin{tabular}{|>{\centering\arraybackslash}m{0.22\textwidth}|p{11em}|p{10 em}|p{3 em}|p{5 em}|p{3em}|}
    
    \hline
    \textbf{Molecule} & \textbf{LLM Contribution} & \textbf{Human Contribution} & \textbf{No. of Regenerations} & \textbf{No. of Retrosynthesis steps} & \textbf{Novel Pathway?} \\
    \hline
     \includegraphics[valign=c, width=0.8\linewidth]{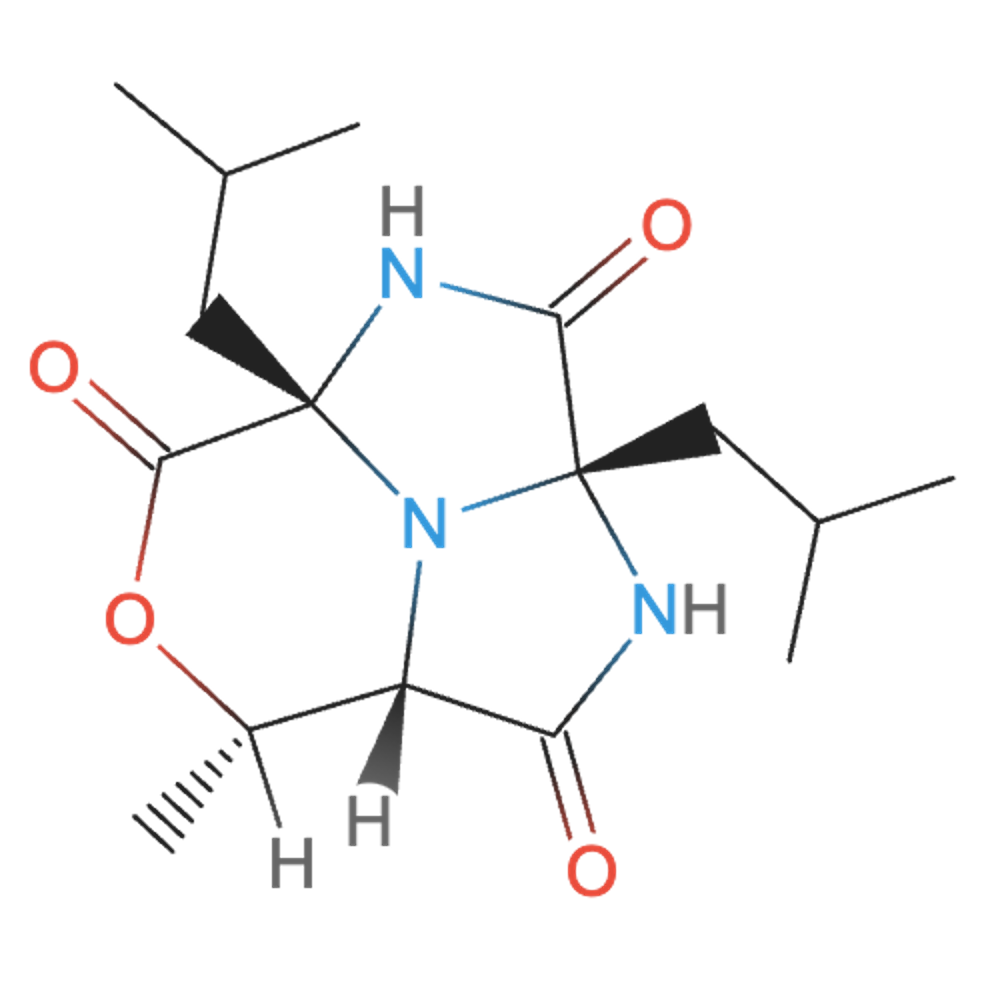}  & Generated a complete and chemically reasonable retrosynthetic pathway based on standard disconnections  &  Identified the basic building blocks that constitute the core of the molecule, helping guide the retrosynthesis & 10 & 4 & Yes \\
     \hline
     \includegraphics[valign=c, width=0.8\linewidth]{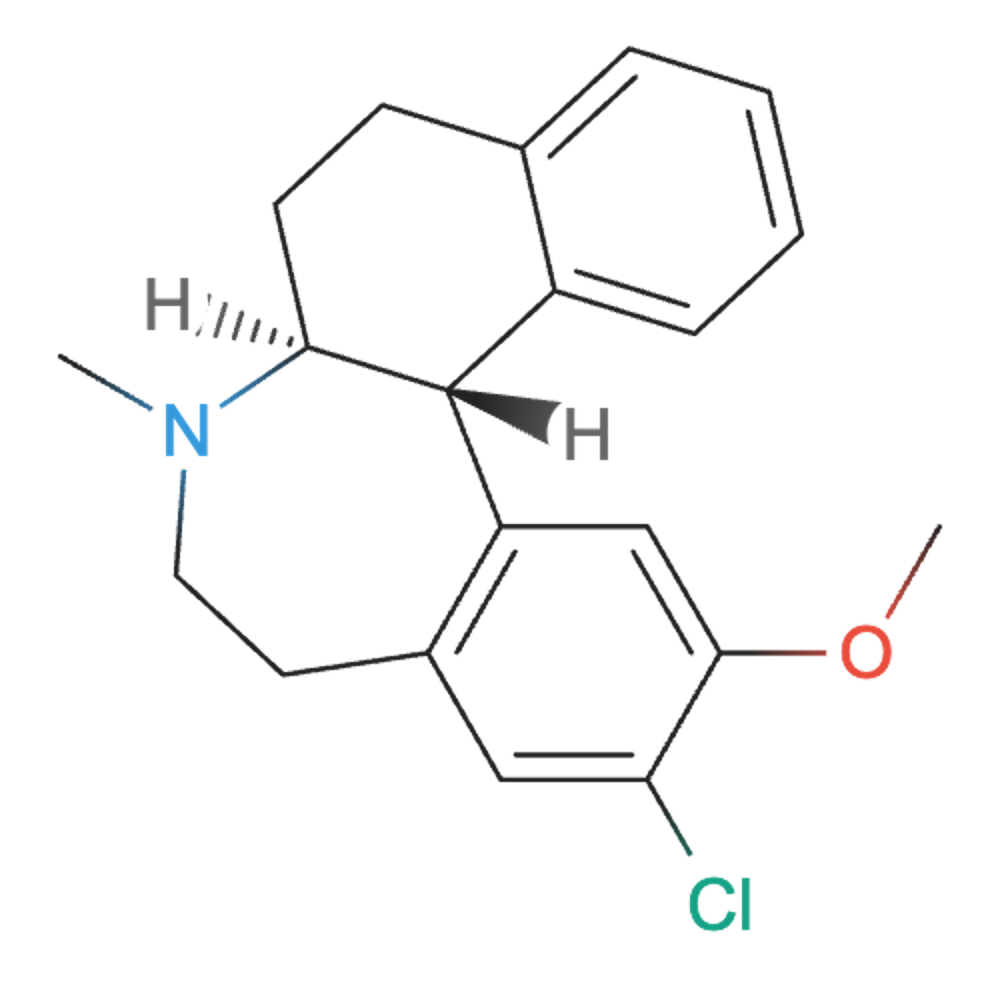}  & Proposed a viable disconnection strategy and correctly identified synthetically relevant intermediates  &  Validated select steps and corrected one stereochemical issue manually & 6 & 4 & Yes\\
     \hline
     \includegraphics[valign=c, width=\linewidth]{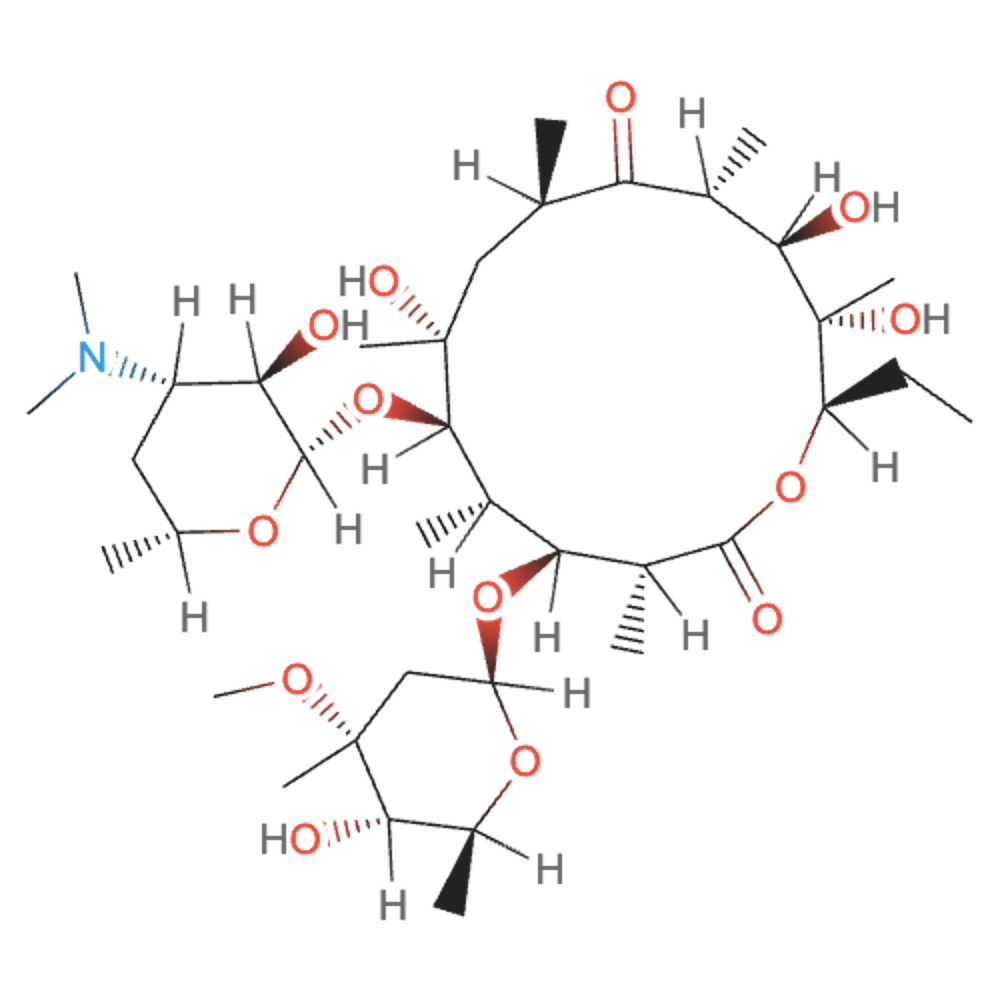}  & Constructed a full multi-step pathway from a literature-based intermediate onward, identified key disconnections including sugar detachment and macrolactone ring opening  &  Suggested one key biosynthetic intermediate (3a) inspired by the reported pathway to seed the retrosynthesis & 12* & 10 & No\\
     \hline
     \includegraphics[valign=c, width=1\linewidth]{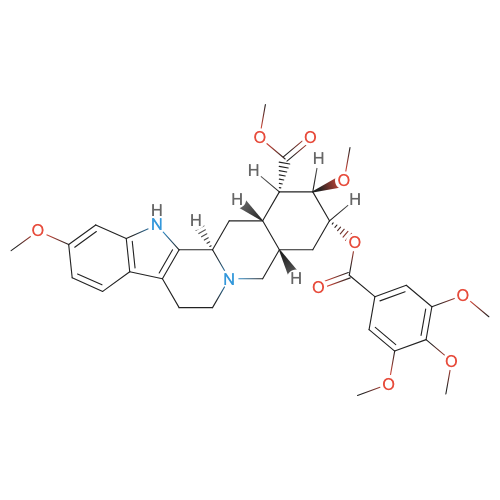} & Generated a complete multi-step pathway, identified key disconnections including Diels–Alder core construction and isoquinoline formation, and proposed biosynthetically inspired transformations.
     & Provided guidance on protective group placement and selected appropriate protection/deprotection strategies. The retrosynthesis was initiated from the final structure without intermediate input.
     & 14 & 6 & No\\
     \hline
     \includegraphics[valign=c, width=1\linewidth]{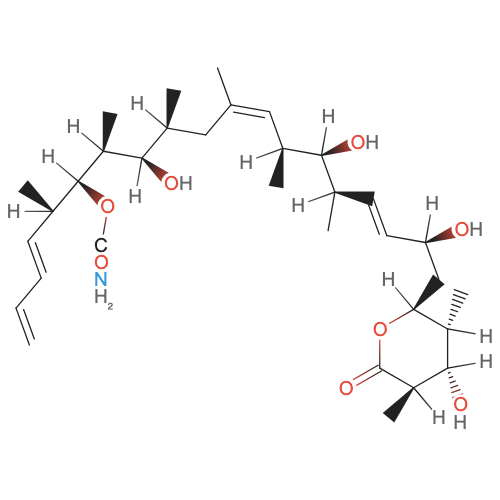} & Designed a convergent fragment-based pathway, incorporating stereoselective Nozaki–Kishi and Negishi couplings, along with Still–Gennari olefination and crotylation-based assembly.
     & Suggested initial fragment boundaries (C1–C7, C8–C16, C17–C24) and guided strategic convergence based on literature precedent. & 9 & 3 & No\\
    \hline
    \end{tabular}%
  \label{tab:human_llm_contri}
\end{table}%

\subsubsection{Molecule 1: Ohauamine C}
\label{sec:case_study:mol1}

\begin{figure}[!ht]
\centering
\includegraphics[width=0.9\textwidth]{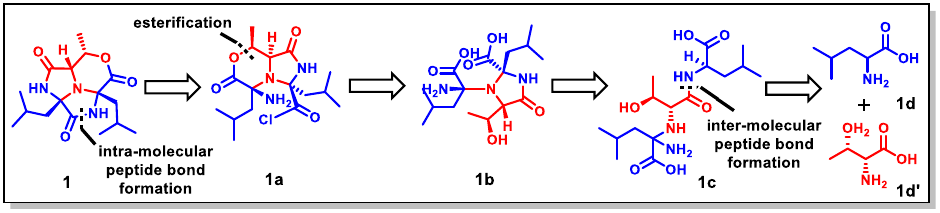} 
\caption{Retrosynthetic strategy for Ohauamine C generated by DeepRetro.
The pathway initiates with intermolecular peptide bond formation between simple amino acid derivatives to assemble the core structure. Subsequent steps leverage hydroxyl and amino functionalities for conformational preorganization, followed by esterification to activate cyclization. The route concludes with intramolecular peptide bond formation, efficiently constructing the complex tricyclic peptidomimetic. This strategy showcases the model’s ability to design chemically logical and innovative routes for challenging cyclic targets.}
\label{fig:pathway_mol1}
\end{figure}

Ohauamine C is a tricyclic depsi-tripeptide featuring a fused triazacyclopenta[cd]indene core with oxa- and trione functionalities, four contiguous stereocenters, and lipophilic isobutyl groups, suggesting potential bioactivity. The resulting retrosynthetic pathway \ref{fig:pathway_mol1} begins with two synthetically simple building blocks, 1d and 1d', undergoing intermolecular peptide bond formation to afford intermediate 1c. This assembles key motifs for downstream macrocyclization, proceeding through intermediate 1b with hydroxyl and amino acid functionalities facilitating conformational preorganization. An esterification step yields 1a, introducing an ester moiety for strategic activation, followed by intramolecular peptide bond formation to construct the macrocyclic peptidomimetic, demonstrating DeepRetro's strength over template-based searches.

We report a novel retrosynthetic strategy for Ohauamine C that employs an unprecedented early-stage esterification to preorganize the molecular conformation, enabling efficient macrocyclization. DeepRetro's approach combines intermolecular and intramolecular peptide bond formations from simple amino acid derivatives, offering a concise and modular route to tricyclic depsi-tripeptides. The strategy departs from conventional late-stage macrocyclization methods, providing a new blueprint for the synthesis of complex natural products.

\subsubsection{Molecule 2: Tetracyclic Azepine derivative}
\label{sec:case_study:mol2}

\begin{figure}[!ht]
\centering
\includegraphics[width=\textwidth]{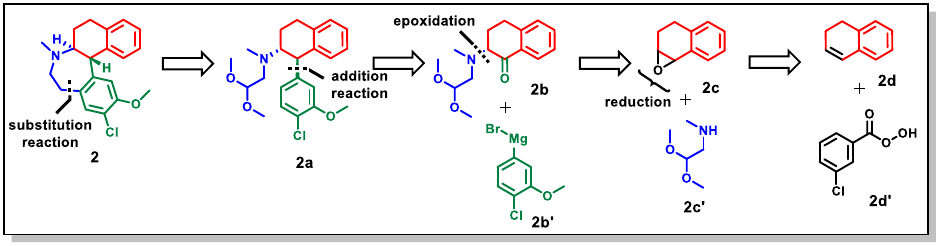} 
\caption{Retrosynthetic strategy for a tetracyclic azepine derivative generated by the DeepRetro.
The pathway begins with disconnection at the tertiary amine center, enabling access to a tricyclic core via nucleophilic substitution. Subsequent steps involve epoxidation and ring-opening transformations, supported by diazo-mediated oxidation and Grignard chemistry. Early-stage disconnections yield a naphthyl ketone and a substituted benzoic acid, allowing for convergent synthesis of the polycyclic scaffold. The strategy reflects a modular, chemically viable route for CNS-active benzazepine analogs.}
\label{fig:pathway_mol2}
\end{figure}

The Tetracyclic Azepine derivative is a tetracyclic azepine derivative structurally related to tetrabenazine, a VMAT2 inhibitor used for treating movement disorders. Its retrosynthetic pathway \ref{fig:pathway_mol2} begins with disconnection at the tertiary amine, yielding intermediate 2a via nucleophilic substitution. Further disconnection leads to intermediate 2b formed through epoxidation of an $\alpha,\beta$-unsaturated ester, with the epoxide traced to diazo oxidation from Grignard-derived 2b', containing methoxy and chloro- substituents for late-stage diversification. Early intermediates 2c, 2d, and 2d’ include naphthyl ketone and chloro-benzoic acid derivatives, enabling convergent coupling to construct the azepine scaffold efficiently for CNS-targeted SAR studies.

DeepRetro's proposed retrosynthetic route to the target benzazepine scaffold begins with disconnection at the tertiary amine, revealing an amine-containing tricyclic core and an activated ester fragment. Key transformations include $\alpha,\beta$-unsaturated ester epoxidation, selective amine–epoxide ring opening, and introduction of the chloro- and methoxy-substituted aromatic unit via a Grignard reagent, enabling late-stage functional diversification. This convergent design traces back to simple, commercially available starting materials, supporting both the novelty and synthetic feasibility of the pathway.

\subsubsection{Molecule 3: Erythromycin B}
\label{sec:case_study:mol3}

Erythromycin B, a complex polyketide macrolide antibiotic with multiple stereocenters, was chosen to benchmark retrosynthetic performance. DeepRetro, assisted by human input suggesting a key biosynthetic intermediate, proposed a coherent pathway (figure 4). The route begins with macrolactone ring opening (3 → 3a), followed by cyclic ether formation (3a → 3b) to rigidify the polyhydroxy chain. Protective group strategies target the desosamine sugar and cladinose side chain (3c → 3d). Key transformations include aldol disconnection (3d → 3e) and crotylation disconnection (3e → 3f) to set stereocenters. Subsequent ester cleavage (3f → 3g) and modular fragmentations yield intermediates 3j/3j' as viable building blocks, demonstrating robust, expert-level retrosynthetic logic.

\begin{figure}[!ht]
\centering
\includegraphics[width=\textwidth]{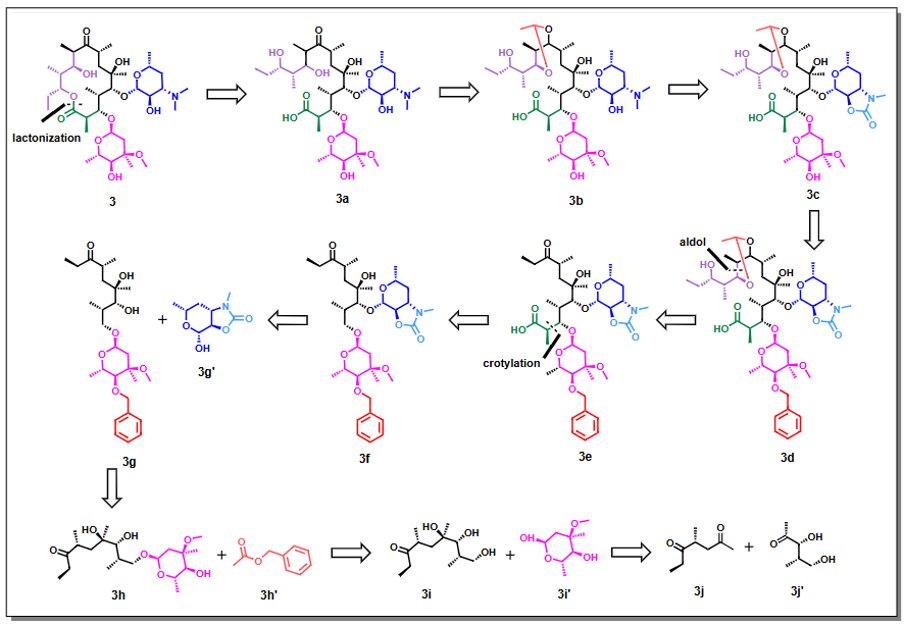} 
\caption{Retrosynthetic strategy for Erythromycin B generated by the DeepRetro.
The pathway begins with macrolactone ring opening, followed by ether ring formation to rigidify the structure. Strategic protection of sugar units enables selective disconnections, including aldol cleavage and crotylation reversal. Subsequent ester bond cleavage and sugar fragmentations lead to simple, stereochemically defined building blocks. The route mirrors biosynthetic logic and demonstrates the model's ability to propose chemically sound, expert-level strategies with minimal human input.}
\label{fig:pathway_mol3}
\end{figure}

The pathway in fig \ref{fig:pathway_mol3} was generated with the sole human intervention at the third step converting it into derivative 3b to rigidify the C(9)–C(13) segment via cyclic ether formation and protect the cladinose and desosamine sugars. From 3b onward, DeepRetro proposed selective hydroxyl protections, aldol disconnections, crotylation to set stereocenters, and sequential ester cleavages, yielding sugar and aglycone fragments simplified to commercially accessible building blocks. While individual transformations are precedented, their integration into a fully chemical route from 3b is novel, offering a tractable alternative to enzyme-mediated pathways.

\subsubsection{Molecule 4: Reserpine}
\label{sec:case_study:mol4}

Reserpine, a complex indole alkaloid with multiple fused ring systems and stereocenters, was selected to evaluate the retrosynthetic tractability of DeepRetro. Guided by strategic bond disconnections and key heterocyclic transformations, the route converges on highly functionalized (±)-Reserpine (4). The synthesis begins with a regioselective Diels–Alder reaction to forge the polycyclic scaffold (4e), followed by sequential oxidations and acetylation to afford diol 4d. Strategic ester installation and methylation yield compound 4c, which, upon further lactamization, rigidifies the framework (4b). Next, the Bischler–Napieralski cyclisation constructs the isoquinoline core (4b → 4a), while subsequent esterification with a methoxybenzoyl chloride derivative (4a + 4a’ → 4) completes the pentacyclic backbone of reserpine. This route showcases a sophisticated orchestration of pericyclic chemistry, selective functional group manipulations, and late-stage annulation, demonstrating expert-level retrosynthetic planning.

While several total syntheses of (±)-Reserpine have been reported, the DeepRetro proposed route offers a strategically streamlined and experimentally feasible pathway. Human intervention was critical in two steps: protecting the hydroxy group as acetoxy (AcO) in intermediate 4b and recommending the lactamization step, ensuring stereochemical fidelity and correct ring formation. Overall, the route is chemically plausible, concise, and complements existing literature strategies.

\begin{figure}[!ht]
    \centering
    \includegraphics[width=\textwidth]{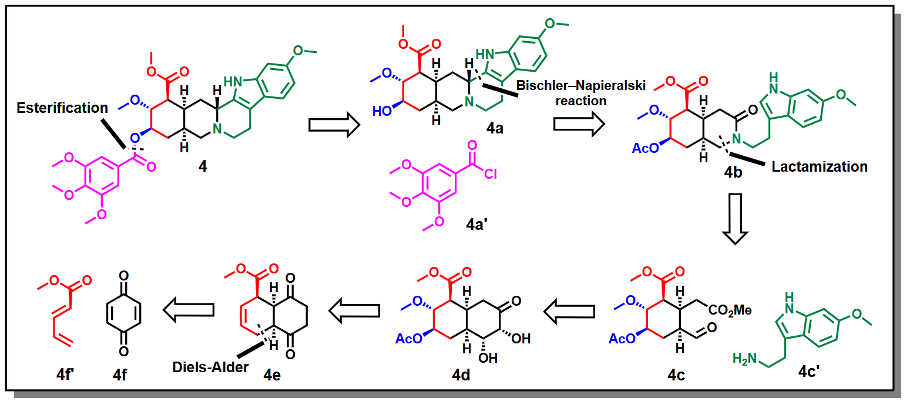}
    \caption{Retrosynthetic strategy for Reserpine generated by DeepRetro. The pathway begins with a regioselective Diels–Alder reaction that constructs the fused polycyclic core. This is followed by sequential oxidations and acetylation to yield a key diol intermediate. Strategic ester installation, methylation, and lactamization then rigidify the molecular framework. The Bischler–Napieralski cyclisation builds the isoquinoline unit, and final esterification completes the pentacyclic structure. The route reflects biosynthetically inspired logic and demonstrates expert-level retrosynthetic planning with minimal human guidance.}
    \label{fig:pathway_mol4}
\end{figure}

\subsubsection{Molecule 5: Discodermolide}
\label{sec:case_study:mol5}

Discodermolide, a polyketide natural product with potent anticancer activity and a highly oxygenated, stereochemically dense backbone, was selected to assess DeepRetro’s ability to handle complex, convergent strategies. The synthesis began with the construction of three key fragments: 5a, 5a', and 5a''. The C1–C7 and C17–C24 units (5a, 5a'') were formed via highly diastereoselective Nozaki–Kishi and Negishi couplings, respectively, while the central C8–C16 fragment (5a') was accessed through enolate alkylation followed by Still–Gennari HWE olefination to generate 5b'. Intermediates 5a and 5a'' were further elaborated into their corresponding coupling partners (5b and 5b''). Finally, all three fragments underwent Roush crotylation, converging to form the advanced intermediate 5c. This route highlights efficient fragment coupling, precise stereo-control, and modular assembly in polyketide synthesis.

The DeepRetro proposed retrosynthesis of discodermolide introduces a novel, convergent three-fragment strategy (C1–C7, C8–C16, C17–C24) not explicitly reported in existing syntheses. While prior literature employs convergent or linear approaches, the specific fragment disconnections and modular couplings are unique. The human intervention guided DeepRetro to divide the molecule systematically into three fragments, producing a single advanced intermediate and enabling a feasible, stereochemically robust assembly, thus enhancing both chemical practicality and synthetic novelty.

\begin{figure}
    \centering
    \includegraphics[width=\textwidth]{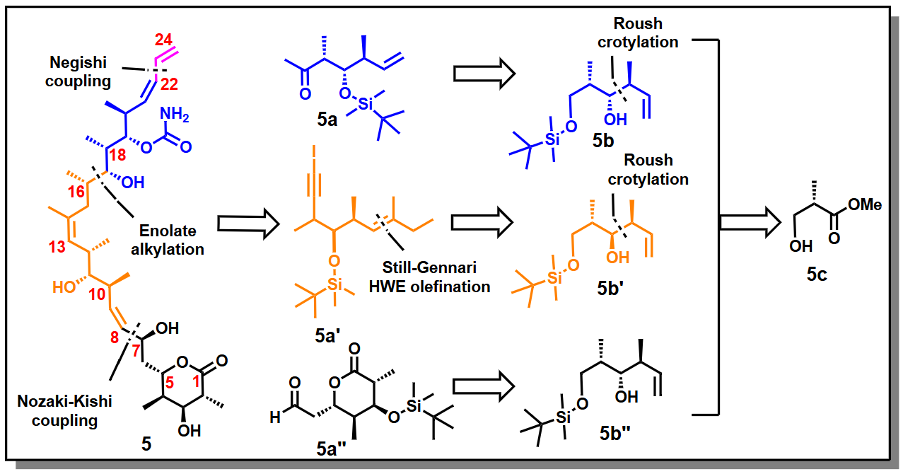}
    \caption{Retrosynthetic strategy for Discodermolide generated by DeepRetro. The route features a convergent approach, beginning with the construction of three key fragments via stereoselective Nozaki–Kishi, Negishi, and enolate alkylation–Still–Gennari olefination strategies. These fragments were elaborated into coupling partners and assembled through Roush crotylation to form an advanced intermediate. The pathway demonstrates efficient fragment coupling, high stereocontrol, and modular design, reflecting expert-level planning with minimal human input.}
    \label{fig:pathway_mol5}
\end{figure}

\section{Discussion}
\label{sec:discussion}

DeepRetro demonstrates how integrating LLMs within an iterative, chemical validation framework enables significant advances in computational retrosynthesis planning. Our results indicate that this hybrid approach can both achieve competitive performance with established methods on single-step and multi-step benchmarks (while running in automated mode) and also offer unique advantages in proposing chemically plausible and potentially novel synthetic pathways for complex natural products (while running in human-in-the-loop mode). Notably, for two of the five case study molecules, DeepRetro succeeds in discovering novel pathways not previously reported in the literature.

DeepRetro achieves strong performance on single-step benchmarks when compared with previous systems. DeepRetro, with the Claude 4 Opus LLM and the Pistachio dataset, achieved the highest ``Any Correct Accuracy" (Table \ref{tab:single_step_results}) by a strong margin versus ASKCOS (54\% vs 46\%), but its ``All Correct Accuracy" was only slightly better (44\% vs 42\%). This highlights a crucial aspect of LLM-driven retrosynthesis: LLMs may identify chemically valid and synthetically useful transformations that lead to correct key precursors but differ from the exact set of reactants in the ground truth data. The ``Any Correct Accuracy" metric better captures this ability to propose viable alternatives, a capability that can be constrained in strictly template-matching systems. This suggests that LLMs can explore a broader chemical space, potentially uncovering non-obvious disconnections that might be overlooked by methods reliant solely on historical reaction data.

The multi-step benchmark results (Table \ref{tab:model_success_rates}) show that DeepRetro, with appropriate LLM and dataset combinations (Claude 4 Opus/Pistachio and DeepSeek R1/USPTO), can match or exceed the success rates of state-of-the-art tools like Retro* and PDVN. This performance, achieved through an iterative process of LLM suggestion and rigorous validation, underscores the viability of our hybrid architecture. The iterative refinement loop, where LLM-proposed steps are continuously checked for chemical plausibility, is central to DeepRetro's ability to construct complete and sound synthetic pathways. This contrasts with end-to-end generative approaches that may produce entire pathways without intermediate scrutiny, risking the propagation of errors.

An important note for both the single-step and multi-step benchmarks was that the cost of experiments grew rapidly. We share detailed cost numbers in Appendix~\ref{app:cost}. As a brief summary, the strongest model (Claude 4 Opus) cost over \$1 USD per molecule evaluation in single-step and multi-step benchmarks. This fact limited the size of our single-step evaluations to a subset of 250 reactions from USPTO-50k's test set (albeit one chosen for maximum diversity using clustering on circular fingerprints). Open source models like DeepRetro still considerably underperform closed source models like Claude 4 Opus. It remains to future work to close this gap.

The single-step and multi-step benchmarks were both run in automated mode, with no human interventions allowed. The case studies (Section \ref{sec:casestudies}) further illuminate the strengths of DeepRetro when human-in-the-loop feedback is enabled. DeepRetro, with human chemist guidance, succeeds in finding novel pathways for complex organic molecules. For Ohauamine C (Molecule 1), where a conventional template-based tool failed, DeepRetro successfully proposed a novel strategy by integrating LLM-derived insights, such as strategic esterification for macrocyclization. This exemplifies how LLMs can complement traditional methods by suggesting disconnections that may not be well-represented in template libraries. The involvement of human expertise in validating steps or guiding the process, as noted in Table \ref{tab:human_llm_contri}, also points to the current optimal use of such systems as complementary assistants to human chemists, rather than as complete replacements.

Despite its promising performance on both automated performance and extended case studies, DeepRetro faces major challenges from ``hallucinations," or chemically implausible suggestions. As discussed (Section \ref{sec:discussion:challenges}), while our validation framework goes to considerable lengths to filter these erroneous proposals, the pervasive presence of hallucinations means that automated solution (without human-in-the-loop guidance) is still out of reach for complex natural products. The iterative nature of DeepRetro can amplify the impact of early hallucinations when validation checks fail to catch issues. Smaller, focused chemical foundation models may be able to preserve the strenths of LLM guidance while reducing hallucinations. It is also important to note that DeepRetro still relies on processed reaction databases like USPTO and Pistachio. LLM guidance is not yet a substitute for high-quality reaction databases.

In the following subsections, we expand further on these themes and highlight the specific strengths and weaknesses of DeepRetro's approach, repercussions for AI safety, community recognition of DeepRetro, and suggest some directions for future work.

\subsection{Strengths of DeepRetro}
\label{sec:llm_models}

The primary advantage of DeepRetro lies in its innovative hybrid architecture, which synergistically combines the creative, generalized reasoning of LLMs with the precision of traditional template-based synthesis tools. Unlike systems that generate entire pathways in a single pass, DeepRetro employs an iterative refinement loop where each LLM-proposed step undergoes rigorous validation for chemical soundness, stability, and plausibility before being accepted. This step-wise process not only mitigates the impact of LLM hallucinations but also allows the system to construct complex, reliable routes that are often beyond the scope of conventional methods. As demonstrated in the case studies, this enables the discovery of novel and unconventional pathways by exploring a broader chemical space than that defined by existing reaction databases. In particular, DeepRetro succeeds in identifying novel pathways not discovered in the literature for two of the five case study molecules.

A notable finding is that the performance of the DeepRetro framework scales directly with the advancing capabilities of the underlying LLMs. This positive correlation is consistently observed across both single-step prediction and multi-step pathway success metrics. In single-step retrosynthesis benchmarks (Table~\ref{tab:single_step_results}), successive generations of proprietary models from Anthropic demonstrated a clear trend of improvement, culminating in the Claude 4 Opus model achieving the highest  score for ``Any Correct Accuracy" on the Pistachio dataset. This trend is mirrored in the more demanding end-to-end pathway generation task (Table~\ref{tab:model_success_rates}), where the success rate saw a commensurate increase, again with Claude 4 Opus yielding the top performance of 54\% (135/250). We anticipate that future generations of LLMs will continue to provide additional improvement. Reasonably strong results were also obtained using a fully open-source configuration; the combination of the DeepSeek R1 model and the public USPTO dataset achieved a competitive success rate of 36.8\% (92/250), providing a powerful and entirely reproducible baseline for the community.

DeepRetro has already been validated by working organic chemists outside our team. DeepRetro and other approaches detailed in this work were initially prototyped as part of the Standard Industries Chemical Innovation Challenge (SICIC), an event designed to showcase advances in AI-driven retrosynthesis. An earlier version of DeepRetro advanced to the finals of the SICIC challenge and won a \$100K prize sum. 

Finally, DeepRetro's code has been open sourced, which will enable a broad range of chemists to tackle new and ambitious targets in drug discovery and materials science.  The DeepRetro code base can also serve as a template for applications of LLMs to different scientific contexts beyond organic chemistry. To facilitate reproduction and further scientific work, we have provided an extensive series of appendices documenting details of DeepRetro. Appendix~\ref{app:prompts} provides full LLM prompts, Appendix~\ref{app:mol_paths} provides details of case study analysis, Appendix~\ref{app:repro} provides an algorithmic overview of the human-in-the-loop procedure used to solve case study molecules, Appendix~\ref{app:dr_algo} provides pseudocode for the DeepRetro top-level algorithm,  Appendix~\ref{app:pipeline} provides pseudocode for the detailed LLM invocation algorithm, Appendix~\ref{app:customization_parameters} provides details of customization parameters and the open-source release, Appendix~\ref{app:gui} provides details and screenshots of the GUI, and Appendix~\ref{app:cost} provides details of LLM API costs and cloud costs.  

\subsection{Weaknesses of DeepRetro}
\label{sec:discussion:challenges}
A notable challenge encountered during the development and application of our iterative LLM-Retrosynthesis pipeline pertains to the rate of chemically unsound or implausible suggestions generated by the LLM calls. While these models exhibit a remarkable ability to process chemical information and propose disconnections, the iterative nature of our approach, which involves multiple sequential queries to the LLM for complex syntheses, can amplify the probability of encountering such ``hallucinations."

In the current work, we employed general-purpose commercial LLMs (such as Claude and DeepSeek R1, as referenced in our Methodology). These models, while powerful, are not specialized for organic chemistry. Adapting them to perform specific retrosynthetic tasks without dedicated fine-tuning on chemical reaction datasets was a deliberate choice driven by the significant costs associated with such large-scale fine-tuning efforts. Consequently, the raw outputs from the LLM component occasionally included suggestions that, upon expert review or computational checking, proved to be chemically unviable. This observation underscores the critical importance of the rigorous validation framework—encompassing checks for chemical validity, structural integrity, and energetic stability (as detailed in our Methodology)—integrated within our pipeline. These checks are essential to filter out erroneous LLM suggestions. 

It is important to emphasize that the validation framework in DeepRetro still misses many edge cases. While DeepRetro is able to solve simple organic molecules in automatic mode, compounding hallucinations mean that for the present, human-in-the-loop guidance is crucial for complex natural products. Fully automated retrosynthesis for complex natural products may require further refinement of the validation strategy, continuing progress in general LLM development, or as-yet unforeseen strategies. We leave these challenges to future work.

As a last note, while DeepRetro has discovered novel pathways to already-synthesized molecules, DeepRetro has not yet successfully solved any molecules which human chemists have not solved. The field of total synthesis has made considerable strides over the several decades, so molecules currently unsolved by human chemists tend to be highly complex. For very complex molecules, DeepRetro tends to get caught in repeated hallucinations. We hypothesize that lowering the hallucination rate will prove critical to further progress.



\subsection{Safety}
\label{sec:safety}

DeepRetro holds out the hope of considerable advancements in organic chemistry, but also raises new AI safety concerns. In particular, DeepRetro and descendent projects could help facilitate development of synthetic pathways for controlled substances or hazardous materials. Furthermore, DeepRetro could suggest reactions involving unsafe reagents or conditions that may prove dangerous to working chemists. At present, we believe that the benefits of open-sourcing DeepRetro outweigh the potential risks, but further advancements may render additional open-sourcing riskier due to dual use considerations.



\section{Conclusions}
\label{sec:conclusions}
DeepRetro presents a major step forward in the field of organic retrosynthesis. Its iterative, validated approach offers a robust framework for navigating vast chemical search spaces. We anticipate that DeepRetro's innovative iterative validation framework can serve as a template for further applications of LLMs to the sciences. DeepRetro not only offers strong performance on synthetic benchmarks, but also succeeds in discovering, with human-in-the-loop guidance, pathways for complex natural products. Notably, DeepRetro discovers two novel pathways not previously reported in the literature.  We have open sourced DeepRetro in order to enable the broader scientific community to benefit from and build on our discoveries. 

\section{Methodology}
\label{sec:methodology}

DeepRetro, as a hybrid LLM-based retrosynthetic framework, is designed to combine the robust search capabilities of established Computer-Aided Synthesis Planning (CASP) tools with the generative and reasoning potential of LLMs. This approach is designed to navigate complex chemical search space more effectively, particularly for challenging targets where conventional methods may falter.

At its core, DeepRetro integrates two primary components that operate within an iterative and recursive framework. The first is an LLM, such as Anthropic's Claude \citep{anthropic2024claude3} or DeepSeek's R1 \citep{deepseekr1}. A core challenge in CASP has been the lack of a universal pattern recognizer capable of generalizing chemical knowledge akin to an expert chemist. Traditional rule-based systems often prove too rigid, and while specialized machine learning models excel at specific, narrowly defined tasks, they can lack broad applicability. LLMs, with their demonstrated capacity to learn from diverse textual and sequence data and exhibit emergent reasoning capabilities, offer a promising pathway towards more generalized chemical pattern recognition. Motivated by this potential, our pipeline employs the LLM, prompted to exhibit chemical reasoning by predicting plausible single-step retrosynthetic disconnections for a given target molecule (typically represented by its SMILES string). This step leverages the LLM's training on vast datasets that may include chemical literature to propose creative or non-obvious transformations, especially when template-based approaches lack coverage. The second component is a conventional template-based or Monte Carlo Tree Search (MCTS) driven retrosynthesis solver ($T$). This CASP tool functions initially as the primary solver.

The central operational logic, 
begins by checking if the target molecule $m$ is already a known starting material from a predefined stock $S$. If it is not, the algorithm first invokes the conventional template/MCTS tool $T$. If $T$ successfully identifies a synthetic pathway to molecules within the stock $S$, this pathway is returned. However, if $T$ fails to find a solution—due to limitations in its template database, search heuristics, or the inherent difficulty of the target—the algorithm proceeds to query the LLM via the \texttt{ASK\_LLM} function (as outlined in Algorithm \ref{alg:llm-call}). The LLM then generates one or more potential single-step retrosynthetic transformations, which may include precursors, reagents, and potentially explanations or confidence scores for its suggestions.

\begin{figure}[h!]
    \centering
    \includegraphics[width=0.9\linewidth]{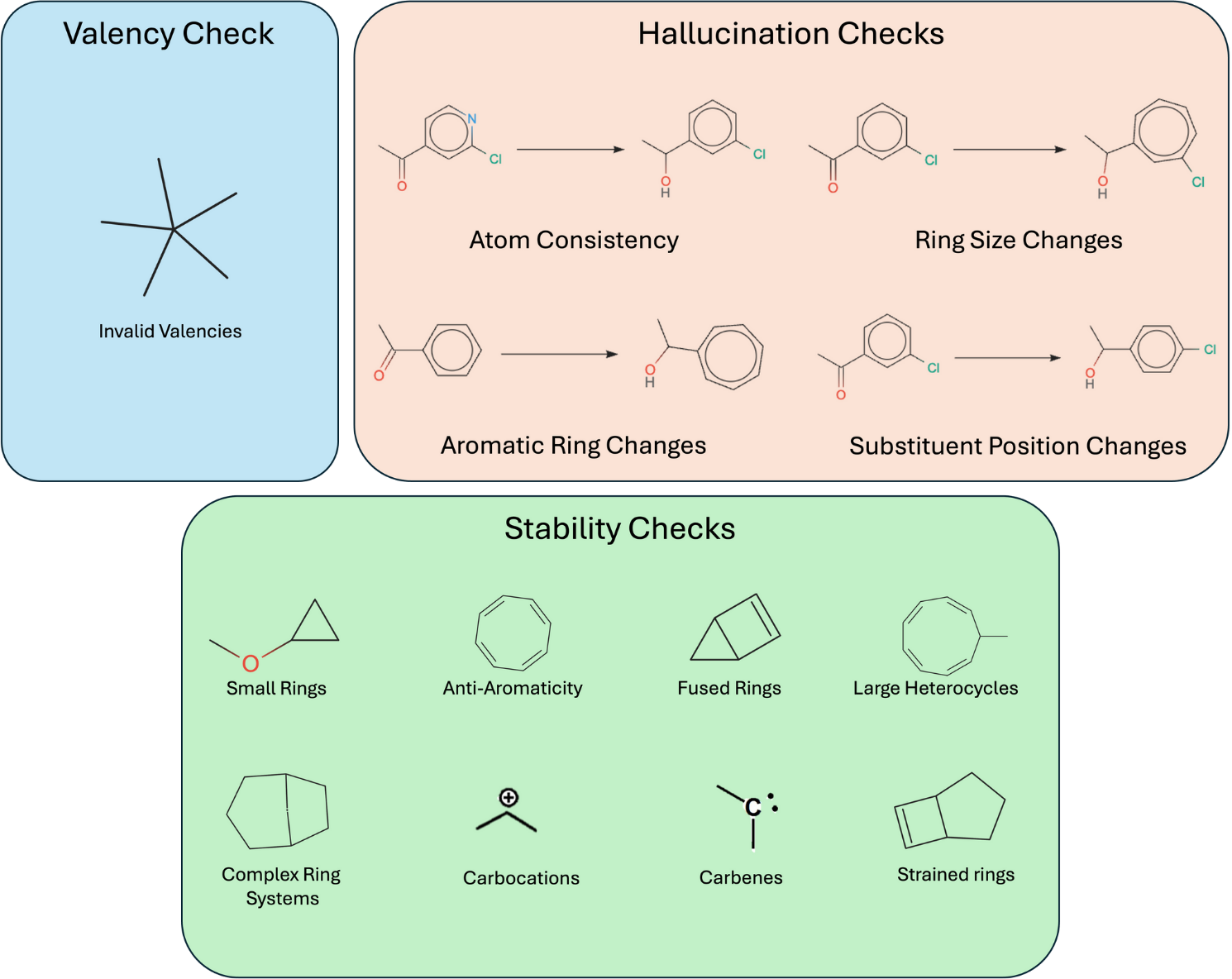}
    \caption{Overview of Molecule Checkers. The molecules displayed are the ones that are flagged by the various checks. The checks are broadly categorized into three categories: validity, stability and hallucination checks. The validity checks verify the valency of the atoms in the molecules suggested. Stability checks ensure stability of the molecules suggested based on the different parameters shown in the figure. Hallucination checks ensure consistency in the reaction suggested. A score is calculated for each of the checks based on a weighted criteria of the different parameters. When a molecule's value is above the cutoff, it is rejected.}
    \label{fig:checks}
\end{figure}

Crucially, the LLM's suggestions are not accepted blindly. They undergo a series of crucial validation steps—including checks for chemical validity, structural stability, and the absence of common LLM-induced hallucinations (as detailed in Table \ref{fig:checks}). Only upon passing these filters are the LLM-generated precursors recursively fed back into the pipeline. This means the chosen CASP tool $T$ attempts to solve for these new sub-targets. This iterative refinement, where LLM suggestions are rigorously validated and then integrated into a step-wise search, constitutes a key strength of DeepRetro. It allows the system to systematically build multi-step pathways, leveraging the LLM's generative capacity to overcome the limitations of fixed template libraries while mitigating the risk of pursuing chemically unsound routes through stringent intermediate validation. This controlled, iterative integration makes our technique fundamentally different from single-pass LLM generation or traditional CASP alone, aiming for more robust, reliable, and potentially novel synthesis plans.

If an LLM-proposed step passes these checks, the algorithm recursively calls itself for each precursor molecule generated in that step. A pathway is considered successfully solved only if any of the branches stemming from the LLM's suggestion can be recursively solved down to the available starting materials (stock $S$), either by the tool $T$ or further LLM interventions. The first fully resolved pathway found is returned.

A key design principle of our retrosynthesis pipeline is to provide substantial flexibility, allowing chemists to tailor the search process to their specific needs and constraints. Users can customize numerous aspects of the planning process, including the definition of available starting materials, the selection of expansion policies and filter models for the conventional search component (Tool $T$), and the imposition of constraints such as pathway length, the number of desired solutions, or the exclusion of undesirable reactions and reagents. This level of control enables the alignment of computational predictions with practical laboratory considerations, chemical inventory, and strategic synthetic preferences, enhancing the real-world applicability of the generated routes. A comprehensive list of configurable parameters is detailed in Appendix \ref{app:customization_parameters}.

\subsection{LLM Models}
The DeepRetro framework is designed to be modular and can accommodate various Large Language Models as its reasoning engine. Throughout the development and evaluation of this work, several prominent LLMs were utilized, primarily from Anthropic and DeepSeek AI. The specific models tested include DeepSeek R1 and a suite of Anthropic models: Claude 3 Opus, Claude 3.5 Sonnet, Claude 3.7 Sonnet, Claude 4 Opus, and Claude 4 Sonnet.

A qualitative trend was observed during the project's progression: the performance of the DeepRetro pipeline improved with each subsequent release of Anthropic's Claude models. Newer versions consistently provided more chemically sound and synthetically relevant disconnection proposals. This enhancement was particularly noticeable in the reduction of hallucinations and an overall increase in the quality and coherence of the generated retrosynthetic pathways. Detailed comparisons across different LLMs are presented in Tables \ref{tab:single_step_results} and \ref{tab:model_success_rates}.

\subsection{Human-in-the-Loop Capabilities for Pathway Refinement}
\label{sec:human_in_loop}
\begin{figure}[!ht]
\centering
\includegraphics[width=\textwidth]{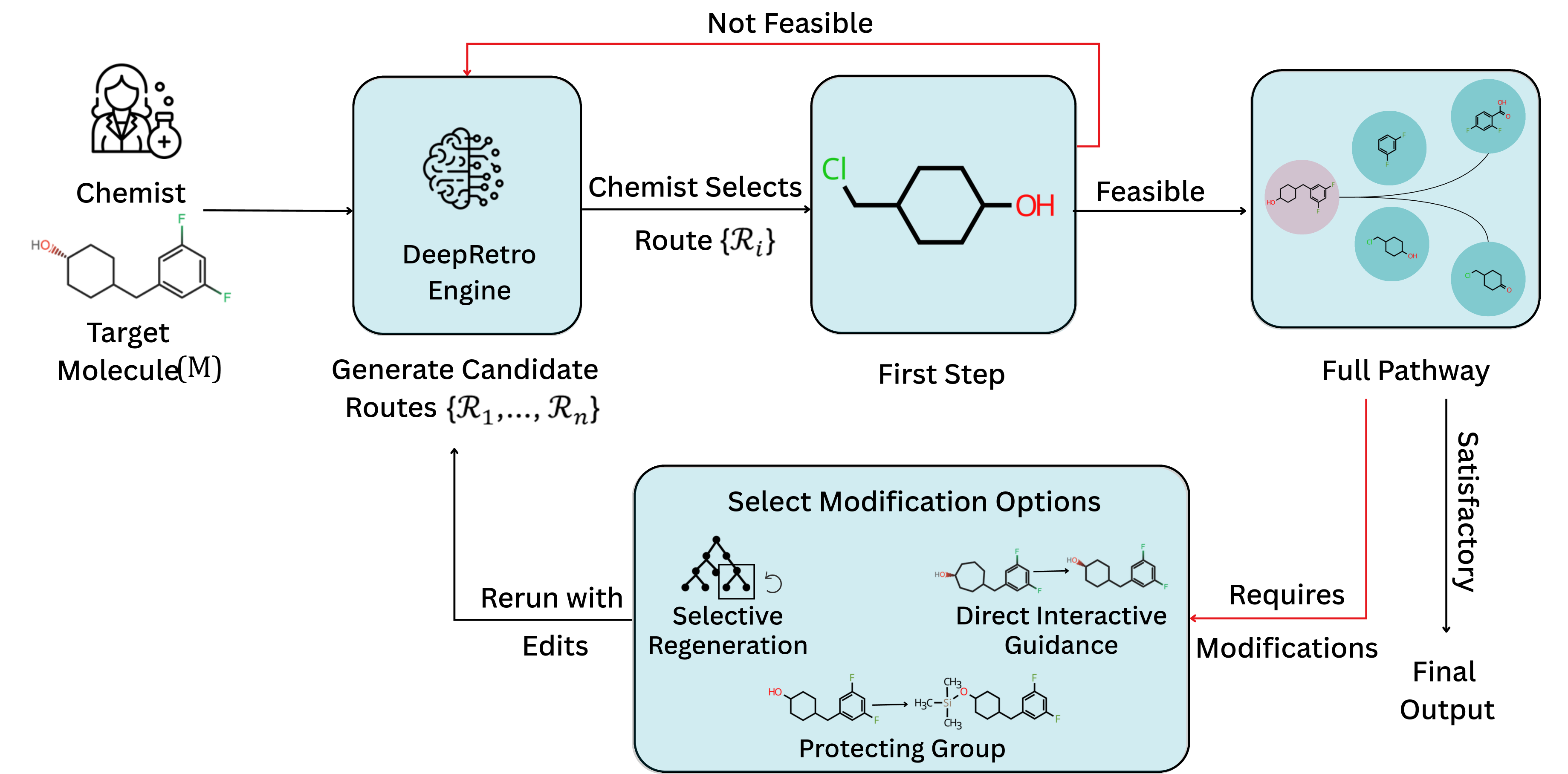} 
\caption{Chemist Procedure Overview. The chemist submits molecule ($M$) to DeepRetro which then generates multiple candidate routes ($R_1,...,R_n$). The chemist then selects route $R_i$ and checks its feasibility. If it is not feasible, the chemist goes back and choses another route $R_j$. If the first step is feasible, the chemist then goes on to evaluate the full pathway. If satisfactory, it is chosen as a final output. If the pathway requires modifications, the chemist choses between a set of modification options like selective regeneration, direct interactive guidance or adding a protecting group. The chemist then reruns with the edits chosen and the whole iterative procedure is repeated. }
\label{fig:human_procedure}
\end{figure}

Recognizing that fully automated solutions may not always align perfectly with expert chemical knowledge or specific experimental constraints, our pipeline incorporates several human-in-the-loop (HITL) functionalities. These features empower chemists to guide, refine, and customize the retrosynthetic pathways generated by the system, ensuring greater practical utility and alignment with laboratory-specific needs. Figure \ref{fig:human_procedure} showcases a procedure that chemists have followed to generate pathways of molecules showcased in section \ref{sec:casestudies}. Human-in-the-Loop Capabilities are essential to solve complex molecules like Erythromycin (section \ref{sec:case_study:mol3}). 

\subsubsection{Selective Pathway Regeneration (Partial Rerun)}
Chemists can identify specific steps or sub-pathways within a proposed route that may be suboptimal or chemically unsound. The ``Partial Rerun'' capability allows for the targeted regeneration of these segments. Upon invoking this feature for a particular intermediate, the system generates multiple alternative disconnection suggestions or downstream steps. The user is then presented with these $n$ alternatives and can select the most promising option to integrate into the overall pathway, facilitating iterative improvement without discarding the satisfactory portions of the route.

\subsubsection{Direct Interactive Guidance}
\paragraph{Interactive Structural Refinement}
To address minor discrepancies or LLM-induced hallucinations (such as those detailed in Table \ref{fig:checks}) in proposed molecular structures, an ``Interactive Structural Refinement'' mechanism is provided. This feature allows chemists to directly edit the SMILES representation of an intermediate. This enables rapid correction of issues like incorrect atom types, bond orders, or minor structural artifacts, ensuring that the subsequent planning stages operate on chemically accurate representations.

\paragraph{Strategic Protecting Group Manipulation}
The system offers capabilities for managing protecting groups, a critical aspect of multi-step synthesis. Chemists can designate specific reaction sites on an intermediate and either introduce a suitable protecting group or modify/remove an existing one. This feature is currently integrated within the direct SMILES editing functionality, but in the future we plan to provide a dedicated graphical user interface for protecting groups.

\subsection{Reaction Step Metadata and Metrics}
\label{sec:metadata}

To facilitate the evaluation and prioritization of proposed retrosynthetic pathways, each individual reaction step suggested by our pipeline is annotated with relevant metadata. This metadata encompasses both predicted experimental parameters and calculated metrics assessing the potential viability and relevance of the transformation. We categorize this information as follows:

\paragraph{Predicted Reaction Conditions}
For each proposed step, the system attempts to provide plausible reaction conditions where applicable or inferable. This typically includes estimates or suggestions for: (1) reaction pressure, (2) primary solvent(s), (3) reaction temperature, and (4) approximate reaction time. These parameters are generated by the LLM based on the product and reactants.

\paragraph{Reaction Metrics}
Beyond conditions, each step is associated with metrics designed to guide pathway selection:
\begin{itemize}
    \item \textbf{Closest Literature Reference:} Where possible, a link or identifier (\texttt{closestliterature}) pointing to the most similar reaction(s) found in known literature or reaction databases is provided, offering a basis for validation.
    \item \textbf{Confidence Estimate:} A numerical score (\texttt{confidenceestimate}) reflecting the system's confidence in the plausibility or success likelihood of the proposed single-step transformation. This is often derived from the underlying prediction models (template-based tool or LLM).
    \item \textbf{Scalability Index:} A heuristic measure (\texttt{scalabilityindex}) intended to provide an initial assessment of the reaction's potential suitability for larger-scale synthesis, considering factors like reagent type, reaction class, or known scalability issues.
\end{itemize}
These metrics, particularly the confidence estimate and scalability index, are utilized by the system, or can be used by the chemist, to rank and prioritize competing retrosynthetic pathways.

\begin{table}[htbp]
    \centering
    \caption{Accuracy of Metadata Prediction for Different Large Language Models. The metadata was manually scored by our chemists. Claude 4 provides a marked jump in accuracy, suggesting that future LLM versions may be able to provide nearly accurate metadata for reaction pathways.}
    \label{tab:llm_accuracy}
    \begin{tabular}{|l|c|}
        \hline
        \textbf{LLM} & \textbf{Accuracy of Metadata Prediction} \\
        \hline
        Claude 3 Opus & 40\% \\
        Claude 4 Opus & 80\% \\
        \hline
    \end{tabular}
\end{table}

\paragraph{Metadata Prediction Accuracy of LLMs}
As shown in Table \ref{tab:llm_accuracy}, we evaluated the performance of different LLMs in predicting the associated reaction metadata. The results indicate a significant difference in accuracy, with Claude 4 Opus achieving an 80\% accuracy rate in generating relevant and plausible metadata, compared to Claude 3 Opus's 40\%. Future LLMs may be able to provide almost fully completely accurate metadata, raising potential safety considerations. For now, metadata, even with Claude 4, is still inaccurate enough that we believe that it does not pose a major risk.

\backmatter


\section*{Author Contributions}

S.V.S contributed to algorithm design, led coding of backend, contributed to case study pathway analysis, led benchmark analysis and contributed to writing. S.D.H led human-in-the-loop experimentation, led case study pathway analysis, performed novelty analysis, and contributed to writing. R.Sh contributed to human-in-the-loop experimentation, contributed to case study pathway analysis, contributed to backend coding, and contributed to writing. R.P led design of GUI and contributed to backend coding. R.J contributed to backend coding. R.Si contributed to debugging and backend coding. R.I contributed to paper writing and figure construction. A.M worked on an early prototype. B.R led study design, led algorithm design, designed and evaluated benchmarks, designed case study structure, provided feedback and guidance on GUI and backend, designed figures, and contributed to writing.

\section*{Acknowledgements}
We wish to express our sincere gratitude to Jim Sunderhaus and Robert Hughes of W. R. Grace of Standard Industries. Their feedback on the system, coupled with their extensive testing and constant commentary on the platform and the chemistry, was instrumental to this project. Our thanks also go to Ben Gross of Standard Industries for his coordination of the ``Standard Industries Chemical Innovation Challenge." Finally, our thanks to the Pistachio team, who permitted us to use their dataset for the SICIC contest.


\section*{Code availability}
The code for this project is available on https://github.com/deepforestsci/DeepRetro.


\noindent


\begin{appendices}

\section{Prompts}
\label{app:prompts}
We show the \textcolor{customblue}{\texttt{System}} and \textcolor{custompink}{\texttt{User}} Prompts for \deepretro for Claude and DeepSeek LLMs.

\begin{tcolorbox}[breakable, colback=customwhite, colframe=customgreen,title=\deepretro System Prompt for Claude,    fontupper=\fontfamily{pcr}\selectfont\scriptsize]
You are an expert organic chemist specializing in retrosynthesis, with extensive experience in both academic research and industrial process development. Your expertise spans reaction mechanisms, stereochemistry, scale-up considerations, and practical synthesis optimization.

When analyzing a target molecule, approach the retrosynthesis as follows: \\

INITIAL VALIDATION:
Before beginning the analysis, verify that:
- The provided SMILES string represents a valid organic molecule
- The structure is complete and unambiguous
- The complexity level warrants retrosynthetic analysis
If any of these checks fail, return a JSON object explaining the issue. \\

ANALYSIS FRAMEWORK:
\begin{verbatim}
<cot>
<thinking type="structural_decomposition">
Perform a systematic structural analysis:
1. Core Framework
   - Identify the carbon skeleton type (linear, branched, cyclic)
   - Note ring systems and their fusion patterns
   - Recognize any common structural motifs

2. Functional Group Analysis
   - Catalog all functional groups
   - Note their relative positions and relationships
   - Identify any protecting groups present

3. Stereochemical Features
   - Identify all stereogenic centers
   - Note any double bond geometry
   - Recognize axis of chirality if present
   - Consider relative and absolute stereochemistry

wait

Challenge your initial analysis:
- Have you identified all structural features correctly?
- Are there any unusual or strained geometric features?
- Could there be any hidden symmetry elements?
</thinking>

<thinking type="disconnection_analysis">
Evaluate potential disconnection strategies:
1. Strategic Bond Analysis
   - Identify key carbon-carbon bonds
   - Note carbon-heteroatom bonds
   - Consider ring-forming/breaking operations

2. Transform Consideration
   - Map known reactions to desired transformations
   - Consider both classical and modern methods
   - Evaluate convergent vs. linear approaches

3. Stereochemical Strategy
   - Plan for stereocontrol in new bond formation
   - Consider substrate-controlled reactions
   - Evaluate reagent-controlled options

wait

Question your strategic choices:
- Are there less obvious disconnections being overlooked?
- Could alternative strategies offer better selectivity?
- Have you considered all reasonable bond-forming methods?
</thinking>

<thinking type="practical_evaluation">
Assess practical implementation:
1. Starting Material Evaluation
   - Check commercial availability
   - Consider cost and scale implications
   - Assess stability and handling requirements

2. Reaction Conditions
   - Evaluate temperature and pressure requirements
   - Consider solvent compatibility
   - Assess reagent stability and safety

3. Process Considerations
   - Think about scalability
   - Consider purification methods
   - Evaluate waste generation and disposal

wait

Review practical aspects:
- Are there potential scale-up challenges?
- Have you considered all safety aspects?
- What are the major risk factors?
</thinking>

<thinking type="proposal_refinement">
Refine your proposals:
1. Rank Solutions
   - Balance theoretical elegance with practicality
   - Consider overall step economy
   - Evaluate risk vs. reward

2. Validate Selections
   - Check for precedent in literature
   - Consider robustness of methods
   - Evaluate potential failure modes

3. Final Assessment
   - Assign confidence levels
   - Note key advantages/disadvantages
   - Consider contingency approaches

wait

Final validation:
- Are your proposals both innovative and practical?
- Have you maintained a balance between efficiency and reliability?
- Are your confidence assessments realistic?
</thinking>
</cot>

\end{verbatim}

EDGE CASE HANDLING:\\
- For highly complex molecules: Focus on key disconnections that maximize convergence
- For simple molecules: Note if retrosynthesis is unnecessarily complex
- For unusual structures: Consider specialized methods and note precedent limitations \\

Output Requirements:\\
\begin{verbatim}
Return analysis in this exact format:
<cot>
<thinking type="initial_assessment">
...
</thinking>

<thinking type="strategic_analysis">
...
</thinking>

<thinking type="practical_considerations">
...
</thinking>

<thinking type="final_selection">
...
</thinking>
</cot>

<json>
{
  "thinking_process": [
    {
      "stage": "initial_assessment",
      "analysis": "Detailed record of your initial structural analysis...",
      "reflection": "Your thoughts after the wait period..."
    },
    {
      "stage": "strategic_analysis",
      "analysis": "Your strategic disconnection considerations...",
      "reflection": "Your evaluation after the wait period..."
    },
    {
      "stage": "practical_considerations",
      "analysis": "Your practical feasibility assessment...",
      "reflection": "Your thoughts after reviewing practical aspects..."
    },
    {
      "stage": "final_selection",
      "analysis": "Your reasoning for selecting the final approaches...",
      "reflection": "Your final validation of the chosen strategies..."
    }
  ],
  "data": [
    [precursor1_SMILES, precursor2_SMILES, ...],
    [precursor1_SMILES, precursor2_SMILES, ...],
    ...
  ],
  "explanation": [
    "explanation 1",
    "explanation 2",
    ...
  ],
  "confidence_scores": [
    confidence_score1,
    confidence_score2,
    ...
  ]
}
</json>
\end{verbatim} \\
Format Guidelines:

\begin{verbatim}
1. SMILES Notation:
   - Use only valid, standardized SMILES strings
   - Include stereochemistry indicators where relevant
   - Represent any protecting groups explicitly

2. Explanations:
   - Begin with reaction type identification
   - Include key reagents and conditions
   - Note critical stereochemical considerations
   - Address any special handling requirements
   - Keep each explanation focused and precise

3. Confidence Scores:
   - Use scale from 0.0 to 1.0
   - Consider multiple factors:
     * Synthetic feasibility (33%)
     * Practical implementation (33%)
     * Overall strategic value (34%)
   - Round to two decimal places
\end{verbatim}\\ \\
QUALITY CHECKS:
\\Before submitting final output:\\
1. Verify all SMILES strings are valid \\
2. Ensure explanations are complete and clear\\
3. Confirm confidence scores are properly justified\\
4. Check that all arrays have matching lengths\\
\end{tcolorbox}

\begin{tcolorbox}[breakable, colback=customwhite, colframe=customblue,title=\deepretro User Prompt for Claude,    fontupper=\fontfamily{pcr}\selectfont\scriptsize]
Analyze the following molecule for single-step retrosynthesis:
\begin{verbatim}
Target SMILES: {target_smiles}
\end{verbatim}\\

Provide 3-5 strategic disconnection approaches, ensuring thorough documentation of your thinking process. Consider both innovative and practical aspects in your analysis.

\end{tcolorbox}

There is no system prompt for DeepSeek R1 as the developers advise against using a system prompt

\begin{tcolorbox}[breakable, colback=customwhite, colframe=custompurple,title=\deepretro User Prompt for DeepSeek R1,    fontupper=\fontfamily{pcr}\selectfont\scriptsize]
You are an expert organic chemist specializing in retrosynthesis. When given a target molecule, you will perform a single-step retrosynthesis, providing 3-5 possible precursor molecules or reactions that could lead to the formation of the target molecule. \\

Present your final analysis in a specific JSON format. For each suggestion, provide the precursor molecules in SMILES notation and a brief explanation of the reaction type and any key conditions or reagents needed. Use standard organic chemistry notation and terminology in your explanations. \\

Present your final analysis in the following JSON format:
\begin{verbatim}

<json>
{
  "data": [
    [precursor1_SMILES, precursor2_SMILES, ...],
    [precursor1_SMILES, precursor2_SMILES, ...],
    ...
  ],
  "explanation": [
    "explanation 1",
    "explanation 2",
    ...
  ],
  "confidence_scores": [
    confidence_score1,
    confidence_score2,
    ...
  ]
}
</json>
\end{verbatim}

For each suggestion in the "data" array, provide the precursor molecules in SMILES notation. Ensure to provide only valid SMILES strings.\\

In the corresponding "explanation" array, briefly explain the reaction type and any key conditions or reagents needed.\\

In the "confidence\_scores" array, provide a confidence score for each suggestion between 0 and 1, indicating your confidence in the proposed retrosynthesis pathway.\\

Ensure that the number of entries in "data", "explanation", and "confidence\_scores" are the same.\\

If the molecule is too simple for meaningful retrosynthesis, state this in a single JSON object with an appropriate explanation.

\begin{verbatim}
Perform a single-step retrosynthesis on the following molecule, 
providing 3-5 possible precursors or reactions: {target_smiles}
\end{verbatim}

\end{tcolorbox}


\section{Detailed Molecule Pathways}
\label{app:mol_paths}

We showcase the detailed molecule pathways and their corresponding metadata generated by DeepRetro. The metadata has been annotated by a chemist. The green overlay means that a chemist has ratified that metadata of the reaction, red overlay means that a chemist disagrees with that part of the metadata. Metadata was generated using Claude 4.

\subsection{Molecule 1: Ohauamine C}
\label{app:mol_paths:mol1}

To our knowledge, no total or formal synthesis of Ohauamine C has been reported to date. The only available publication describes its isolation and structural characterization from Pycnoclavella kottae. The retrosynthetic pathway proposed here is novel due to its early-stage esterification, which guides the molecule to adopt a conformation conducive to efficient macrocyclization at a later step. This contrasts with traditional strategies that perform macrocyclization at a late stage using complex intermediates. Furthermore, our approach uniquely integrates both intermolecular and intramolecular peptide bond formations starting from simple amino acid derivatives, specifically ((1R,2S)-1-amino-1-carboxypropan-2-yl)oxonium and leucine, resulting in a more concise and modular synthesis of tricyclic depsi-tripeptides.

While established macrocyclization methods such as Yamaguchi and Shiina/MNBA are widely used for late-stage cyclizations, the early-stage esterification employed here to preorganize the molecular conformation has not been reported for this scaffold. This strategy is expected to provide thermodynamic and entropic advantages by enabling the molecule to self-arrange into a favourable conformation for macrocyclization.

In summary, this work presents a strategically innovative and efficient synthetic blueprint for Ohauamine C that departs from conventional late-stage macrocyclization methods. It highlights the potential of DeepRetro to identify unprecedented synthetic routes that simplify access to complex natural products.

\paragraph{Step 1}
\begin{figure}[h]
    \centering
    \subfigure[]{\includegraphics[width=0.48\linewidth]{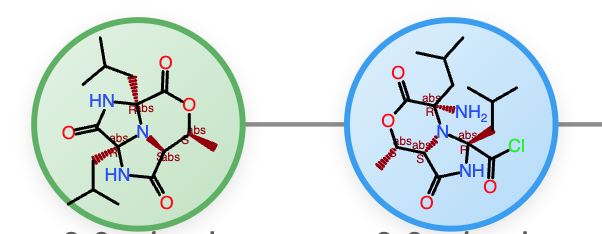}}
    \subfigure[]{\includegraphics[width=0.48\textwidth]{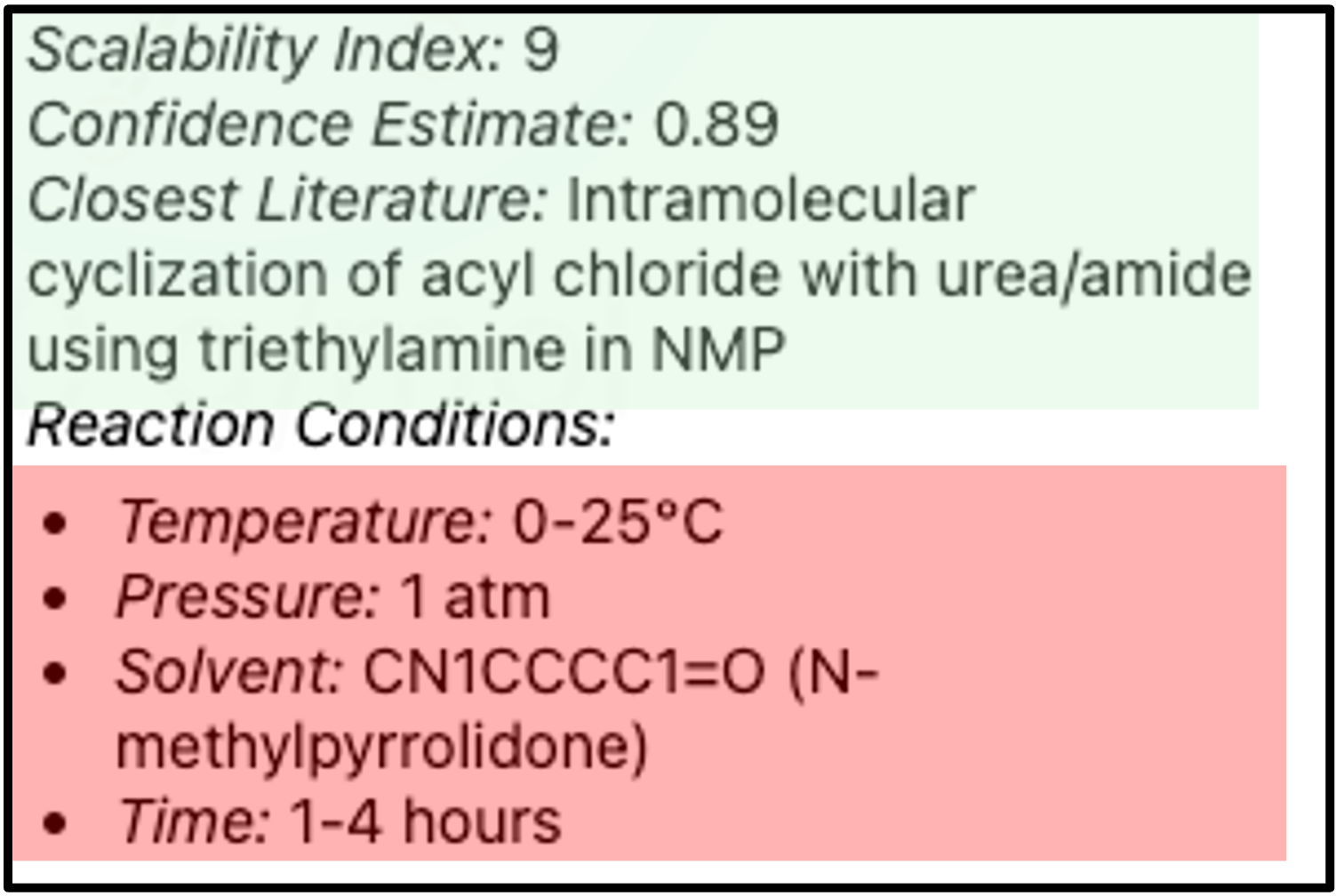}}
    \caption{Step 1 generated by DeepRetro. (a) Shows the pathway and (b) shows the Reaction Metrics} 
    \label{fig:mol_paths:mol1:s1}
\end{figure}

For step 1, we select one of the 10 pathways generated by DeepRetro. The selected pathway is shown in \ref{fig:mol_paths:mol1:s1}. The SMILES and reaction metrics for step 1 are shared below. \\
\begin{verbatim}
Smiles:
    Product: O=C1[C@@]2(CC(C)C)N([C@]3([H])C(N2)=O)[C@](CC(C)C)(C(O[C@H]3C)=O)N1
    Reactant: O=C(Cl)[C@@]1(CC(C)C)N([C@]2([H])C(=O)N1)[C@](CC(C)C)(C(O[C@H]2C)=O)N
\end{verbatim}

\paragraph{Step 2}

\begin{figure}[h!]
    \centering
    \subfigure[]{\includegraphics[width=0.48\linewidth]{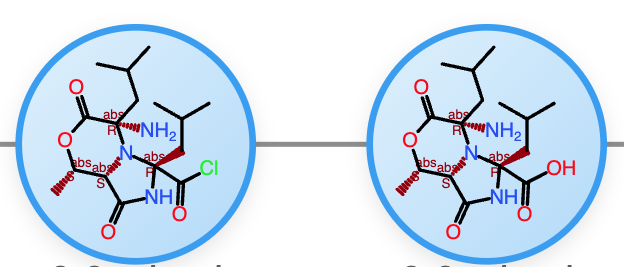}}
    \subfigure[]{\includegraphics[width=0.48\textwidth]{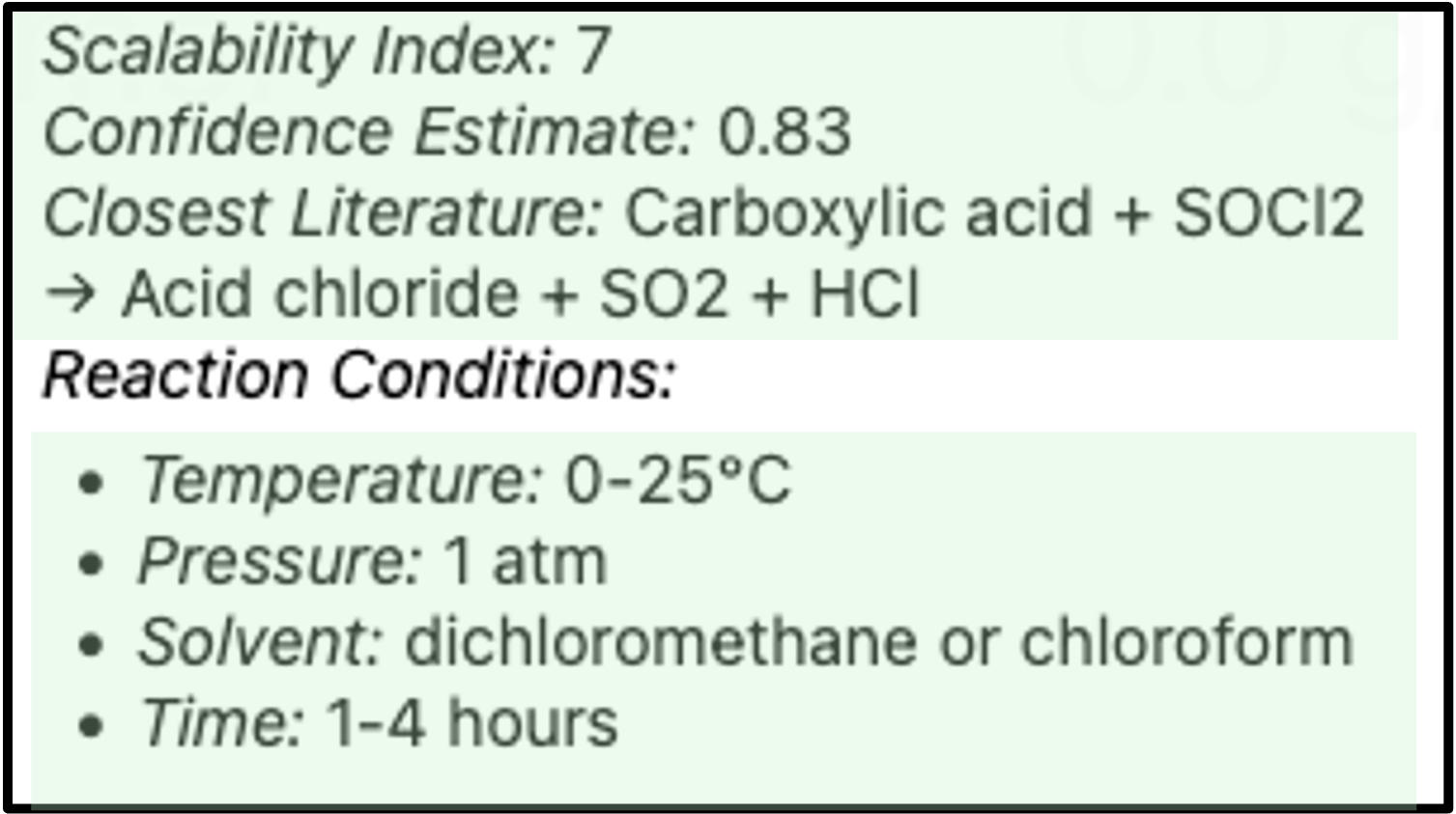}}
    \caption{Step 2 generated by DeepRetro. (a) Shows the pathway and (b) shows the Reaction Metrics} 
    \label{fig:mol_paths:mol1:s2} 
\end{figure}

For step 2, we rerun the system using the "partial-rerun" capability. We then selected one of the 10 pathways generated by DeepRetro.The selected pathway is shown in \ref{fig:mol_paths:mol1:s2}. The SMILES and reaction metrics for step 2 are shared below. \\
\begin{verbatim}
Smiles:
    Product: O=C(Cl)[C@@]1(CC(C)C)N([C@]2([H])C(=O)N1)[C@](CC(C)C)(C(O[C@H]2C)=O)N
    Reactant: O=C(O)[C@@]1(CC(C)C)N([C@]2([H])C(=O)N1)[C@](CC(C)C)(C(O[C@H]2C)=O)N 
\end{verbatim}


\paragraph{Step 3}

\begin{figure}[h!]
    \centering
    \subfigure[]{\includegraphics[width=0.48\linewidth]{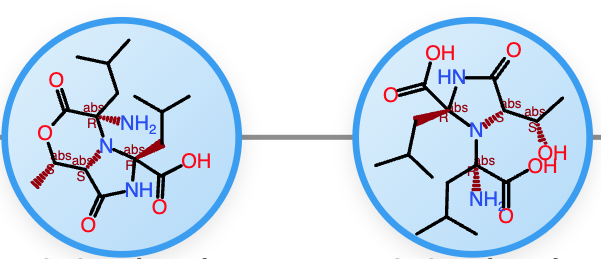}}
    \subfigure[]{\includegraphics[width=0.48\textwidth]{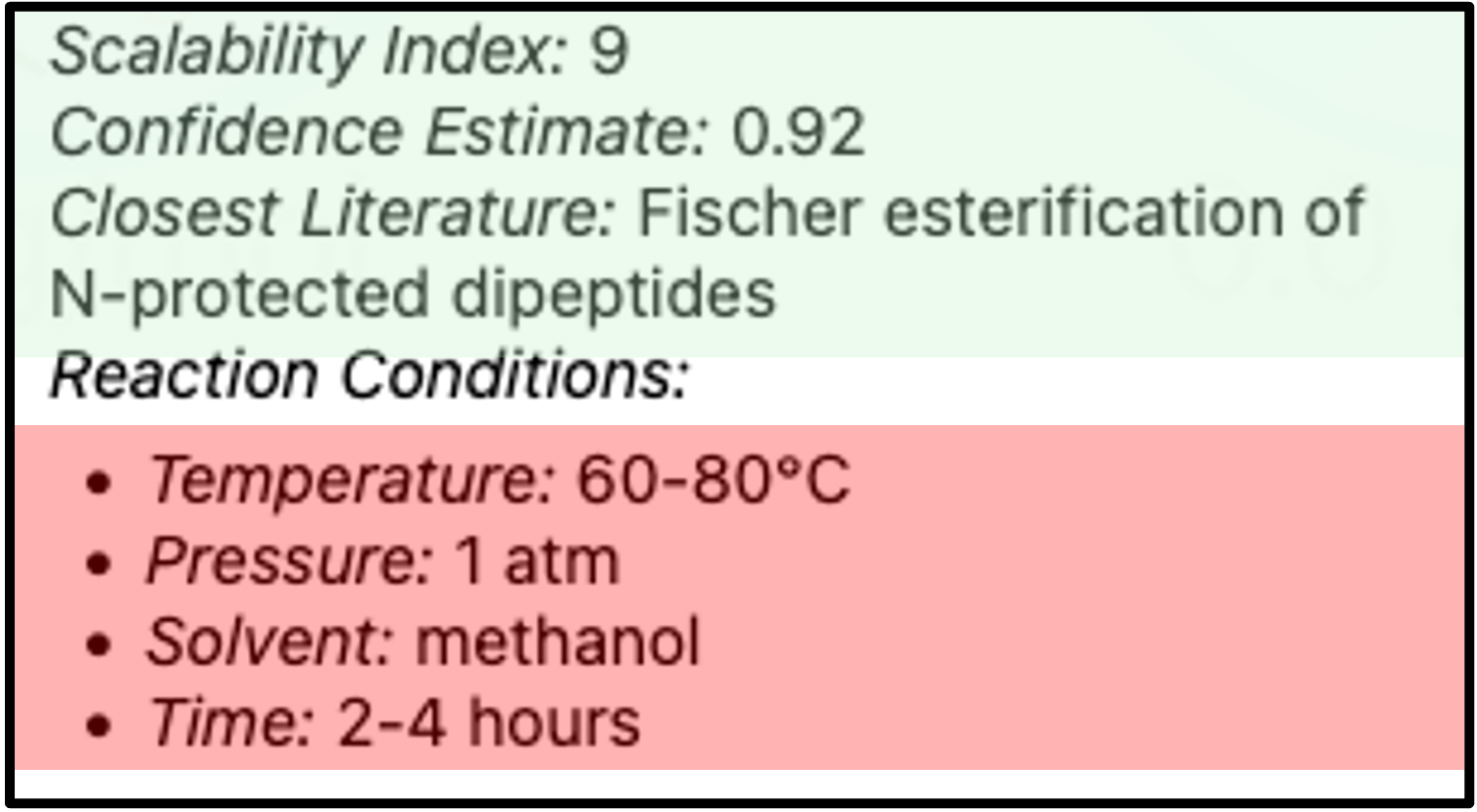}}
    \caption{Step 3 generated by DeepRetro. (a) Shows the pathway and (b) shows the Reaction Metrics} 
    \label{fig:mol_paths:mol1:s3_1} 
\end{figure}

For step 3, we then selected selected the reaction that involves breaking of the Ester bond.The selected pathway is shown in \ref{fig:mol_paths:mol1:s3_1}. The SMILES and reaction metrics for step 3 are shared below. \\
\begin{verbatim}
Smiles:
    Product: O=C(O)[C@@]1(CC(C)C)N([C@]2([H])C(=O)N1)[C@](CC(C)C)(C(O[C@H]2C)=O)N
    Reactant: O=C([C@]1(N([C@](C(O)=O)(N)CC(C)C)[C@](C(N1)=O)([C@@H](O)C)[H])CC(C)C)O
\end{verbatim}



\begin{figure}[h!]
    \centering
    \subfigure[]{\includegraphics[width=0.48\linewidth]{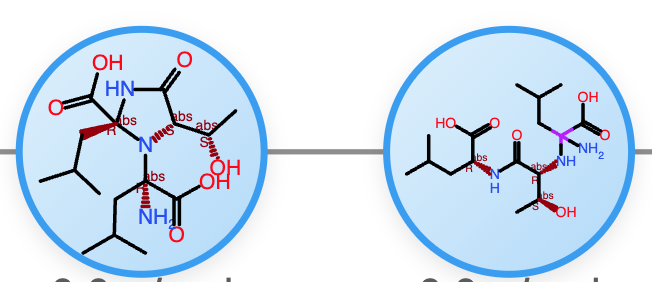}}
    \subfigure[]{\includegraphics[width=0.48\textwidth]{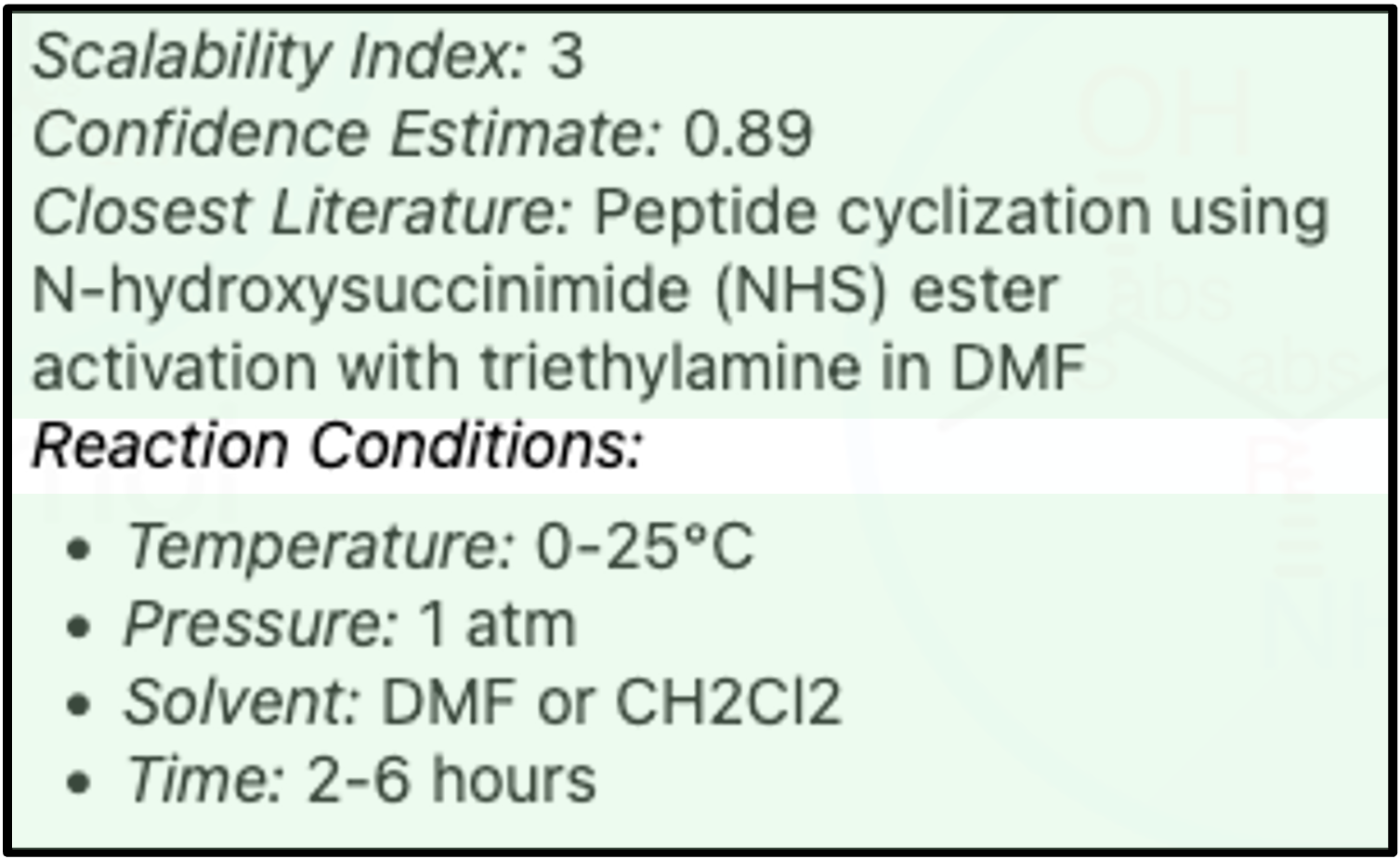}}
    \caption{Step 4 generated by DeepRetro. (a) Shows the pathway and (b) shows the Reaction Metrics} 
    \label{fig:mol_paths:mol1:s4} 
\end{figure}

\paragraph{Step 4}
We obtain the following steps. Steps 4,5 were generated by DeepRetro without any human intervention and the same pathway was generated 3 out of 10 times. The selected pathway is shown in \ref{fig:mol_paths:mol1:s4}. The SMILES and reaction metrics for step 4 are shared below. \\

\begin{verbatim}
Smiles:
    Product: O=C([C@]1(N([C@](C(O)=O)(N)CC(C)C)[C@](C(N1)=O)([C@@H](O)C)[H])CC(C)C)O
    Reactant: O=C(O)C(N)(CC(C)C)N[C@@H](C(=O)N[C@](CC(C)C)(C(=O)O)[H])[C@@H](O)C
\end{verbatim}

\paragraph{Step 5}

\begin{figure}[h!]
    \centering
    \subfigure[]{\includegraphics[width=0.48\linewidth]{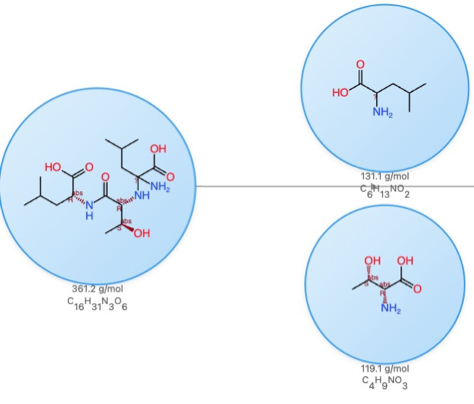}}
    \subfigure[]{\includegraphics[width=0.48\textwidth]{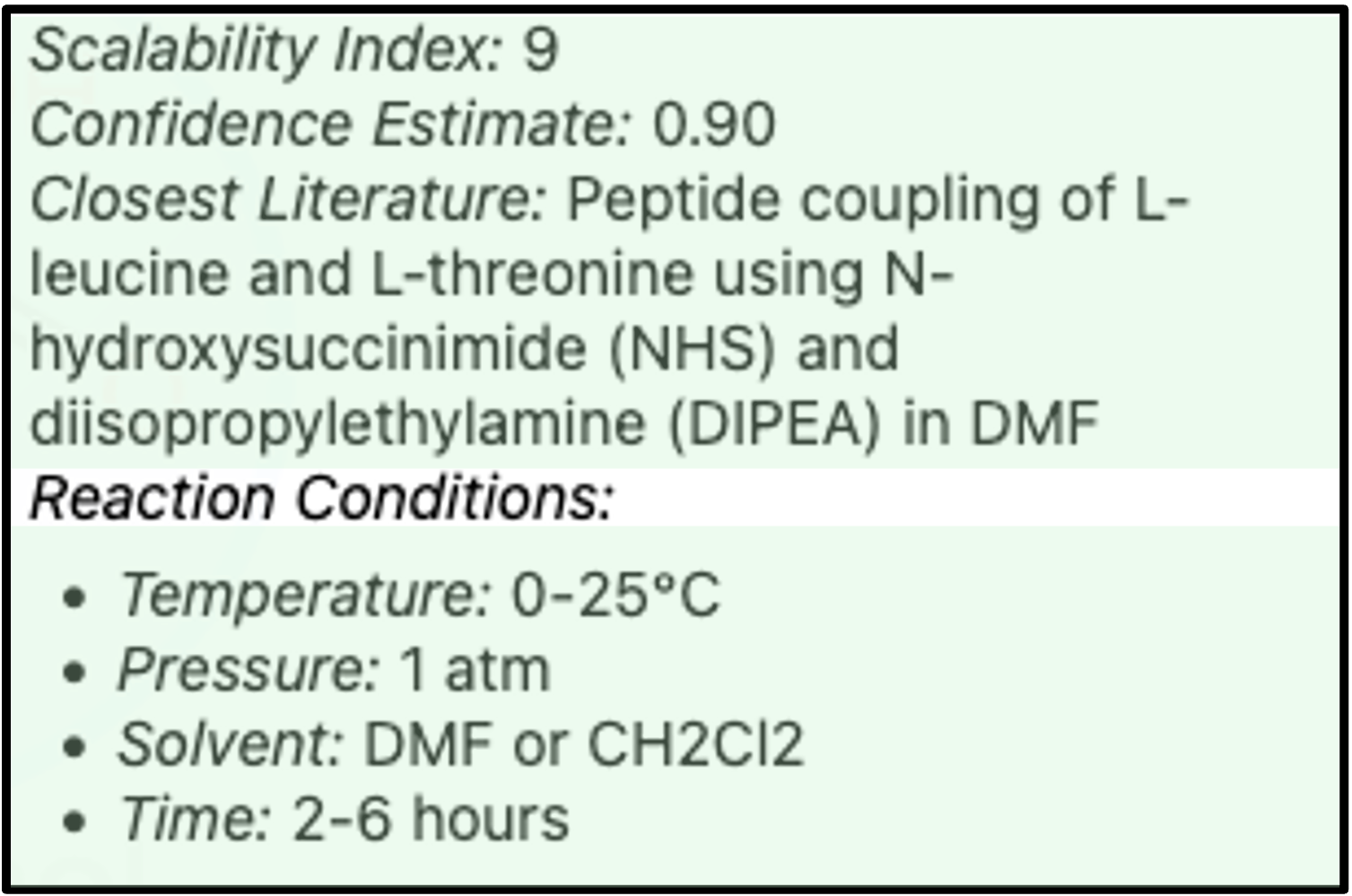}}
    \caption{Step 5 generated by DeepRetro. (a) Shows the pathway and (b) shows the Reaction Metrics} 
    \label{fig:mol_paths:mol1:s5} 
\end{figure}
The SMILES and reaction metrics for step 5 are shared below. \\
\begin{verbatim}
Smiles:
    Product: O=C(O)C(N)(CC(C)C)N[C@@H](C(=O)N[C@](CC(C)C)(C(=O)O)[H])[C@@H](O)C
    Reactant: CC(C)CC(N)C(=O)O
              N[C@@H](C(=O)O)[C@@H](O)C
    
\end{verbatim}

We stop at step 5 as the suggested reactants are available amino acids in the market. But DeepRetro generated further steps breaking down the amino acids into simpler molecules as DeepRetro was not configured to stop at amino acids

\subsection{Molecule 2: Tetracyclic Azepine derivative}
\label{app:mol_paths:mol2}

The proposed retrosynthetic pathway for tetracyclic azepine derivative is distinct from strategies reported for related tetracyclic benzazepines such as tetrabenazine. Literature syntheses of these scaffolds typically employ pre-assembled polycyclic amines followed by late-stage functionalization, often via intramolecular Friedel–Crafts cyclizations or reductive aminations. In contrast, the present route introduces a novel early disconnection at the tertiary amine, enabling scaffold construction from simpler fragments. The sequence employs an $\alpha,\beta$-unsaturated ester epoxidation, followed by epoxide opening with an amine-containing tricyclic core, an approach not commonly applied to benzazepine frameworks. Strategic use of a Grignard intermediate bearing methoxy and chloro substituents allows for late-stage functional diversification, which is advantageous for medicinal chemistry optimization.
The pathway is convergent, tracing back to readily available starting materials such as naphthyl ketone and a chloro-substituted benzoic acid derivative; thereby enhancing synthetic accessibility. Each key transformation is supported by precedent, including high-yield epoxidation and selective amine–epoxide ring opening, lending confidence to the feasibility of the route. Overall, the design represents a conceptual shift from existing literature, combining convergent fragment assembly with late-stage diversification, offering both novelty in bond disconnections and practical validity for accessing a synthetically challenging, pharmacologically relevant scaffold.

\paragraph{Step 1}

\begin{figure}[h!]
    \centering
    \subfigure[]{\includegraphics[width=0.48\linewidth]{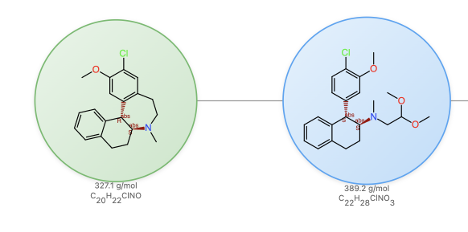}}
    \subfigure[]{\includegraphics[width=0.48\textwidth]{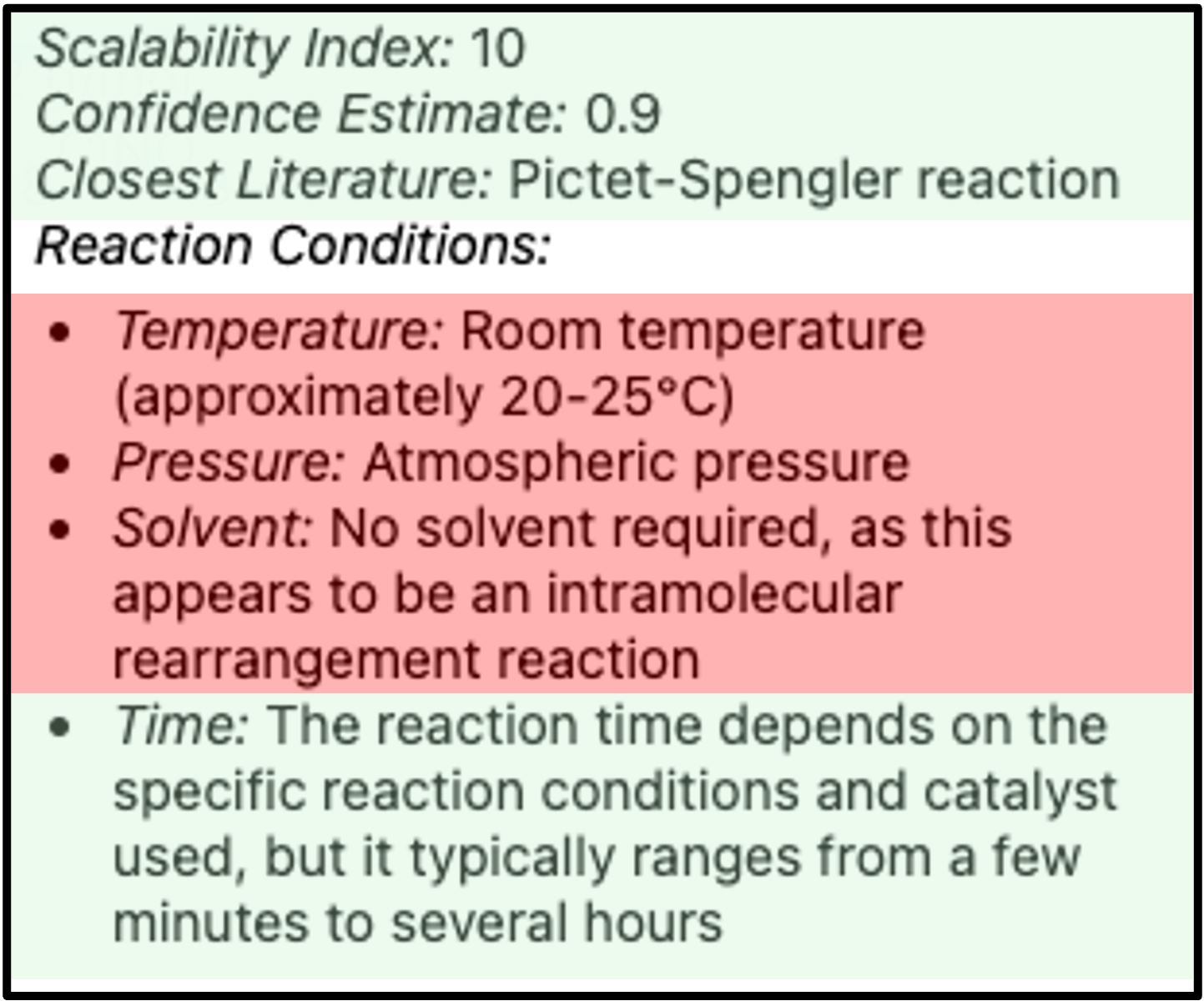}}
    \caption{Step 1 generated by DeepRetro. (a) Shows the pathway and (b) shows the Reaction Metrics} 
    \label{fig:mol_paths:mol2:s1} 
\end{figure}

Retro ring-opening of a fused N-methylazepane via cleavage of the CH2-CH2 linkage, yielding a tertiary amine bearing a pendant methoxy side chain. The pathway is shown in \ref{fig:mol_paths:mol2:s1}

The SMILES and reaction metrics for step 1 are shared below. \\
\begin{verbatim}
Smiles:
    Product: ClC(C(OC)=C1)=CC2=C1[C@@H]3[C@@H](N(C)CC2)CCC4=CC=CC=C43
    Reactant: N(CC(OC)OC)(C)[C@@H]1[C@H](C=2C(CC1)=CC=CC2)C3=CC(OC)=C(Cl)C=C3    
\end{verbatim}


\paragraph{Step 2}

\begin{figure}[h!]
    \centering
    \subfigure[]{\includegraphics[width=0.48\linewidth]{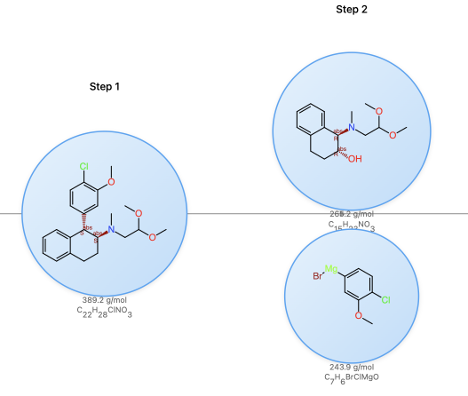}}
    \subfigure[]{\includegraphics[width=0.48\textwidth]{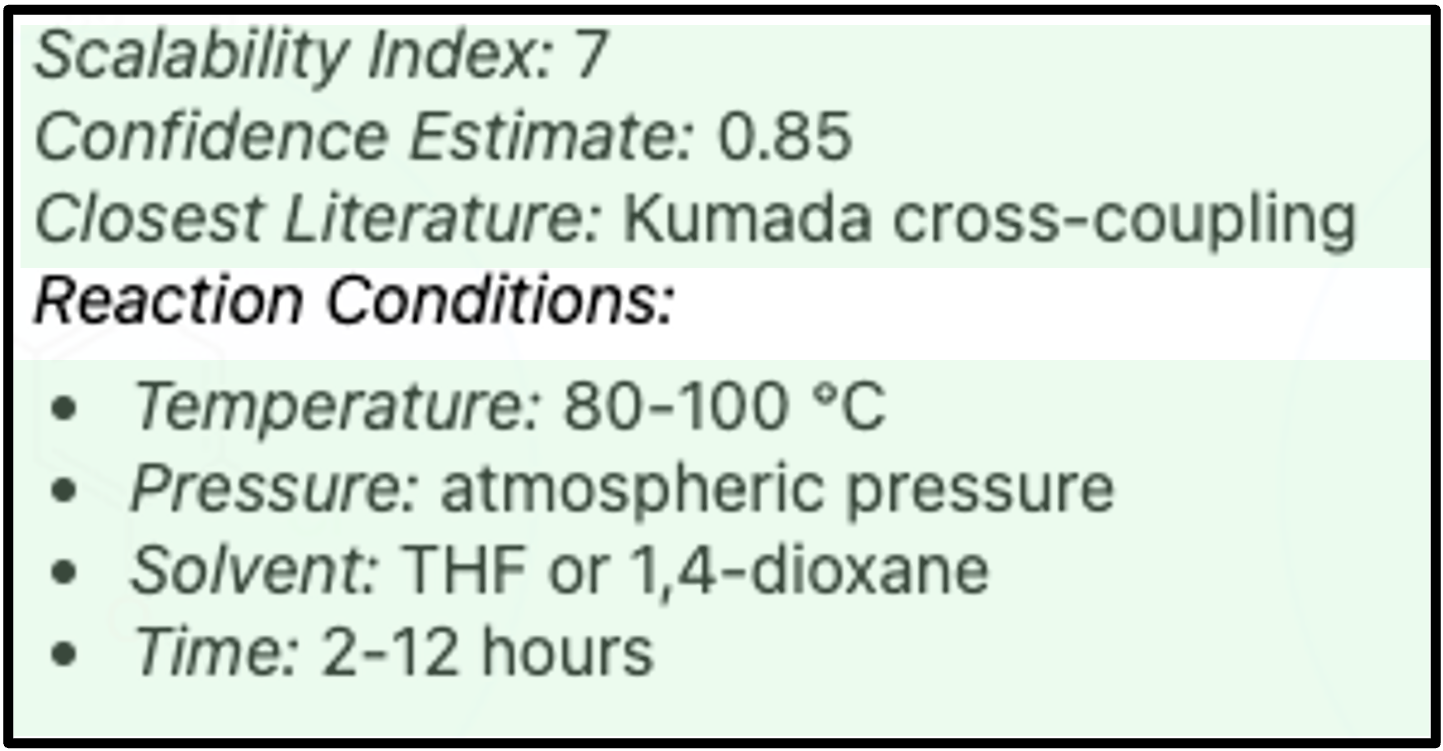}}
    \caption{Step 2 generated by DeepRetro. (a) Shows the pathway and (b) shows the Reaction Metrics} 
    \label{fig:mol_paths:mol2:s2} 
\end{figure}

Further cleavage at the C-1 position of the tetralin moiety to generate a Grignard reagent and the remaining fused tetralin structure paired with the tertiary amine fragment. The pathway is shown in \ref{fig:mol_paths:mol2:s2}
The SMILES and reaction metrics for step 2 are shared below. \\
\begin{verbatim}
Smiles:
    Product: N(CC(OC)OC)(C)[C@@H]1[C@H(C=2C(CC1)=CC=CC2)C3=CC(OC)=C(Cl)C=C3
    Reactant: COC(CN(C)[C@@H]1c2ccccc2CC[C@H]1O)OC  (upper)
              ClC1=CC=C([Mg]Br)C=C1OC  (lower)
\end{verbatim}


\paragraph{Step 3}

\begin{figure}[h!]
    \centering
    \subfigure[]{\includegraphics[width=0.48\linewidth]{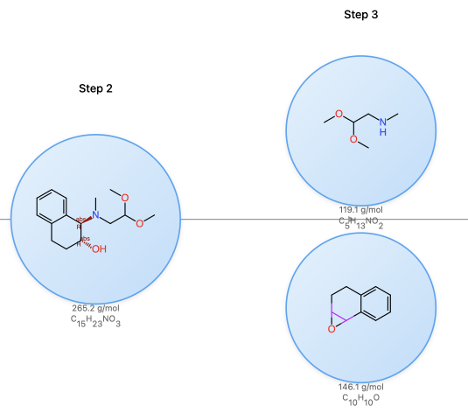}}
    \subfigure[]{\includegraphics[width=0.48\textwidth]{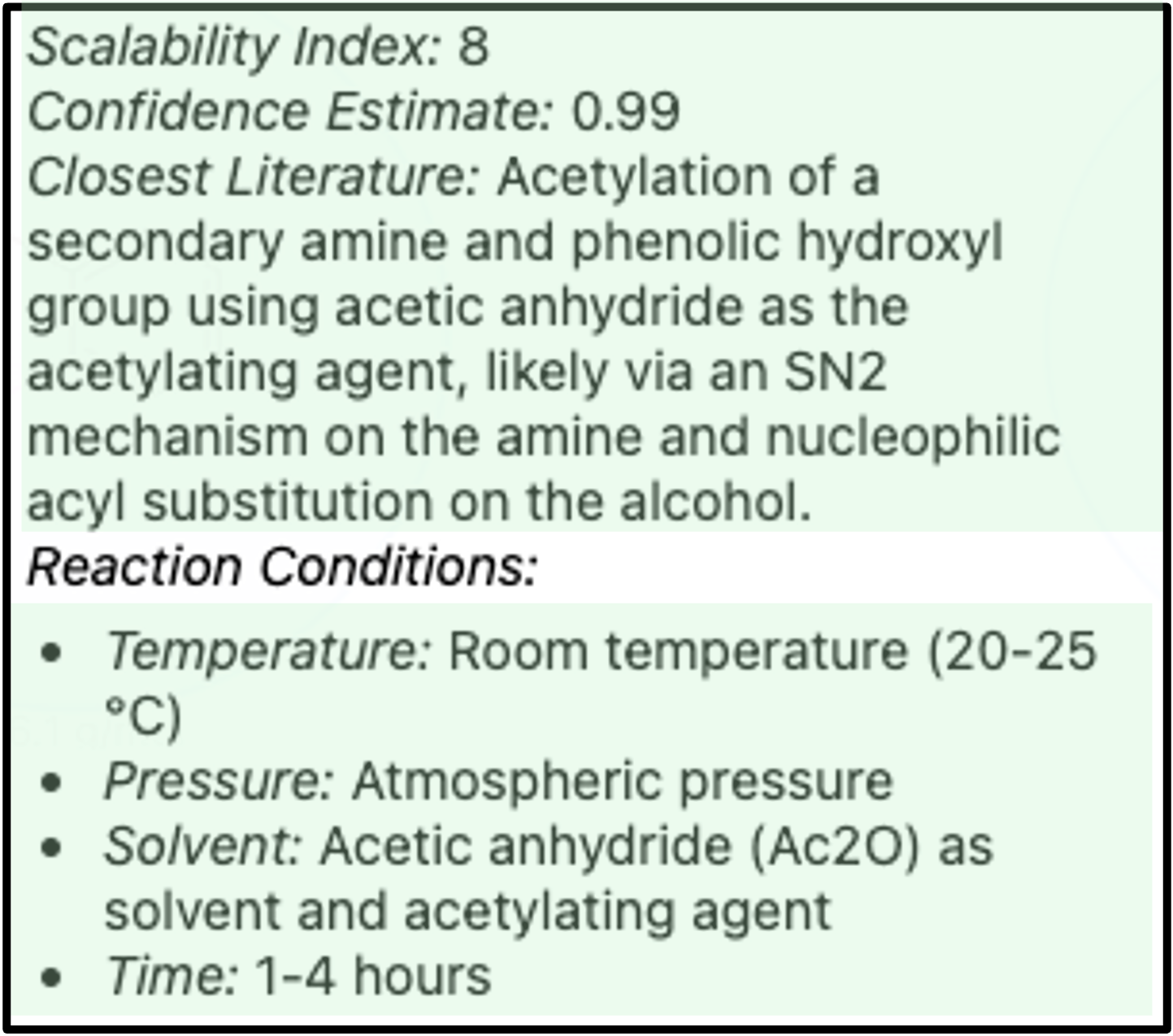}}
    \caption{Step 3 generated by DeepRetro. (a) Shows the pathway and (b) shows the Reaction Metrics} 
    \label{fig:mol_paths:mol2:s3} 
\end{figure}

Cleavage of the tertiary amine chain with the tetralin moiety resulting into epoxy tetralin and (Methylamino)acetaldehyde dimethyl acetal. The pathway is shown in \ref{fig:mol_paths:mol2:s3}
The SMILES and reaction metrics for step 3 are shared below. \\
\begin{verbatim}
Smiles:
    Product: COC(CN(C)[C@@H]1c2ccccc2CC[C@H]1O)OC
    Reactant: CNCC(OC)OC  (upper)
              c1ccc2c(c1)CCC1OC21  (lower)
\end{verbatim}


\paragraph{Step 4}

\begin{figure}[h!]
    \centering
    \subfigure[]{\includegraphics[width=0.48\linewidth]{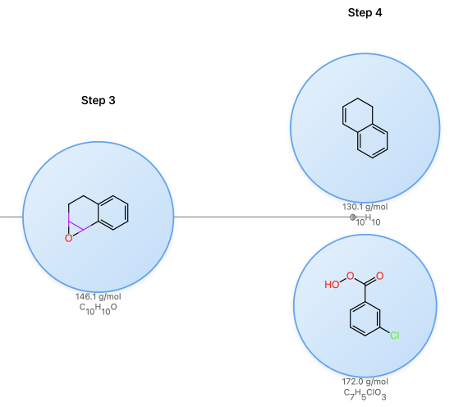}}
    \subfigure[]{\includegraphics[width=0.48\textwidth]{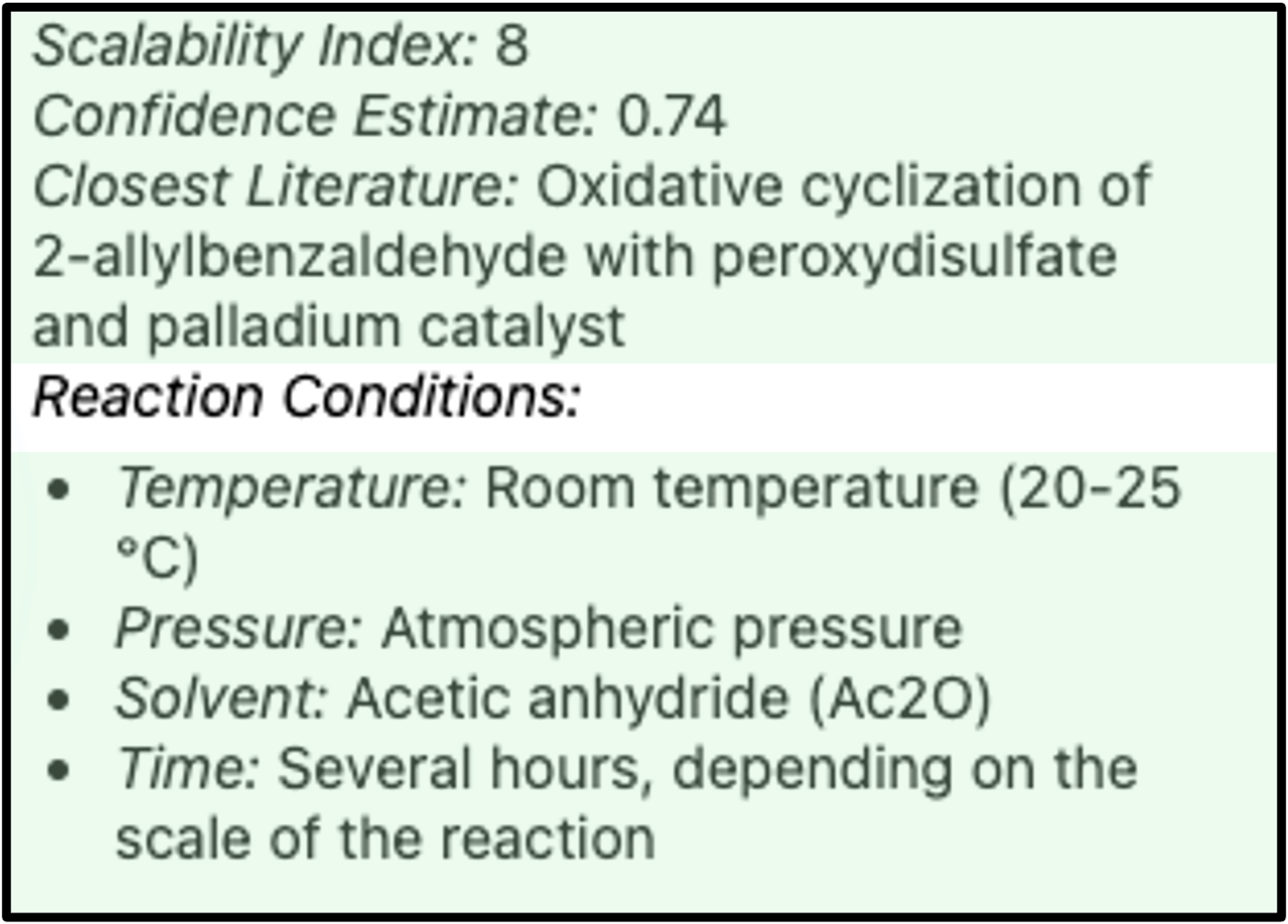}}
    \caption{Step 4 generated by DeepRetro. (a) Shows the pathway and (b) shows the Reaction Metrics} 
    \label{fig:mol_paths:mol2:s4} 
\end{figure}

The epoxytetralin intermediate was retrosynthetically traced to tetralin via an oxidative epoxidation strategy, with further disconnection revealing 4-chlorophenyl chloroformate as the electrophilic carbonate source facilitating intramolecular cyclization. The pathway is shown in \ref{fig:mol_paths:mol2:s4}

The SMILES and reaction metrics for step 4 are shared below. \\
\begin{verbatim}
Smiles:
    Product:  c1ccc2c(c1)CCC1OC21
    Reactant: C1=Cc2ccccc2CC1  (upper)
              O=C(OO)c1cccc(Cl)c1  (lower)
\end{verbatim}

\subsection{Molecule 3: Erythromycin}
\label{app:mol_paths:mol3}

Erythromycin B, a complex polyketide macrolide, was selected to test LLM-driven retrosynthesis on a highly functionalized, stereochemically dense target. Using DeepRetro with minimal human input, the pathway was developed through iterative model–tool interactions. The sole human intervention was the insertion of a biosynthetically inspired intermediate \cite{erythromycin_total}, chosen to anchor the route in chemically realistic territory.
From this point, DeepRetro proposed a coherent sequence: macrolactone ring opening, cyclic ether formation for chain rigidification, selective protection of desosamine and cladinose hydroxyls, aldol disconnection of a $\beta$-hydroxy ketone motif, crotylation to establish stereocenters, and successive ester cleavages to yield sugar and aglycone fragments. These were further simplified to commercially accessible building blocks.
The pathway reflects known biosynthetic logic (e.g., macrolactone cleavage, sugar separation) yet differs from any single published laboratory synthesis. While individual transformations are well-precedented, their ordering and integration form a novel, fully chemical alternative to enzyme-mediated routes. Each disconnection yields chemically tractable intermediates, validating feasibility.

\paragraph{Exact Human Intervention}
The only human intervention occurred at the third step, where erythromycin B was converted into derivative 3b. This modification rigidified the C(9)–C(13) segment through cyclic ether formation between the C(11) and C(9S) hydroxyl groups and introduced protective groups on the cladinose and desosamine sugars. This guided the retrosynthetic analysis, simplifying subsequent disconnections. From 3b onward, the LLM independently proposed a coherent pathway, including selective hydroxyl protections, aldol disconnections of $\beta$-hydroxy ketones, crotylation to establish stereocenters, and sequential ester cleavages. These steps led to sugar and aglycone fragments, further simplified to commercially accessible building blocks, demonstrating effective human–AI collaboration.

\paragraph{Step 1}

\begin{figure}[!h]
    \centering
    \subfigure[]{\includegraphics[width=0.48\linewidth]{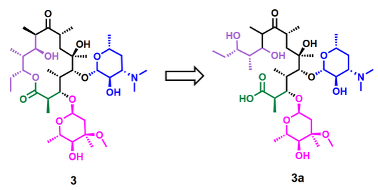}}
    \subfigure[]{\includegraphics[width=0.48\textwidth]{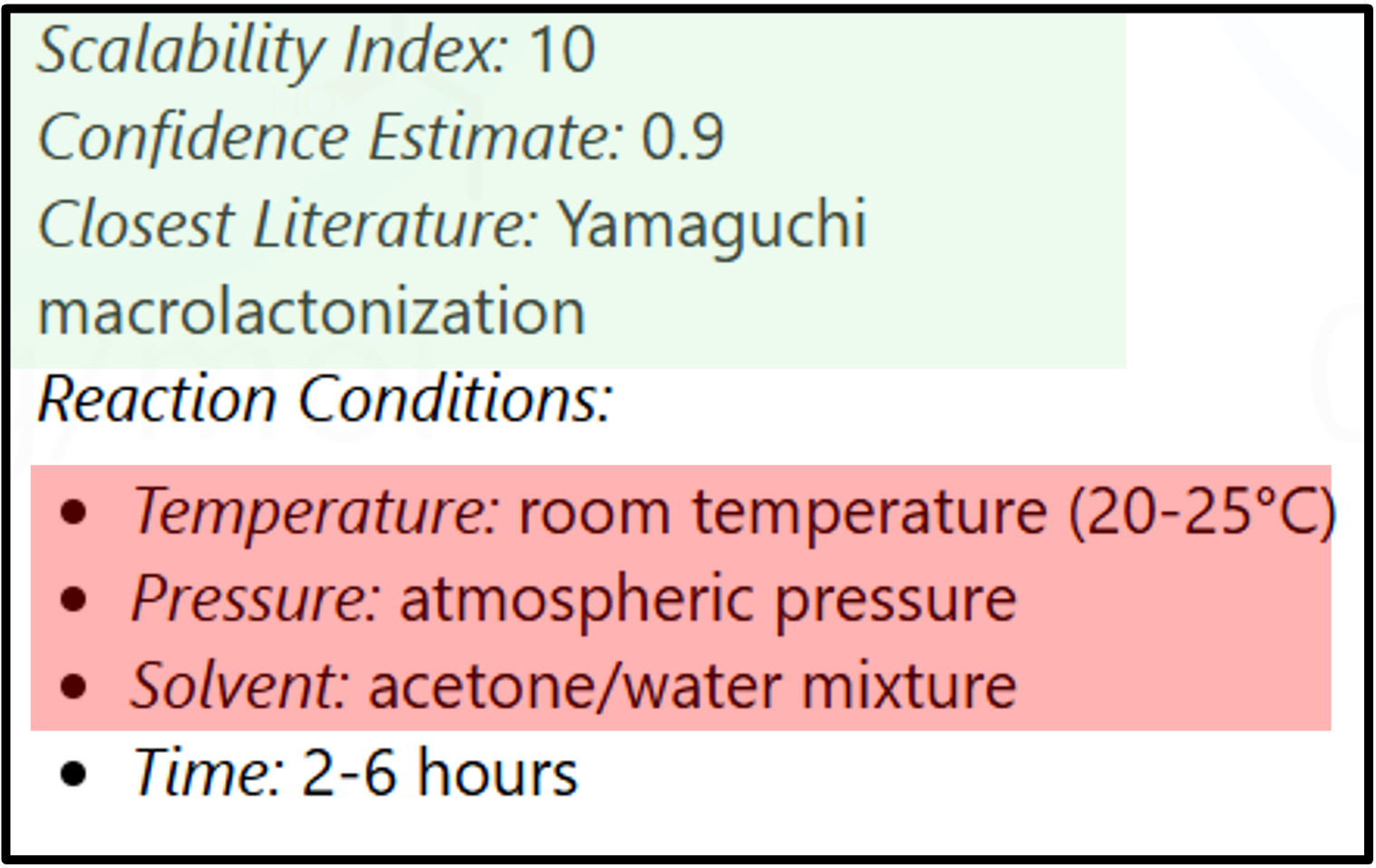}}
    \caption{Step 1 generated by DeepRetro. (a) Shows the pathway and (b) shows the Reaction Metrics} 
    \label{fig:mol_paths:mol3:s1} 
\end{figure}

For step 1, the DeepRetro was suggested to break the ester group of the lactone ring to initiate the retrosynthesis step. This is shown in figure \ref{fig:mol_paths:mol3:s1}

The SMILES and reaction metrics for step 1 are shared below. \\
\begin{verbatim}
Smiles:
    Product: O[C@@](C[C@@H](C)C([C@@H]1C)=O)(C)[C@@H]([C@@H](C)[C@@H]
    ([C@@H](C)C(O[C@H](CC)[C@H](C)[C@@H]1O)=O)O[C@@H]2O[C@@H](C)[C@@H]
    ([C@](C2)(C)OC)O)O[C@@H]([C@@H]3O)O[C@H](C)C[C@@H]3N(C)C
    Reactant: O[C@@](C[C@@H](C)C(C(C)C([C@H](C)[C@H](CC)O)O)=O)(C)
    [C@@H]([C@@H](C)[C@@H]([C@@H](C)C(O)=O)O[C@@H]1O[C@@H](C)[C@@H]
    ([C@](C1)(C)OC)O)O[C@@H]([C@@H]2O)O[C@H](C)C[C@@H]2N(C)C
\end{verbatim}


\paragraph{Step 2}

\begin{figure}[h!]
    \centering
    \subfigure[]{\includegraphics[width=0.48\linewidth]{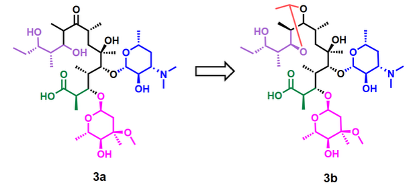}}
    \subfigure[]{\includegraphics[width=0.48\textwidth]{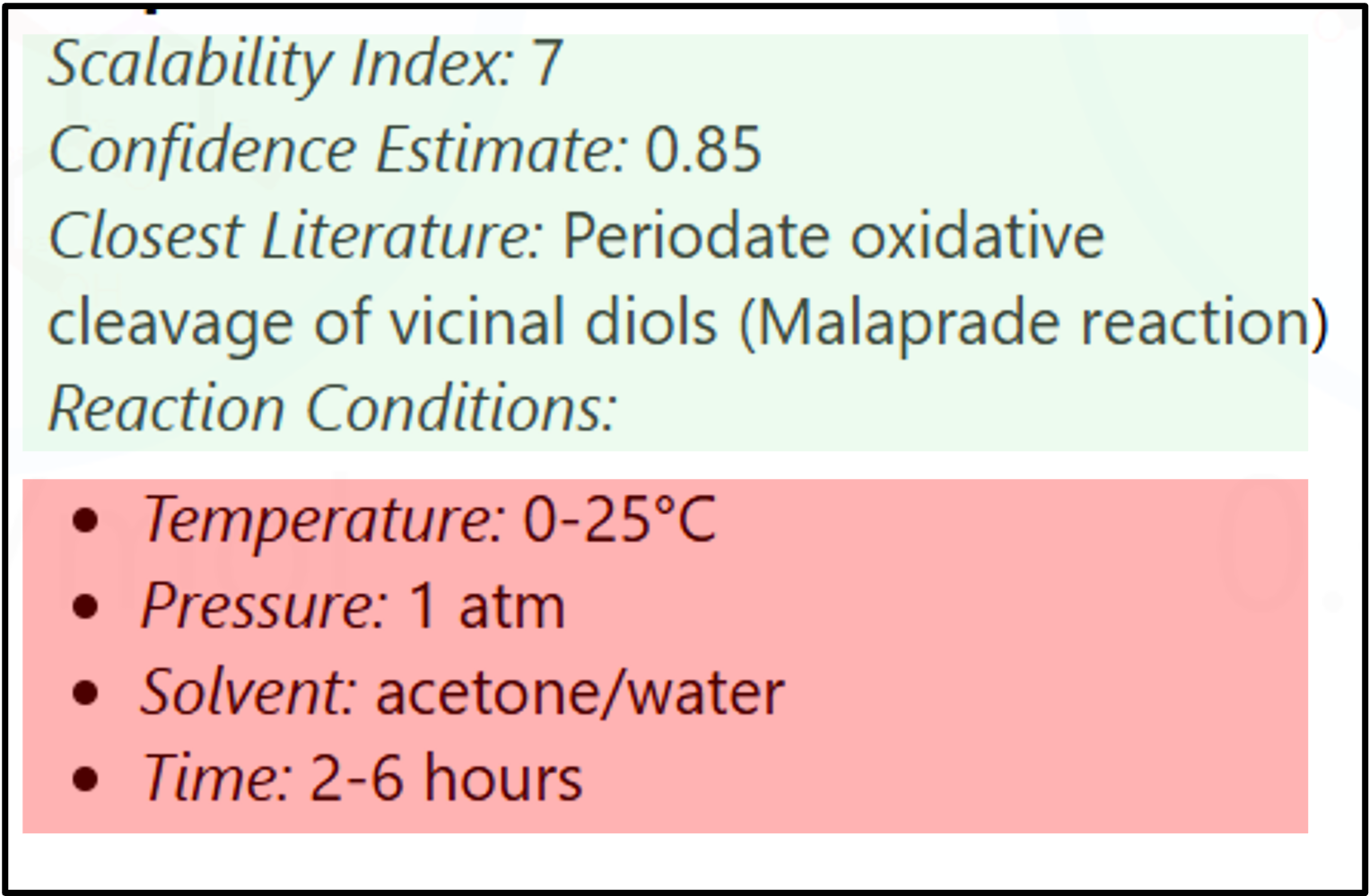}}
    \caption{Step 2 generated by DeepRetro. (a) Shows the pathway and (b) shows the Reaction Metrics} 
    \label{fig:mol_paths:mol3:s2} 
\end{figure}

For step 2, the DeepRetro was again suggested add a necessary protective group between the ketone and adjacent hydroxyl group to main rigidity in the bulky chain. This also separates possible reactive hydroxy groups of the lactone ring from interested part of the intermediate. This is shown in figure \ref{fig:mol_paths:mol3:s2}

The SMILES and reaction metrics for step 2 are shared below. \\
\begin{verbatim}
Smiles:
    Product: O[C@@](C[C@@H](C)C(C(C)C([C@H](C)[C@H](CC)O)O)=O)(C)
    [C@@H]([C@@H](C)[C@@H]([C@@H](C)C(O)=O)O[C@@H]1O[C@@H](C)[C@@H]
    ([C@](C1)(C)OC)O)O[C@@H]([C@@H]2O)O[C@H](C)C[C@@H]2N(C)C
    Reactant: O[C@@](C[C@@H](C)C1O[C@@H](C)O[C@@H]([C@H](C)[C@H](CC)O)
    [C@H]1C)(C)[C@@H]([C@@H](C)[C@@H]([C@@H](C)C(O)=O)O[C@@H]2O[C@@H]
    (C)[C@@H]([C@](C2)(C)OC)O)O[C@@H]([C@@H]3O)O[C@H](C)C[C@@H]3N(C)C
\end{verbatim}


\paragraph{Step 3}

\begin{figure}[h!]
    \centering
    \subfigure[]{\includegraphics[width=0.48\linewidth]{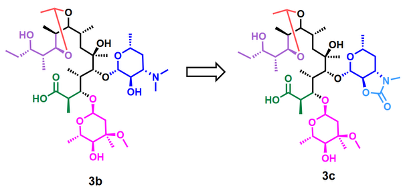}}
    \subfigure[]{\includegraphics[width=0.48\textwidth]{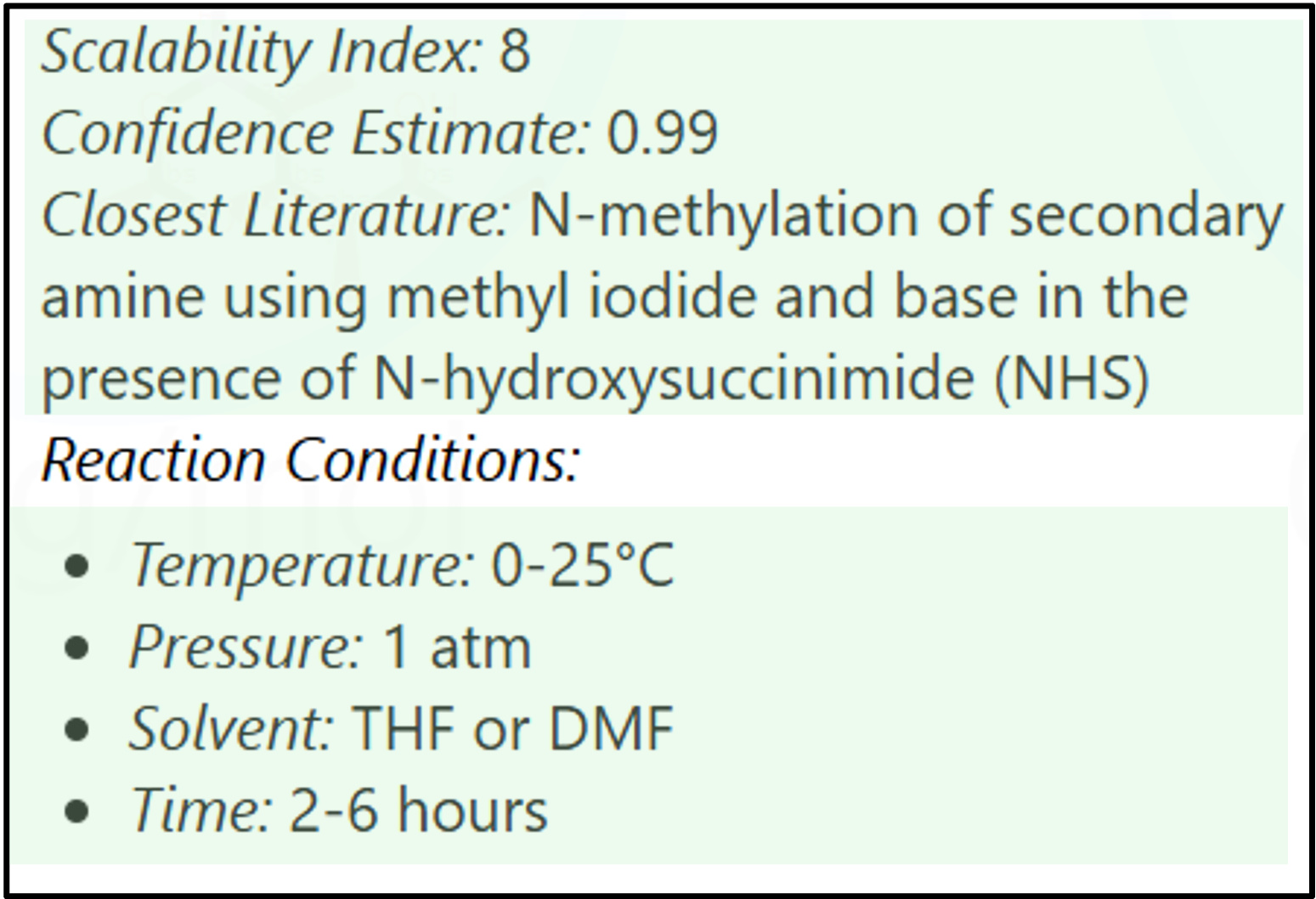}}
    \caption{Step 3 generated by DeepRetro. (a) Shows the pathway and (b) shows the Reaction Metrics} 
    \label{fig:mol_paths:mol3:s3} 
\end{figure}

For step 3, human intervened to further protect reactive sites in the side glucose moiety. This is shown in figure \ref{fig:mol_paths:mol3:s3}

The SMILES and reaction metrics for step 3 are shared below. \\
\begin{verbatim}
Smiles:
    Product: O[C@@](C[C@@H](C)C1O[C@@H](C)O[C@@H]([C@H](C)[C@H](CC)O)
    [C@H]1C)(C)[C@@H]([C@@H](C)[C@@H]([C@@H](C)C(O)=O)O[C@@H]2O[C@@H]
    (C)[C@@H]([C@](C2)(C)OC)O)O[C@@H]([C@@H]3O)O[C@H](C)C[C@@H]3N(C)C
    Reactant: O[C@@](C[C@@H](C)C1O[C@@H](C)O[C@@H]([C@H](C)[C@H](CC)O)
    [C@H]1C)(C)[C@@H]([C@@H](C)[C@@H]([C@@H](C)C(O)=O)O[C@@H]2O[C@@H]
    (C)[C@@H]([C@](C2)(C)OC)O)O[C@@H]([C@@H]3OC4=O)O[C@H]
    (C)C[C@@H]3N4C
\end{verbatim}


\paragraph{Step 4}

\begin{figure}[h!]
    \centering
    \subfigure[]{\includegraphics[width=0.48\linewidth]{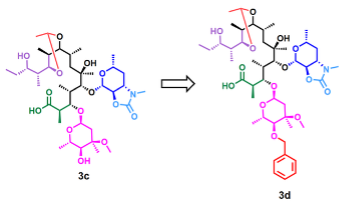}}
    \subfigure[]{\includegraphics[width=0.48\textwidth]{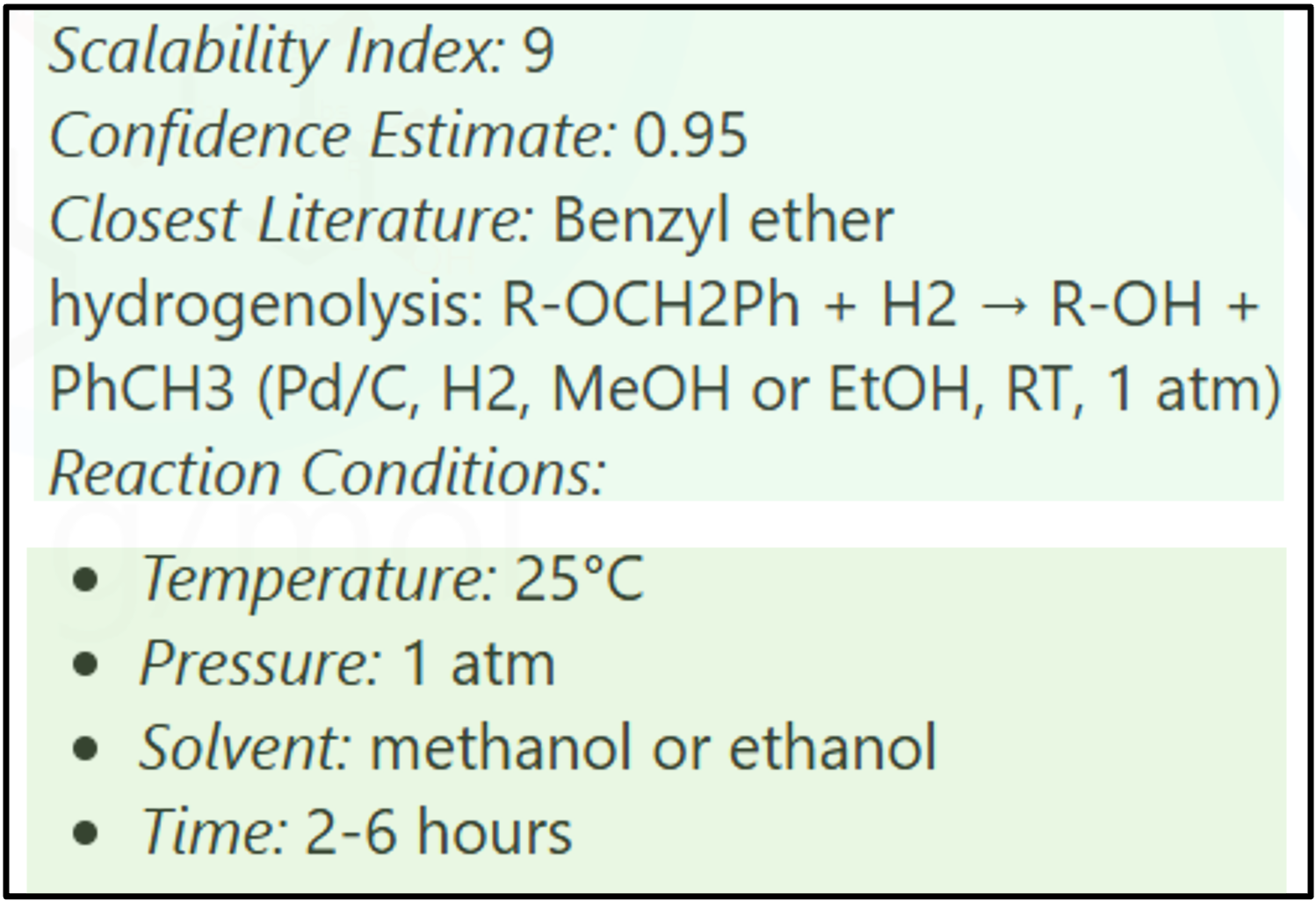}}
    \caption{Step 4 generated by DeepRetro. (a) Shows the pathway and (b) shows the Reaction Metrics} 
    \label{fig:mol_paths:mol3:s4} 
\end{figure}

For step 4, a similar protective group is added onto the other glucose moiety. This is shown in figure \ref{fig:mol_paths:mol3:s4}

The SMILES and reaction metrics for step 4 are shared below. \\
\begin{verbatim}
Smiles:
    Product: O[C@@](C[C@@H](C)C1O[C@@H](C)O[C@@H]([C@H](C)[C@H](CC)O)
    [C@H]1C)(C)[C@@H]([C@@H](C)[C@@H]([C@@H](C)C(O)=O)O[C@@H]2O[C@@H]
    (C)[C@@H]([C@](C2)(C)OC)O)O[C@@H]([C@@H]3OC4=O)O[C@H]
    (C)C[C@@H]3N4C
    Reactant:  O[C@@](C[C@@H](C)C1O[C@@H](C)O[C@@H]([C@H](C)[C@H](CC)O)
    [C@H]1C)(C)[C@@H]([C@@H](C)[C@@H]([C@@H](C)C(O)=O)O[C@@H]2O[C@@H]
    (C)[C@@H]([C@](C2)(C)OC)OCC3=CC=CC=C3)O[C@@H]([C@@H]4OC5=O)O[C@H]
    (C)C[C@@H]4N5C
\end{verbatim}


\paragraph{Step 5}

\begin{figure}[h!]
    \centering
    \subfigure[]{\includegraphics[width=0.48\linewidth]{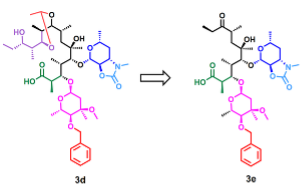}}
    \subfigure[]{\includegraphics[width=0.48\textwidth]{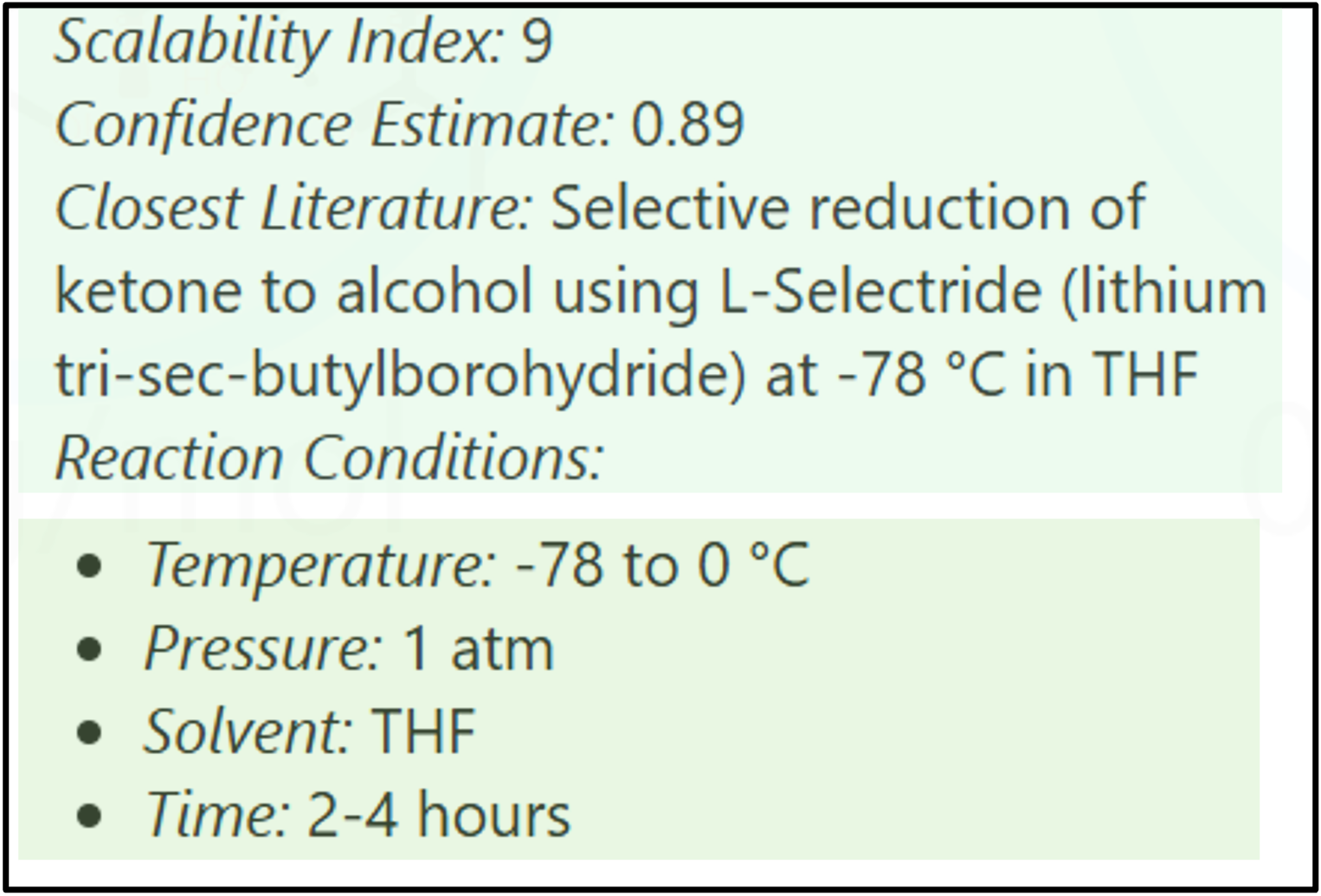}}
    \caption{Step 5 generated by DeepRetro. (a) Shows the pathway and (b) shows the Reaction Metrics} 
    \label{fig:mol_paths:mol3:s5} 
\end{figure}

For step 5, for the first time DeepRetro generated an intermediate 3e without human intervention. This is shown in figure \ref{fig:mol_paths:mol3:s5}

The SMILES and reaction metrics for step 5 are shared below. \\
\begin{verbatim}
Smiles:
    Product: O[C@@](C[C@@H](C)C1O[C@@H](C)O[C@@H]([C@H](C)[C@H](CC)O)
    [C@H]1C)(C)[C@@H]([C@@H](C)[C@@H]([C@@H](C)C(O)=O)O[C@@H]2O[C@@H](C)
    [C@@H]([C@](C2)(C)OC)OCC3=CC=CC=C3)O[C@@H]([C@@H]4OC5=O)O[C@H]
    (C)C[C@@H]4N5C
    Reactant: O[C@@](C[C@@H](C)C(CC)=O)(C)[C@@H]([C@@H](C)[C@@H]([C@@H]
    (C)C(O)=O)O[C@@H]1O[C@@H](C)[C@@H]([C@](C1)(C)OC)OCC2=CC=CC=C2)O[C@@H]
    ([C@@H]3OC4=O)O[C@H](C)C[C@@H]3N4C
\end{verbatim}


\paragraph{Step 6}

\begin{figure}[h!]
    \centering
    \subfigure[]{\includegraphics[width=0.48\linewidth]{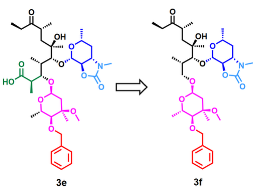}}
    \subfigure[]{\includegraphics[width=0.48\textwidth]{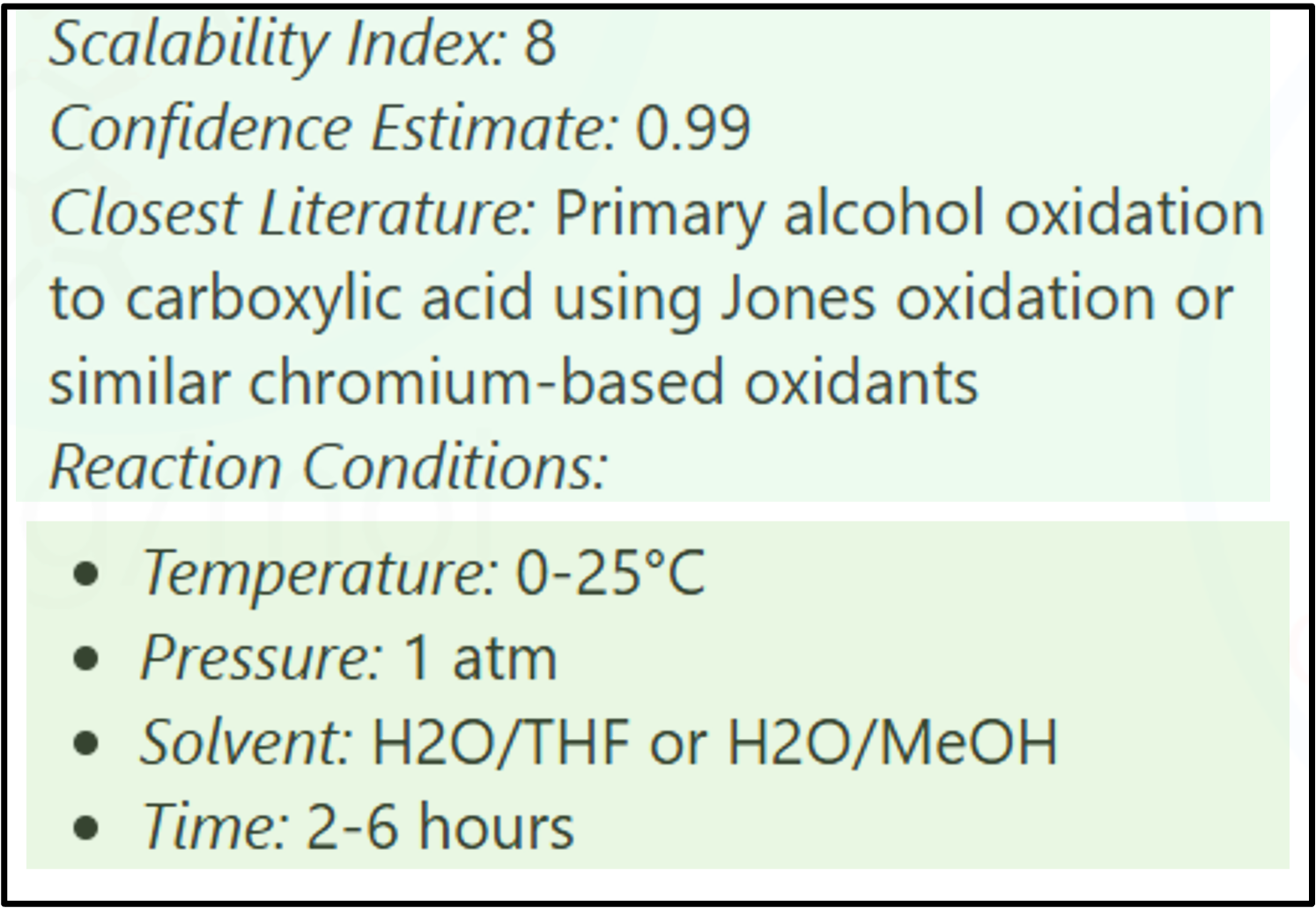}}
    \caption{Step 6 generated by DeepRetro. (a) Shows the pathway and (b) shows the Reaction Metrics} 
    \label{fig:mol_paths:mol3:s6} 
\end{figure}

For step 6, a decarboxylation was carried out by DeepRetro to generate 3f. This is shown in figure \ref{fig:mol_paths:mol3:s6}

The SMILES and reaction metrics for step 6 are shared below. \\
\begin{verbatim}
Smiles:
    Product: O[C@@](C[C@@H](C)C(CC)=O)(C)[C@@H]([C@@H](C)[C@@H]([C@@H]
    (C)C(O)=O)O[C@@H]1O[C@@H](C)[C@@H]([C@](C1)(C)OC)OCC2=CC=CC=C2)O[C@@H]
    ([C@@H]3OC4=O)O[C@H](C)C[C@@H]3N4C
    Reactant: O[C@@](C[C@@H](C)C(CC)=O)(C)[C@@H]([C@@H](C)CO[C@@H]1O[C@@H]
    (C)[C@@H]([C@](C1)(C)OC)OCC2=CC=CC=C2)O[C@@H]([C@@H]3OC4=O)O[C@H]
    (C)C[C@@H]3N4C
\end{verbatim}


\paragraph{Step 7}

\begin{figure}[h!]
    \centering
    \subfigure[]{\includegraphics[width=0.48\linewidth]{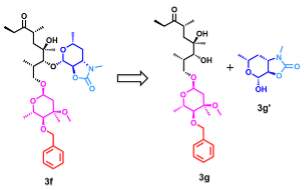}}
    \subfigure[]{\includegraphics[width=0.48\textwidth]{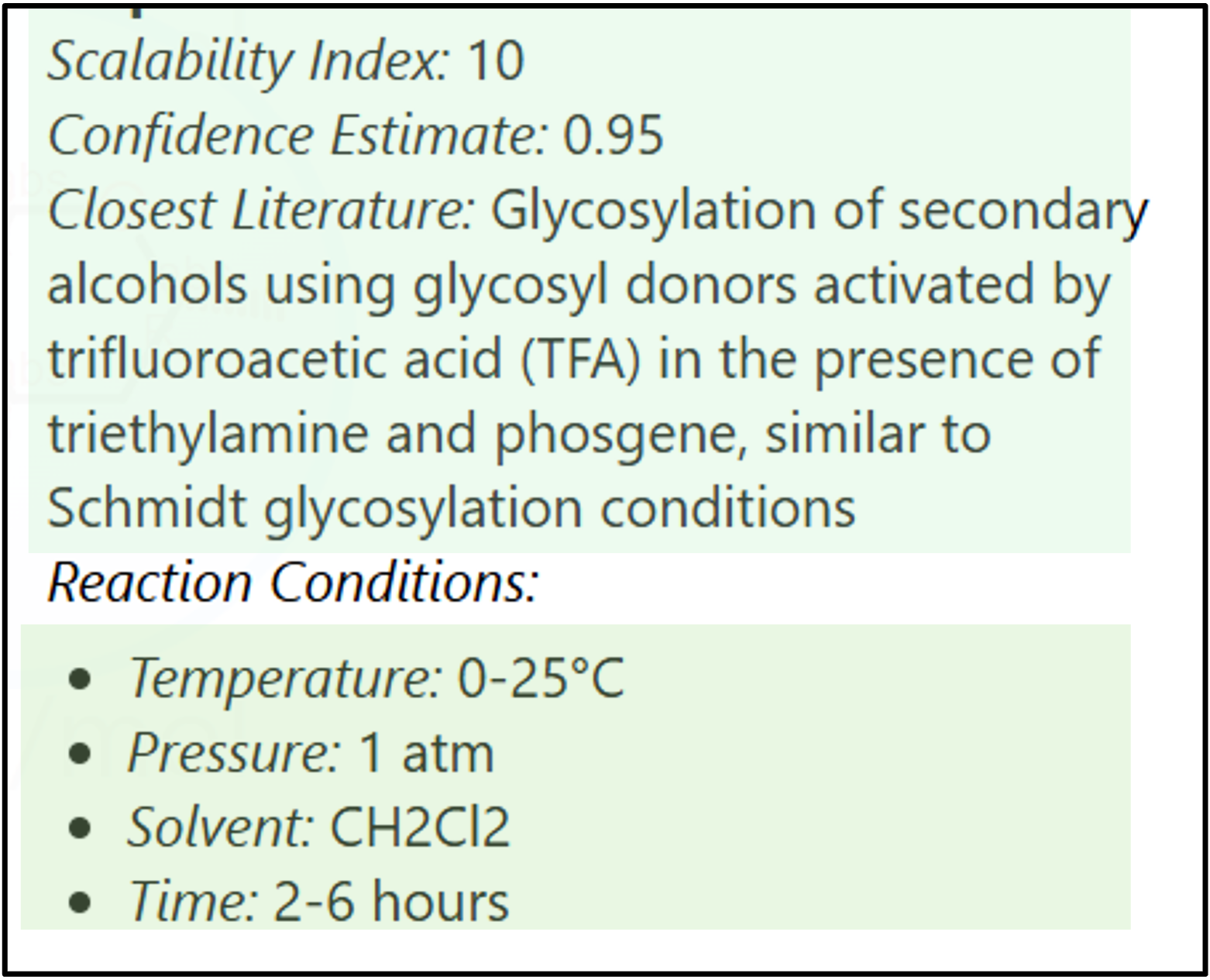}}
    \caption{Step 7 generated by DeepRetro. (a) Shows the pathway and (b) shows the Reaction Metrics} 
    \label{fig:mol_paths:mol3:s7} 
\end{figure}

For Step 7, DeepRetro generated the below pathway without human intervention. This is shown in figure \ref{fig:mol_paths:mol3:s7}

The SMILES and reaction metrics for step 7 are shared below. \\
\begin{verbatim}
Smiles:
    Product: O[C@@](C[C@@H](C)C(CC)=O)(C)[C@@H]([C@@H](C)CO[C@@H]1O[C@@H]
    (C)[C@@H]([C@](C1)(C)OC)OCC2=CC=CC=C2)O[C@@H]([C@@H]3OC4=O)O[C@H]
    (C)C[C@@H]3N4C
    Reactant: (3g) O[C@H]([C@@H](C)CO[C@@H]1O[C@@H](C)[C@@H]([C@](C1)
    (C)OC)OCC2=CC=CC=C2)[C@](C[C@@H](C)C(CC)=O)(C)O
              (3g')O[C@@H]1O[C@@H](C[C@H]2[C@H]1OC(N2C)=O)C
\end{verbatim}


\paragraph{Step 8}

\begin{figure}[h!]
    \centering
    \subfigure[]{\includegraphics[width=0.48\linewidth]{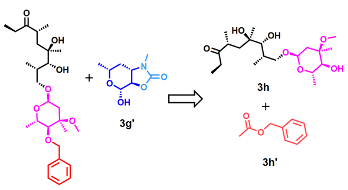}}
    \subfigure[]{\includegraphics[width=0.48\textwidth]{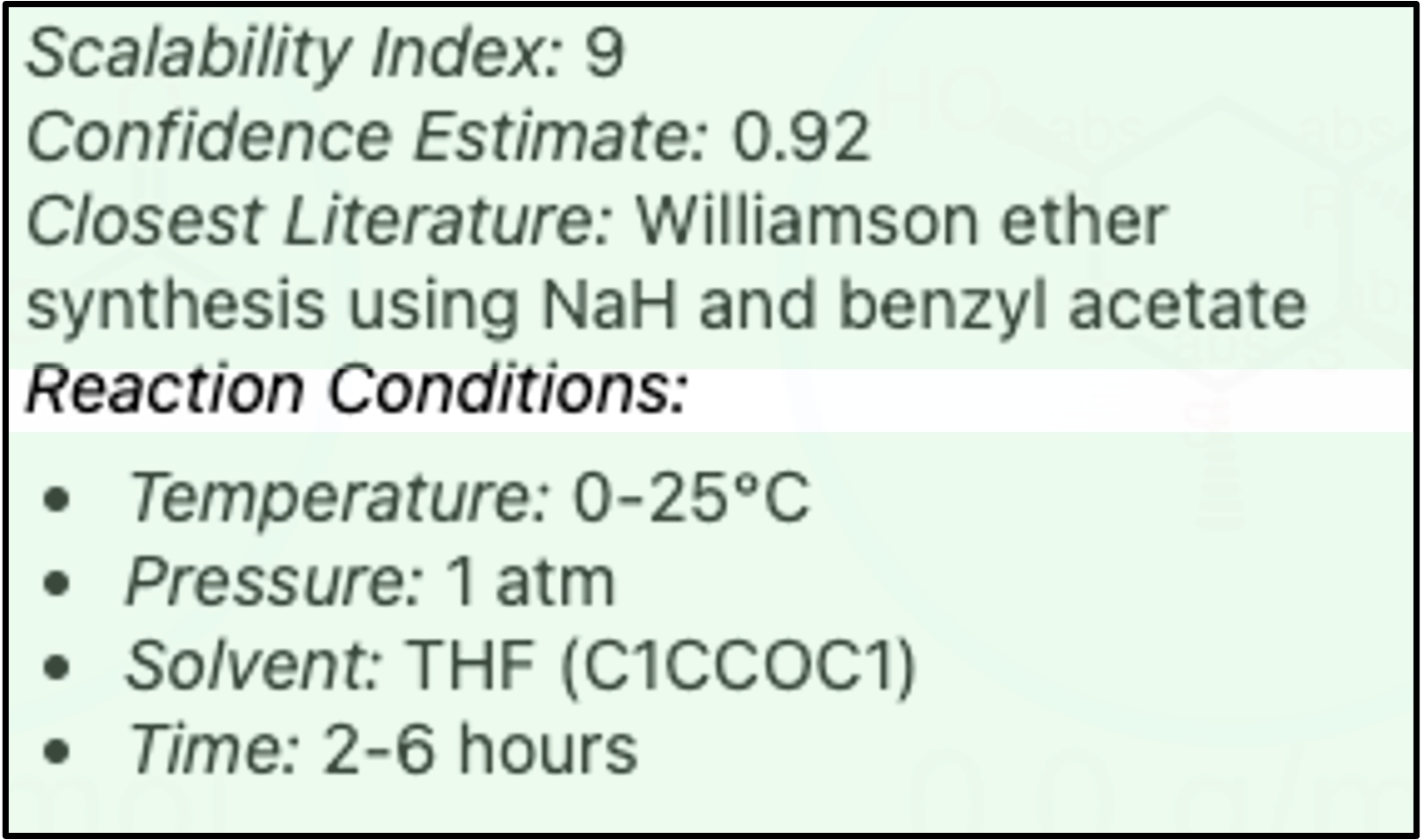}}
    \caption{Step 8 generated by DeepRetro. (a) Shows the pathway and (b) shows the Reaction Metrics} 
    \label{fig:mol_paths:mol3:s8} 
\end{figure}

For Step 8, DeepRetro generated the below pathway without human intervention. This is shown in figure \ref{fig:mol_paths:mol3:s8}

The SMILES and reaction metrics for step 8 are shared below. \\
\begin{verbatim}
Smiles:
    Product:  O[C@H]([C@@H](C)CO[C@@H]1O[C@@H](C)[C@@H]([C@](C1)
    (C)OC)OCC2=CC=CC=C2)[C@](C[C@@H](C)C(CC)=O)(C)O
    Reactant: (3h) O[C@H]([C@@H](C)CO[C@@H]1O[C@@H](C)[C@@H]([C@](C1)(C)OC)O)
    [C@](C[C@@H](C)C(CC)=O)(C)O
              (3h')CC(OCC1=CC=CC=C1)=O
\end{verbatim}


\paragraph{Step 9}

\begin{figure}[h!]
    \centering
    \subfigure[]{\includegraphics[width=0.48\linewidth]{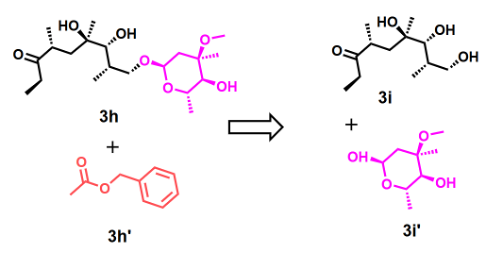}}
    \subfigure[]{\includegraphics[width=0.48\textwidth]{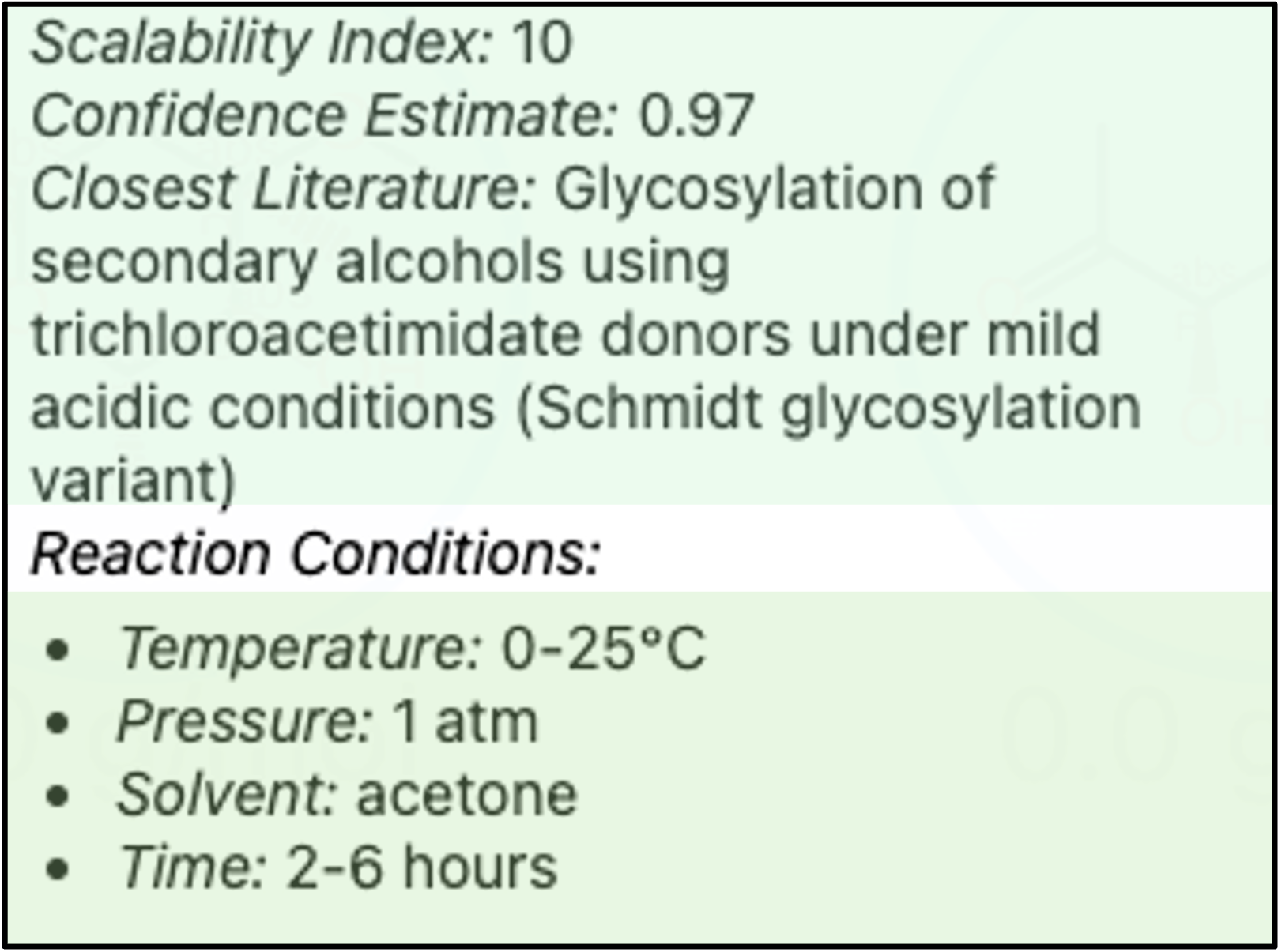}}
    \caption{Step 9 generated by DeepRetro. (a) Shows the pathway and (b) shows the Reaction Metrics} 
    \label{fig:mol_paths:mol3:s9} 
\end{figure}

For Step 9, DeepRetro generated the below pathway without human intervention. This is shown in figure \ref{fig:mol_paths:mol3:s9}

The SMILES and reaction metrics for step 9 are shared below. \\
\begin{verbatim}
Smiles:
    Product:  O[C@H]([C@@H](C)CO[C@@H]1O[C@@H](C)[C@@H]([C@](C1)(C)OC)O)
    [C@](C[C@@H](C)C(CC)=O)(C)O
    Reactant: (3i)  O[C@H]([C@@H](C)CO)[C@](C[C@@H](C)C(CC)=O)(C)O
              (3i') O[C@@H]1O[C@@H](C)[C@@H]([C@](C1)(C)OC)O
\end{verbatim}

\paragraph{Step 10}

\begin{figure}[h!]
    \centering
    \subfigure[]{\includegraphics[width=0.48\linewidth]{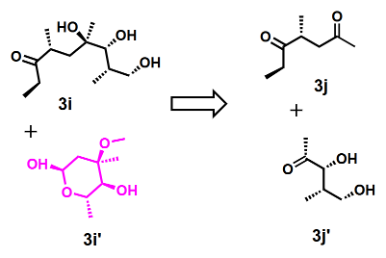}}
    \subfigure[]{\includegraphics[width=0.48\textwidth]{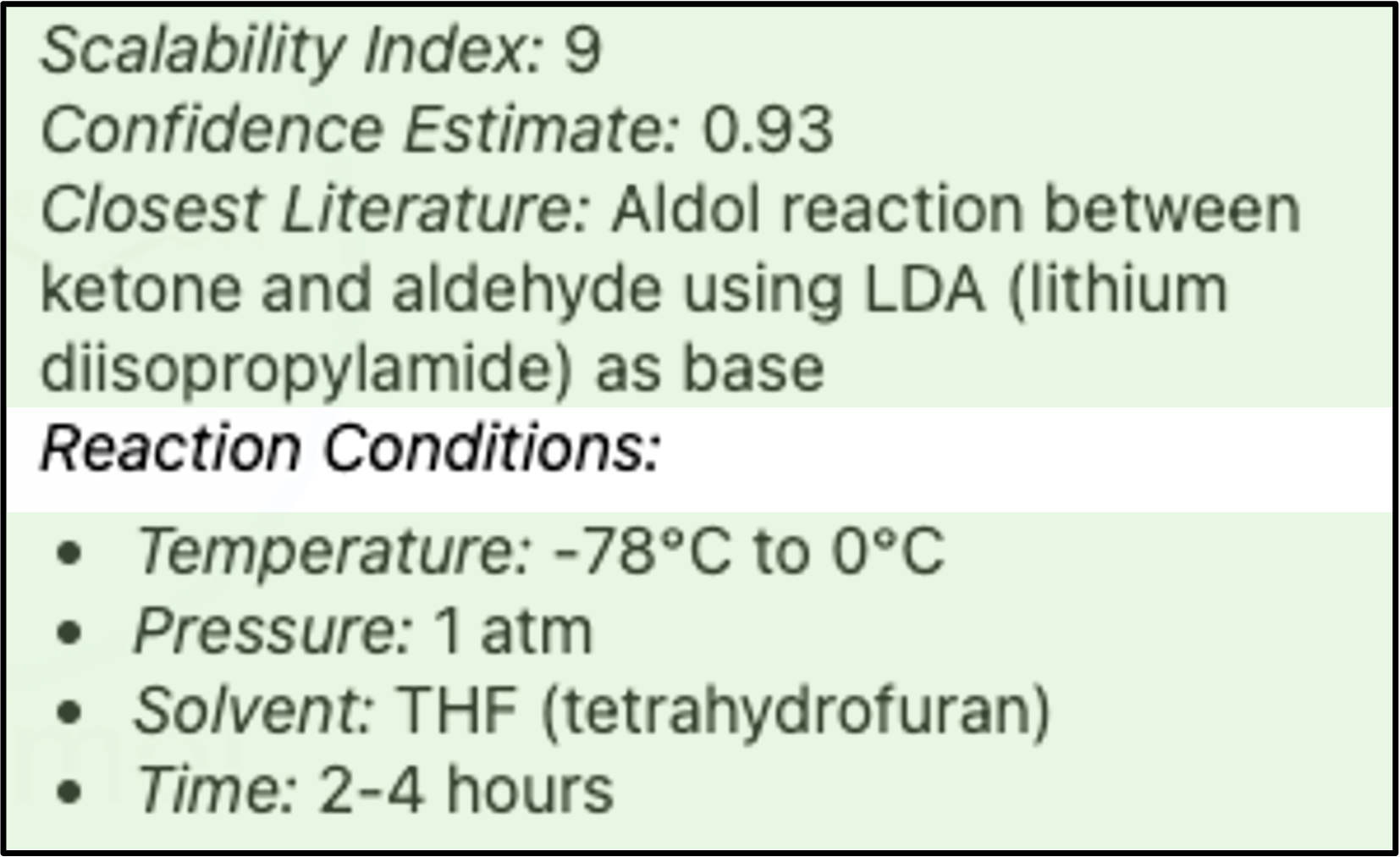}}
    \caption{Step 10 generated by DeepRetro. (a) Shows the pathway and (b) shows the Reaction Metrics} 
    \label{fig:mol_paths:mol3:s10} 
\end{figure}

For Step 10, DeepRetro generated the below pathway without human intervention. This is shown in figure \ref{fig:mol_paths:mol3:s10}

The SMILES and reaction metrics for Step 10 are shared below. \\
\begin{verbatim}
Smiles:
    Product:  O[C@H]([C@@H](C)CO)[C@](C[C@@H](C)C(CC)=O)(C)O
    Reactant: (3j)  O=C(CC)[C@H](C)CC(C)=O
              (3j') O[C@H]([C@@H](C)CO)C(C)=O
\end{verbatim}

\subsection{Molecule 4: Reserpine}
\label{app:mol_paths:mol4}
Reserpine, (methyl (1S,2R,3R,4aS,13bR,14aS)-2,11-dimethoxy-3-((3,4,5-trimethoxybenzoyl)oxy)-1,2,3,4,4a,5,7,8,13,13b,14,14a-dodecahydroindolo[2',3':3,4]pyrido[1,2-b]isoquinoline-1-carboxylate) a highly oxygenated indole alkaloid, features a pentacyclic core with multiple stereocenters and a rich array of functional groups, making it a formidable synthetic target. To assess its retrosynthetic accessibility, DeepRetro was deployed to map a strategic disconnection pathway, with emphasis on modular transformations and stereocontrolled ring construction.

\paragraph{Exact human intervention}

Human intervention involved protecting the active hydroxy group as acetoxy (AcO) in intermediate 4b and suggesting the crucial lactamization step to ensure correct ring formation and stereochemical fidelity.

\paragraph{Step 1}

\begin{figure}[!h]
    \centering
    \subfigure[]{\includegraphics[width=0.48\linewidth]{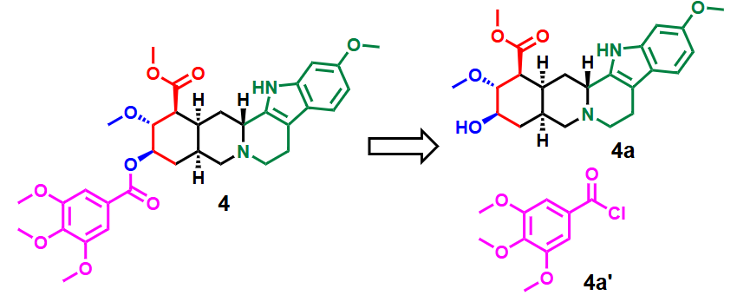}}
    \subfigure[]{\includegraphics[width=0.38\textwidth]{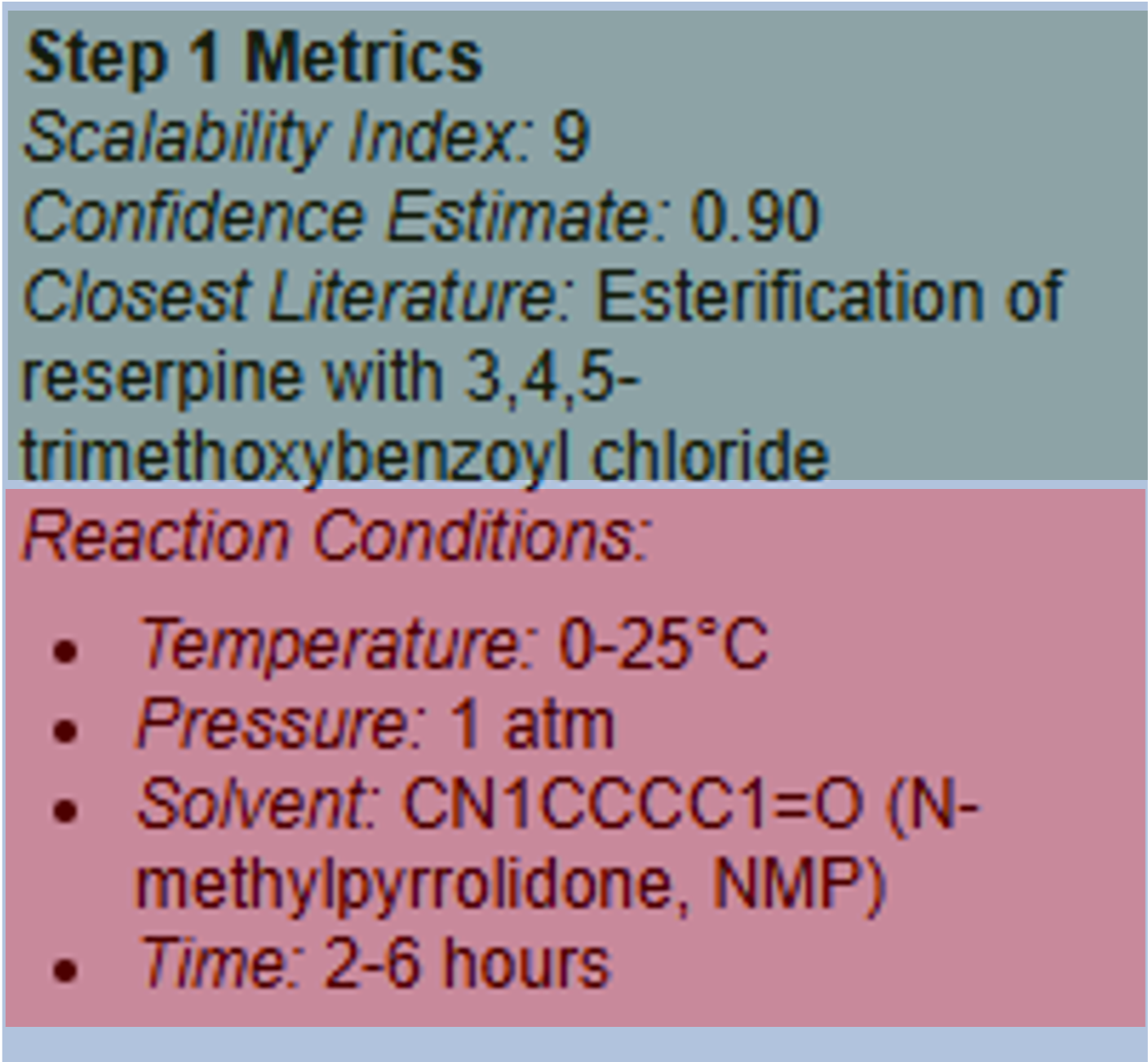}}
    \caption{Step 1 generated by DeepRetro. (a) Shows the pathway and (b) shows the Reaction Metrics} 
    \label{fig:mol_paths:mol4:s1} 
\end{figure}

Step 1 involves the breakage of the ester bond to form intermediate 4a, bearing the isoquinoline framework, with an activated methoxybenzoyl chloride derivative (4a') . This is shown in figure \ref{fig:mol_paths:mol4:s1}

The SMILES and reaction metrics for step 1 are shared below. \\
\begin{verbatim}
Smiles:
    Product: [H][C@]12C[C@H]([C@@H]([C@H]([C@]1(C[C@]3(N(C2)CCC4=C3
    NC5=C4C=CC(OC)=C5)[H])[H])C(OC)=O)OC)OC(C6=CC(OC)=C(C(OC)=C6)OC)=O
    Reactant: [H][C@]12C[C@H]([C@@H]([C@H]([C@]1(C[C@]3(N(C2)CCC4=C3
    NC5=C4C=CC(OC)=C5)[H])[H])C(OC)=O)OC)O
               ClC(C1=CC(OC)=C(C(OC)=C1)OC)=O
\end{verbatim}

\paragraph{Step 2}

\begin{figure}[!h]
    \centering
    \subfigure[]{\includegraphics[width=0.48\linewidth]{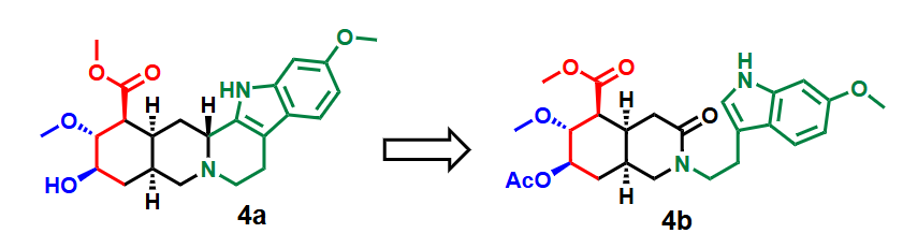}}
    \subfigure[]{\includegraphics[width=0.36\textwidth]{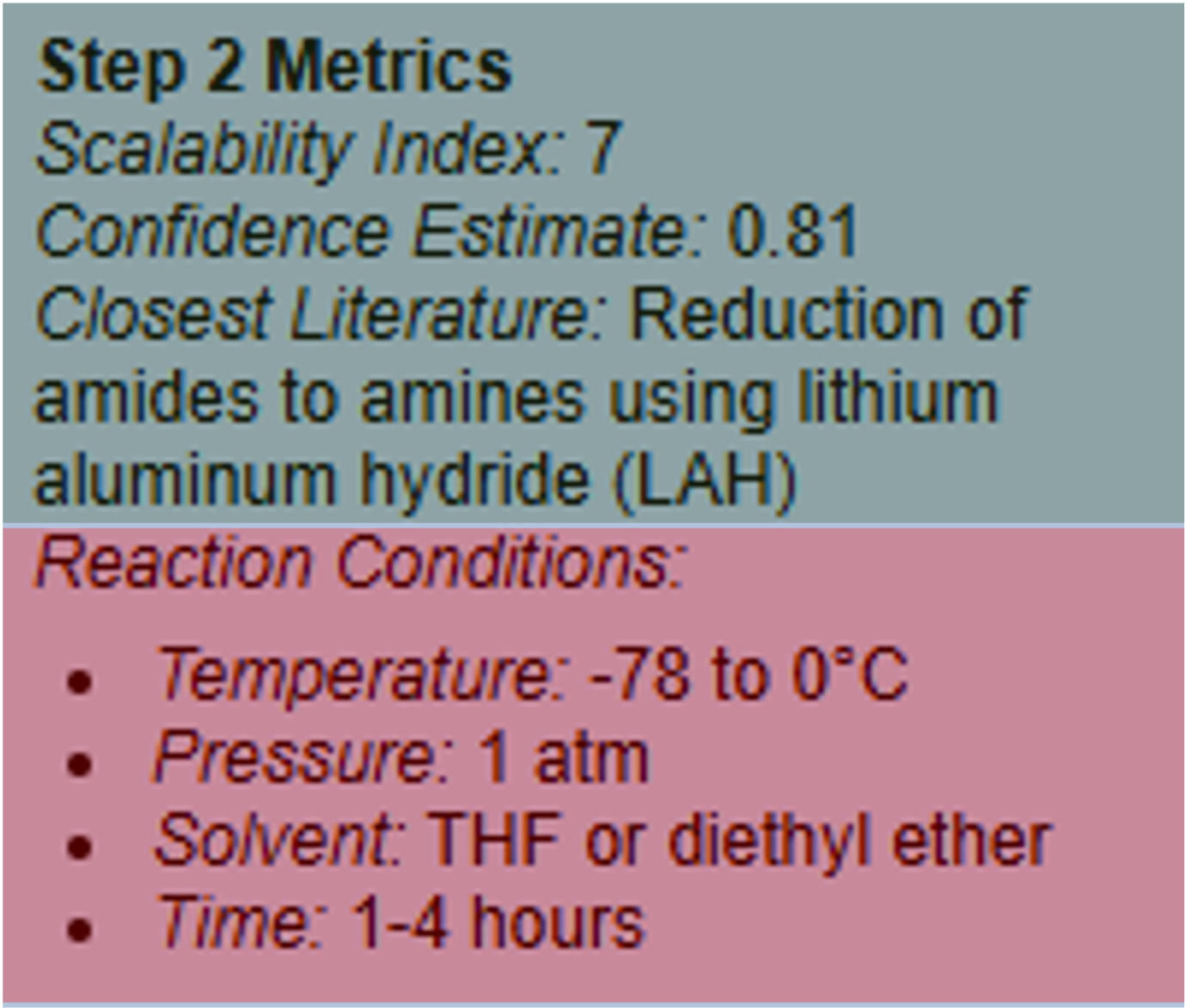}}
    \caption{Step 2 generated by DeepRetro. (a) Shows the pathway and (b) shows the Reaction Metrics} 
    \label{fig:mol_paths:mol4:s2} 
\end{figure}

Retrosynthetic simplification of 4a reveals its origin via a Bischler–Napieralski cyclisation, a pivotal heterocyclization that constructs the isoquinoline nucleus (4a ← 4b). This is shown in figure \ref{fig:mol_paths:mol4:s2}

The SMILES and reaction metrics for step 2 are shared below. \\
\begin{verbatim}
Smiles:
    Product: [H][C@]12C[C@H]([C@@H]([C@H]([C@]1(C[C@]3(N(C2)CCC4=C3
    NC5=C4C=CC(OC)=C5)[H])[H])C(OC)=O)OC)O
    Reactant: [H][C@]12C[C@H]([C@@H]([C@H]([C@]1(CC(N(C2)CCC3=CNC4=C3
    C=CC(OC)=C4)=O)[H])C(OC)=O)OC)OC(C)=O
\end{verbatim}

\paragraph{Step 3}

\begin{figure}[!h]
    \centering
    \subfigure[]{\includegraphics[width=0.48\linewidth]{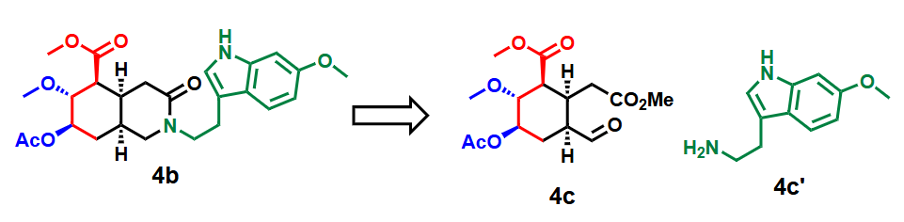}}
    \subfigure[]{\includegraphics[width=0.36\textwidth]{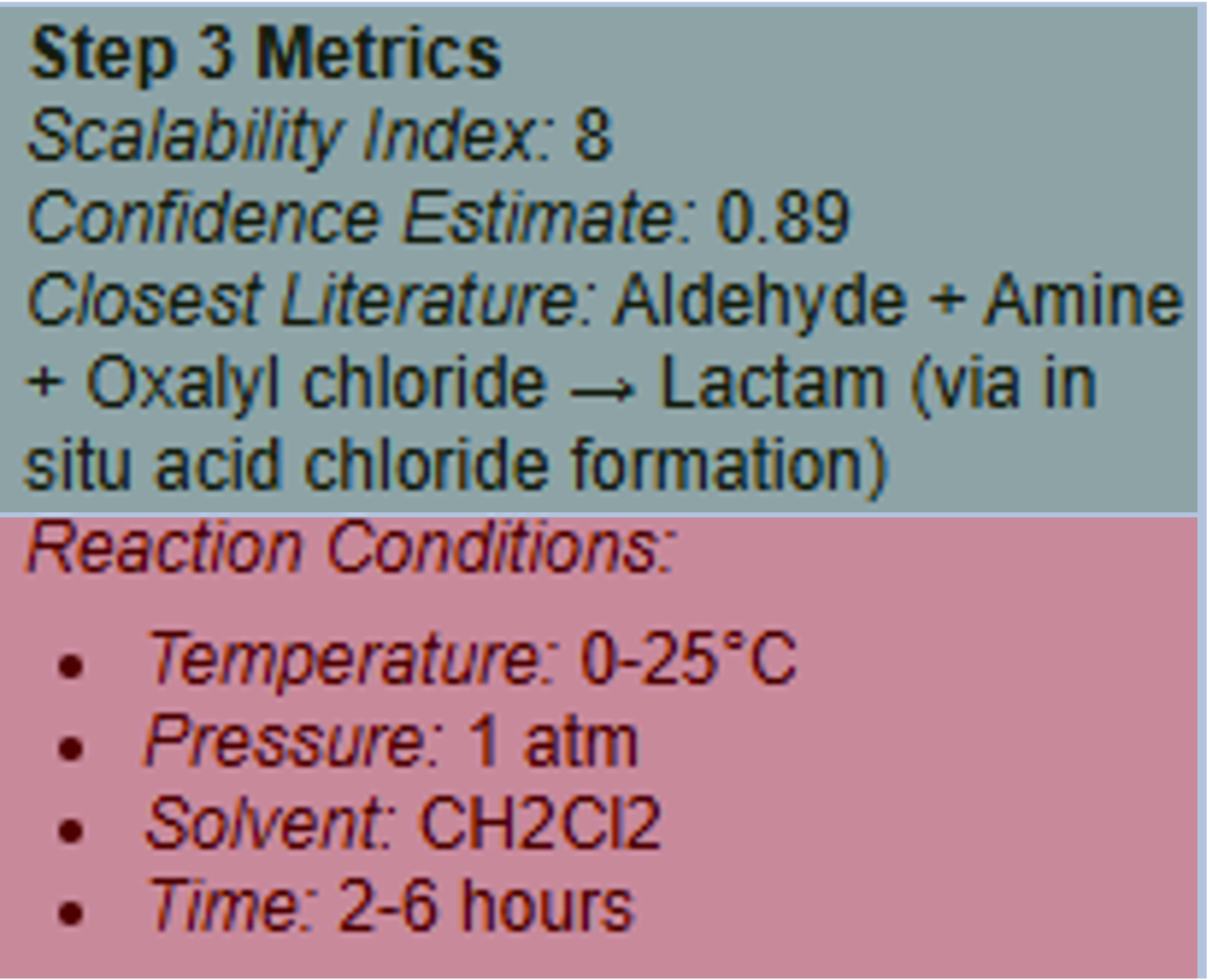}}
    \caption{Step 3 generated by DeepRetro. (a) Shows the pathway and (b) shows the Reaction Metrics} 
    \label{fig:mol_paths:mol4:s3} 
\end{figure}

Precursor 4b is derived through lactamization of the methyl ester-containing intermediate 4c. This is shown in figure \ref{fig:mol_paths:mol4:s3}

The SMILES and reaction metrics for step 3 are shared below. \\
\begin{verbatim}
Smiles:
    Product: [H][C@]12C[C@H]([C@@H]([C@H]([C@]1(CC(N(C2)CCC3=CNC4=C3
    C=CC(OC)=C4)=O)[H])C(OC)=O)OC)OC(C)=O
    Reactant: [H][C@]1(C=O)C[C@H]([C@@H]([C@H]([C@]1(CC(OC)=O)[H])
    C(OC)=O)OC)OC(C)=O
               NCCC1=CNC2=C1C=CC(OC)=C2
\end{verbatim}


\paragraph{Step 4}

\begin{figure}[!h]
    \centering
    \subfigure[]{\includegraphics[width=0.48\linewidth]{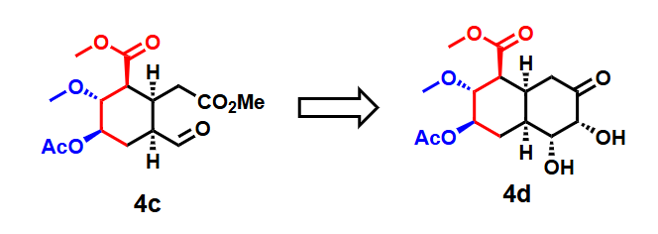}}
    \subfigure[]{\includegraphics[width=0.36\textwidth]{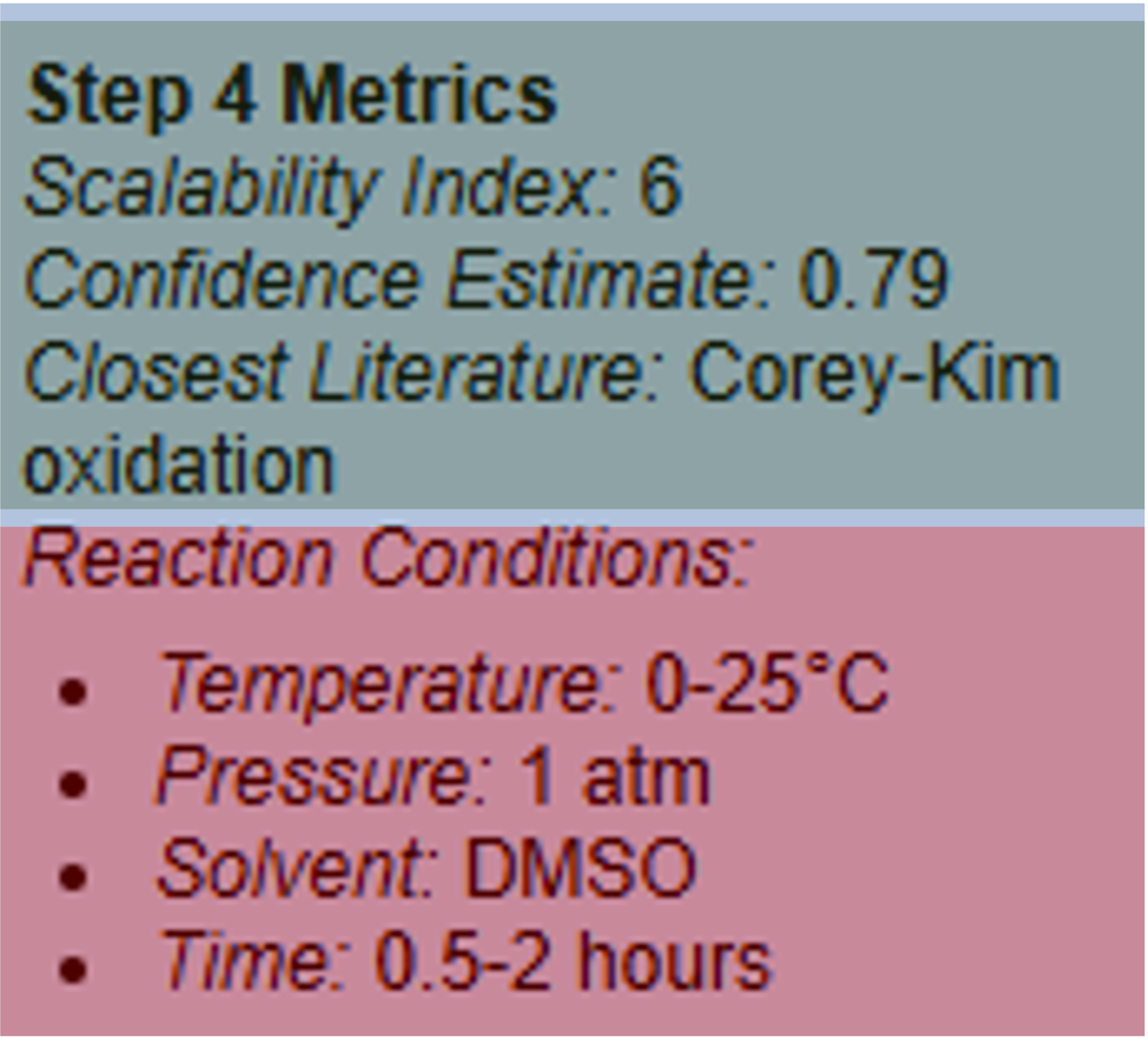}}
    \caption{Step 4 generated by DeepRetro. (a) Shows the pathway and (b) shows the Reaction Metrics} 
    \label{fig:mol_paths:mol4:s4} 
\end{figure}

Precursor 4c is derived from strategic esterification and methylation of diol 4d. This is shown in figure \ref{fig:mol_paths:mol4:s4}

The SMILES and reaction metrics for step 4 are shared below. \\
\begin{verbatim}
Smiles:
    Product: [H][C@]1(C=O)C[C@H]([C@@H]([C@H]([C@]1(CC(OC)=O)[H])
    C(OC)=O)OC)OC(C)=O
    Reactant: [H][C@]1([C@@H](O)[C@H]2O)C[C@H]([C@@H]([C@H]([C@]1
    (CC2=O)[H])C(OC)=O)OC)OC(C)=O
\end{verbatim}


\paragraph{Step 5}

\begin{figure}[!h]
    \centering
    \subfigure[]{\includegraphics[width=0.48\linewidth]{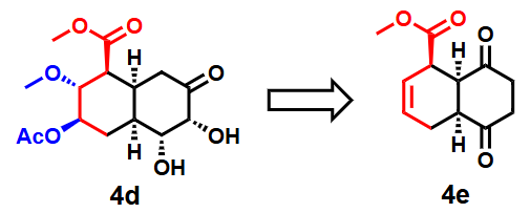}}
    \subfigure[]{\includegraphics[width=0.36\textwidth]{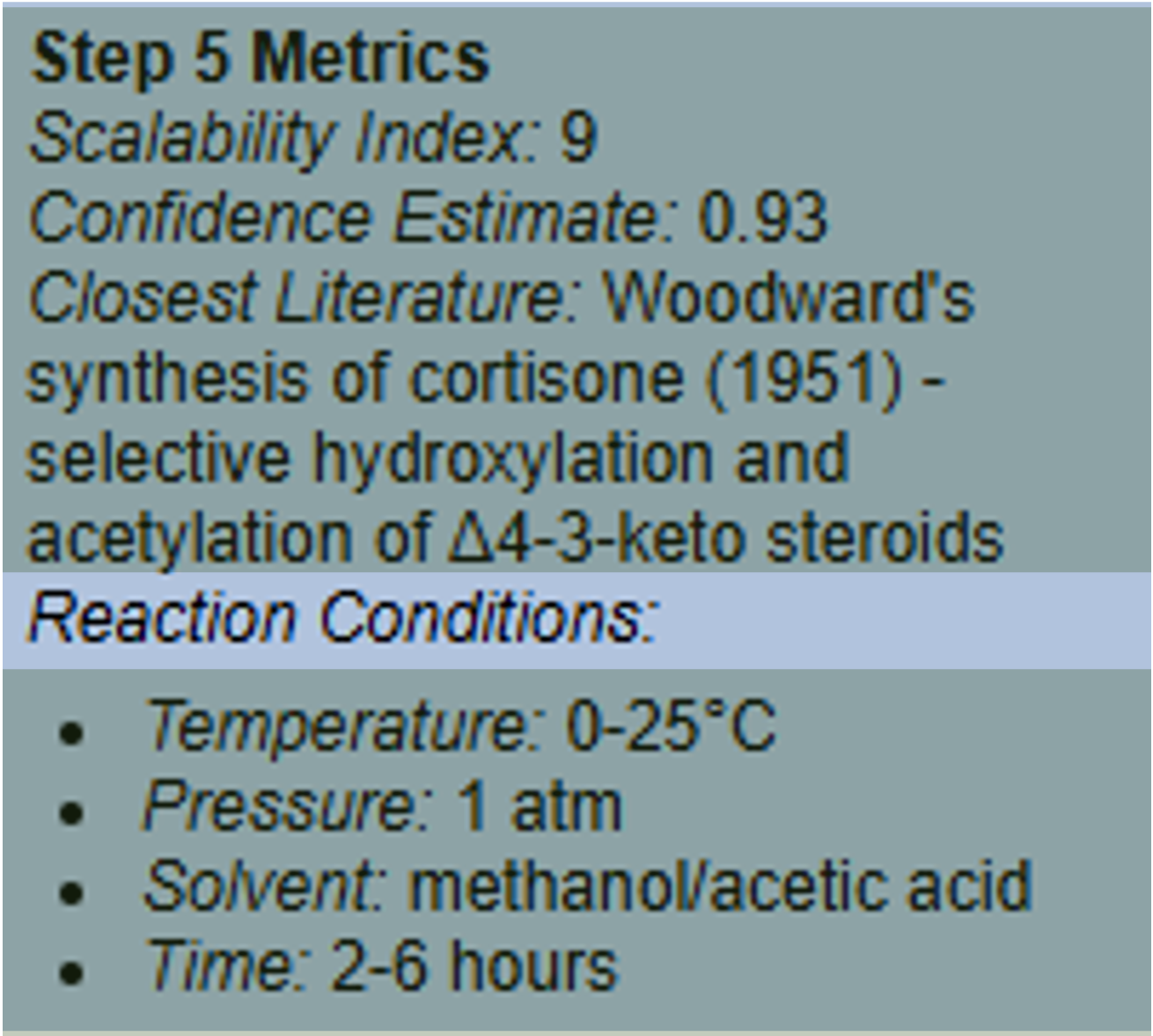}}
    \caption{Step 5 generated by DeepRetro. (a) Shows the pathway and (b) shows the Reaction Metrics} 
    \label{fig:mol_paths:mol4:s5} 
\end{figure}

The diol 4d arises from controlled oxidation of 4e. This is shown in figure \ref{fig:mol_paths:mol4:s5}

The SMILES and reaction metrics for step 5 are shared below. \\
\begin{verbatim}
Smiles:
    Product: [H][C@]1([C@@H](O)[C@H]2O)C[C@H]([C@@H]([C@H]([C@]1
    (CC2=O)[H])C(OC)=O)OC)OC(C)=O
    Reactant: [H][C@]12CC=C[C@H]([C@]1(C(CCC2=O)=O)[H])C(OC)=O
\end{verbatim}

\paragraph{Step 6}

\begin{figure}[!h]
    \centering
    \subfigure[]{\includegraphics[width=0.48\linewidth]{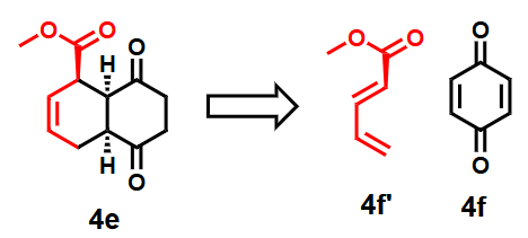}}
    \subfigure[]{\includegraphics[width=0.36\textwidth]{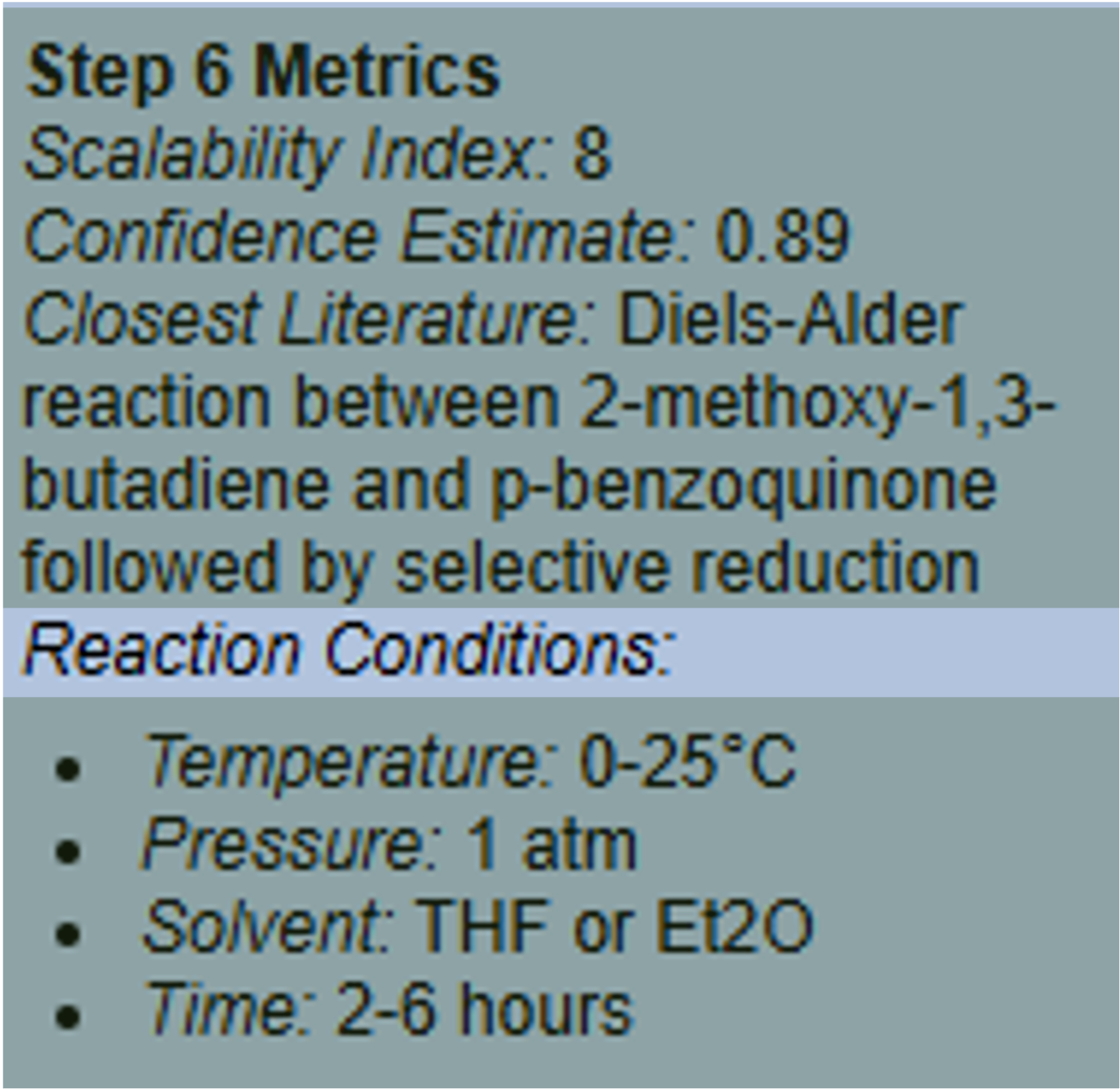}}
    \caption{Step 6 generated by DeepRetro. (a) Shows the pathway and (b) shows the Reaction Metrics} 
    \label{fig:mol_paths:mol4:s6} 
\end{figure}

The precursor 4e, a key intermediate forged via a regioselective Diels–Alder reaction between a substituted diene and dienophile (4f and 4f', respectively). This is shown in figure \ref{fig:mol_paths:mol4:s6}

The SMILES and reaction metrics for step 6 are shared below. \\
\begin{verbatim}
Smiles:
    Product: [H][C@]12CC=C[C@H]([C@]1(C(CCC2=O)=O)[H])C(OC)=O
    Reactant: O=C(C=C1)C=CC1=O
              C=C/C=C/C(OC)=O
\end{verbatim}

\subsection{Molecule 5: Discodermolide}
\label{app:mol_paths:mol4}
Discodermolide, (3Z,5S,6S,7S,8R,9S,11Z,13S,14S,15S,16Z,18S)-8,14,18-trihydroxy-19-((2S,3R,4S,5R)-4-hydroxy-3,5-dimethyl-6-oxotetrahydro-2H-pyran-2-yl)-5,7,9,11,13,15-hexamethylnonadeca-1,3,11,16-tetraen-6-yl carbamate a highly oxygenated polyketide natural product with potent anticancer properties, presents a formidable synthetic challenge due to its extended carbon framework and dense array of stereocenters. To evaluate DeepRetro’s ability to navigate such complexity, a convergent retrosynthetic pathway was devised, showcasing a modular strategy grounded in fragment coupling and stereocontrolled transformations.

\paragraph{Exact human intervention}

The human contribution guided the LLM to systematically divide discodermolide into three key fragments (5a, 5a', 5a''), enabling their convergent assembly into a single advanced intermediate (5c). This strategic fragmentation ensured stereochemical control, minimized synthetic complexity, and produced a practical pathway that the LLM alone might not have proposed.

\paragraph{Step 1}

\begin{figure}[!h]
    \centering
    \subfigure[]{\includegraphics[width=0.48\linewidth]{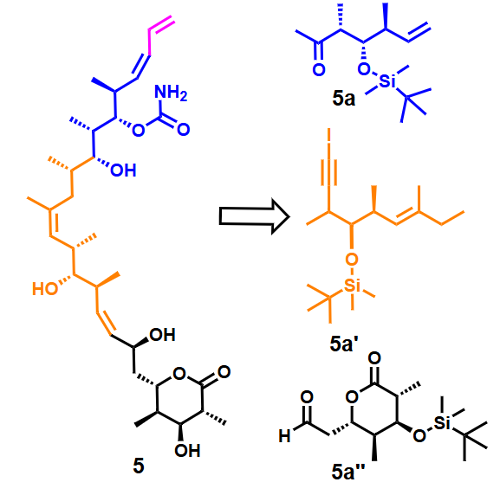}}
    \subfigure[]{\includegraphics[width=0.48\textwidth]{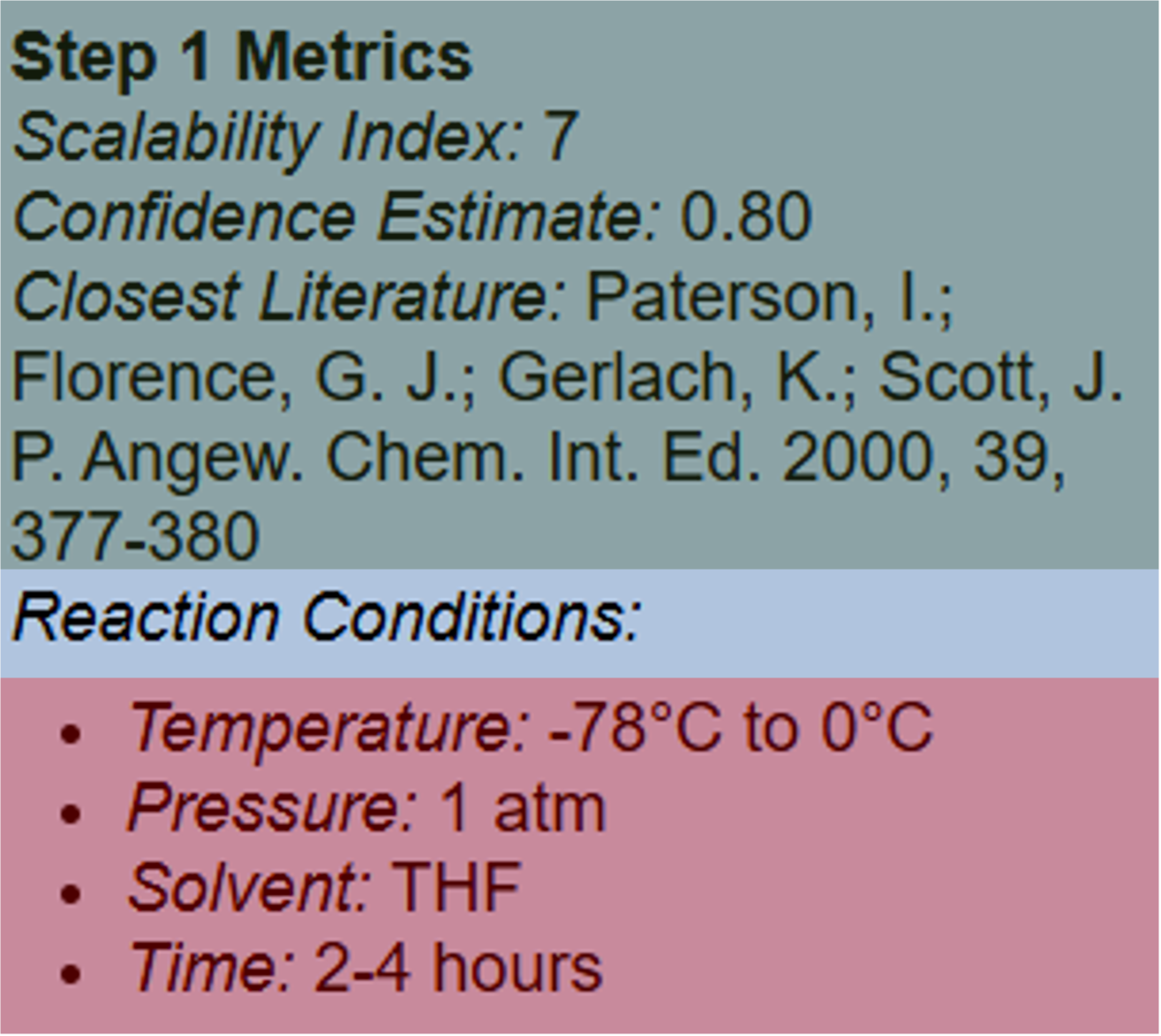}}
    \caption{Step 1 generated by DeepRetro. (a) Shows the pathway and (b) shows the Reaction Metrics} 
    \label{fig:mol_paths:mol5:s1} 
\end{figure}

The target molecule is broken down into 3 key fragments representing the C1–C7, C8–C16, and C17–C24 segments of discodermolide. 
Fragment 5a is constructed via a Nozaki–Kishi coupling, enabling precise installation of the C1–C7 polyol motif. In parallel, fragment 5a'' is synthesised through a Negishi coupling, efficiently forming the terminal C17–C24 unit while preserving stereochemical integrity. The central segment, 5a', is accessed through an enolate alkylation strategy, enabling controlled formation of the C13–C16 segment.
This is shown in figure \ref{fig:mol_paths:mol5:s1}

The SMILES and reaction metrics for step 1 are shared below. \\
\begin{verbatim}
Smiles:
    Product: C[C@H]1[C@H](C[C@H](O)/C=C\[C@@H]([C@H](O)[C@H](/C=C(C[C@@H]
    ([C@@H](O)[C@@H]([C@@H](OC(N)=O)[C@H](/C=C\C=C)C)C)C)/C)C)C)OC
    ([C@H](C)[C@H]1O)=O
    Reactant: C[C@H]([C@@H](O[Si](C(C)(C)C)(C)C)[C@H](C=C)C)C(C)=O
              CC/C(C)=C/[C@H](C)[C@@H](O[Si](C)(C)C(C)(C)C)C(C#CI)C
              O=C1O[C@@H](CC([H])=O)[C@H](C)[C@H](O[Si](C)(C)C(C)(C)C)[C@H]1C
\end{verbatim}

\begin{figure}[!h]
    \centering
    \subfigure[]{\includegraphics[width=0.48\linewidth]{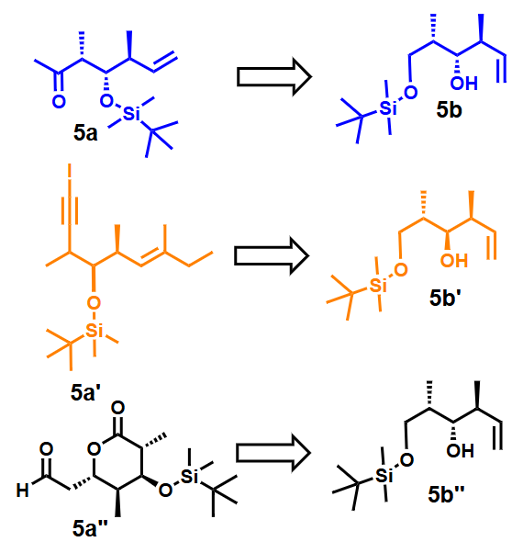}}
    \subfigure[]{\includegraphics[width=0.23\textwidth]{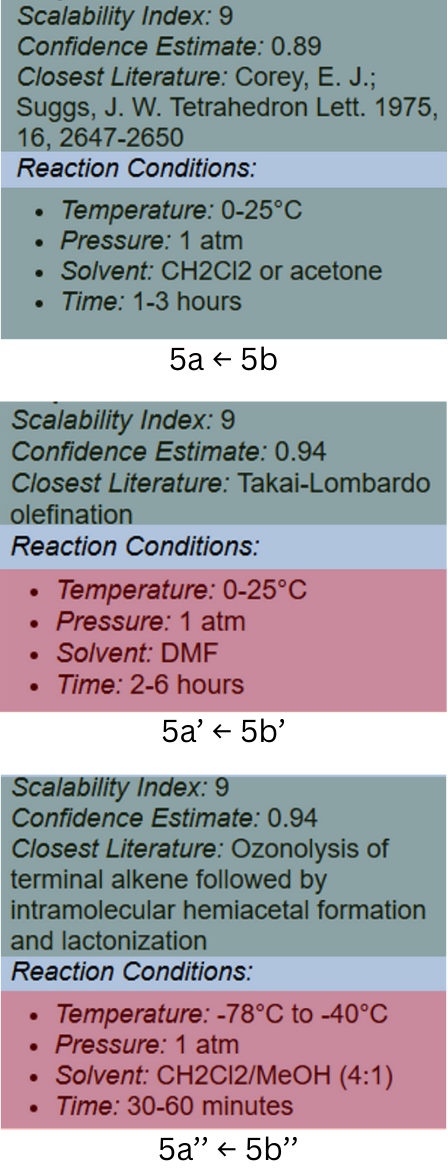}}
    \caption{Step 2 generated by DeepRetro. (a) Shows the pathway and (b) shows the Reaction Metrics} 
    \label{fig:mol_paths:mol5:s2} 
\end{figure}

\paragraph{Step 2}
Fragments 5a and 5a'' are elaborated through Roush crotylation, delivering advanced intermediates 5b and 5b'', respectively. Intermediate 5b' was obtained by a Still–Gennari Horner–Wadsworth–Emmons (HWE) olefination to forge the crucial Z-alkene. 
This is shown in figure \ref{fig:mol_paths:mol5:s2}

The SMILES and reaction metrics for step 2 are shared below. \\
\begin{verbatim}
Smiles:
5a ← 5b
    Product: C[C@H]([C@@H](O[Si](C(C)(C)C)(C)C)[C@H](C=C)C)C(C)=O
    Reactant: C=C[C@H](C)[C@H](O)[C@@H](C)CO[Si](C(C)(C)C)(C)C

5a' ← 5b'
    Product: CC/C(C)=C/[C@H](C)[C@@H](O[Si](C)(C)C(C)(C)C)C(C#CI)C
    Reactant: C=C[C@H](C)[C@@H](O)[C@@H](C)CO[Si](C(C)(C)C)(C)C

5a'' ← 5b''
    Product: O=C1O[C@@H](CC([H])=O)[C@H](C)[C@H](O[Si](C)(C)C(C)(C)C)[C@H]1C
    Reactant: C=C[C@H](C)[C@H](O)[C@@H](C)CO[Si](C(C)(C)C)(C)C
\end{verbatim}

\paragraph{Step 3}

\begin{figure}[!h]
    \centering
    \subfigure[]{\includegraphics[width=0.48\linewidth]{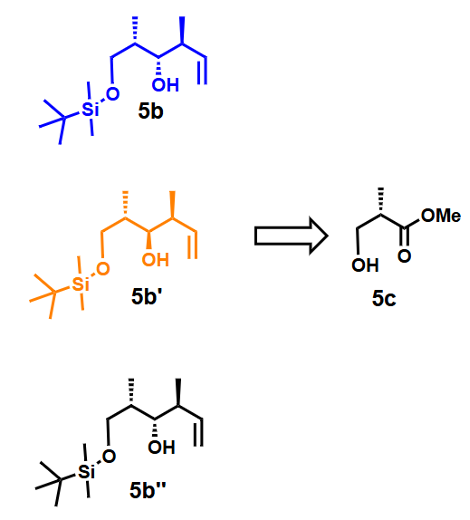}}
    \subfigure[]{\includegraphics[width=0.23\textwidth]{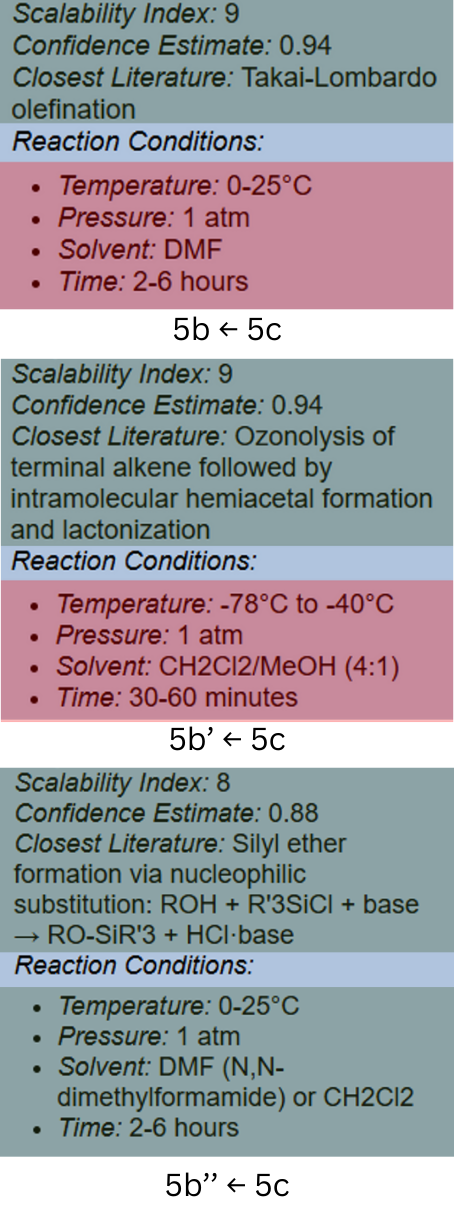}}
    \caption{Step 3 generated by DeepRetro. (a) Shows the pathway and (b) shows the Reaction Metrics} 
    \label{fig:mol_paths:mol5:s3} 
\end{figure}

Fragments 5b, 5b', and 5b'' are broken down into the intermediate 5c, bearing the full carbon skeleton and key stereochemical features of discodermolide.
This is shown in figure \ref{fig:mol_paths:mol5:s3}

The SMILES and reaction metrics for step 3 are shared below. \\
\begin{verbatim}
Smiles:
    Product: C=C[C@H](C)[C@H](O)[C@@H](C)CO[Si](C(C)(C)C)(C)C
             C=C[C@H](C)[C@@H](O)[C@@H](C)CO[Si](C(C)(C)C)(C)C
             C=C[C@H](C)[C@H](O)[C@@H](C)CO[Si](C(C)(C)C)(C)C
    Reactant: C[C@H](C(OC)=O)CO
\end{verbatim}


\section{Reproducibility}
\label{app:repro}

Algorithm \ref{alg:Chemist_in_loop} gives the pseudocode for Chemist-in-the-Loop Retrosynthetic Route Generation that may serve as a base for a future fully-automated algorithm that does not require human-in-the-loop intervention.

Procedure \ref{alg:Chemist_in_loop_procedure} gives a lab like procedure that the Chemist can follow to perform analyses with DeepRetro


\begin{algorithm}
\caption{Chemist-in-the-Loop Retrosynthetic Route Generation}
\label{alg:Chemist_in_loop}
\begin{algorithmic}[1]
\REQUIRE Target molecule $M$ (SMILES representation)
\ENSURE Validated retrosynthetic route $R$

\STATE $\text{satisfied} \leftarrow \text{False}$
\WHILE{$\text{satisfied} = \text{False}$}
    \STATE Input molecule $M$ into DeepRetro system
    \STATE $\mathcal{R} \leftarrow \text{GenerateRetrosynthesis}(M)$ \COMMENT{Generate candidate routes}
    
    \STATE $R \leftarrow \text{ChemistSelect}(\mathcal{R})$ \COMMENT{Chemist selects most feasible route}
    
    \STATE $\text{step\_valid} \leftarrow \text{ChemistValidate}(R[0])$ \COMMENT{Check first step}
    \IF{$\text{step\_valid} = \text{False}$}
        \STATE \textbf{continue} \COMMENT{Rerun entire route generation}
    \ENDIF
    
    \STATE $\text{route\_satisfaction} \leftarrow \text{ChemistEvaluate}(R)$
    \IF{$\text{route\_satisfaction} = \text{True}$}
        \STATE Download and save route $R$
        \STATE $\text{satisfied} \leftarrow \text{True}$
    \Else
        \STATE Choose refinement strategy:
        \IF{partial route needs modIFication}
            \STATE $R \leftarrow \text{RerunPartialRoute}(R, \text{specIFied\_steps})$
        \ELSIF{molecular structure needs correction}
            \STATE $M \leftarrow \text{ChemistEditSMILES}(M)$ \COMMENT{Correct minor mistakes}
        \ELSIF{protecting groups needed}
            \STATE $M \leftarrow \text{AddProtectingGroups}(M)$
        \ENDIF
    \ENDIF
\ENDWHILE

\Return $R$ \COMMENT{Validated retrosynthetic route}
\end{algorithmic}
\end{algorithm}

\begin{algorithm}[H]
\caption{Reproducibility Procedure for Chemists with Step-wise Validation}
\label{alg:Chemist_in_loop_procedure}
\begin{algorithmic}[1]
\REQUIRE Input molecule $M$
\ENSURE Satisfactory retrosynthetic route $R$
\STATE Chemist enters molecule $M$ into DeepRetro system.
\STATE DeepRetro generates a set of retrosynthetic routes $\mathcal{R} = \{R_1, R_2, \ldots, R_n\}$.
\STATE Chemist selects the most promising route $R^* \in \mathcal{R}$.
\STATE Chemist examines the first retrosynthetic step of the selected route $R^*$.
\IF{the first step is deemed unsatisfactory}
    \STATE Go to Step 2 to generate a new set of routes.
\ELSE
    \IF{Chemist is satisfied with the entire route $R^*$}
        \STATE Chemist downloads route $R^*$.
    \ELSE
        \STATE Chemist selects one of the following options for route modification:
        \STATE \hspace{1em} \textbf{Option a:} Request rerun of a partial route segment.
        \STATE \hspace{1em} \textbf{Option b:} Edit molecule SMILES to correct minor errors and rerun.
        \STATE \hspace{1em} \textbf{Option c:} Add a protecting group to the molecule and rerun.
        \STATE Execute the selected option and go to Step 2.
    \ENDIF
\ENDIF
\STATE Repeat the process until a satisfactory route is confirmed and downloaded.
\RETURN $R^*$
\end{algorithmic}
\end{algorithm}


    

\section{Iterative DeepRetro Algorithm}
\label{app:dr_algo}

Algorithm \ref{alg:recursive_prithvi} showcases the iterative algorithm used in DeepRetro

\begin{algorithm}
\caption{Recursive Retrosynthesis with DeepRetro}
\label{alg:recursive_prithvi}
\begin{algorithmic}[1]
\REQUIRE Target molecule $M$, LLM model $\mathcal{L}$, AZ model $\mathcal{A}$
\ENSURE Synthesis tree $\mathcal{T}$, solved status $\sigma$

\STATE $\sigma, \mathcal{T} \leftarrow \text{AiZynthFinder}(M, \mathcal{A})$

\IF{$\sigma = \text{False}$}
    \STATE $\mathcal{P}, \mathcal{E}, \mathcal{C} \leftarrow \text{LLMPipeline}(M, \mathcal{L})$; //{AZ failed, use LLM for retrosynthesis}
    
    \STATE Initialize synthesis tree $\mathcal{T}$ with molecule $M$ and confidence $\mathcal{C}$
    
    \FOR{each pathway $p \in \mathcal{P}$}
        \IF{$p$ is a reaction pathway (list of precursors)}
            \STATE $\text{all\_solved} \leftarrow \text{True}$
            \FOR{each precursor molecule $m \in p$}
                \STATE $\mathcal{T}_{\text{sub}}, \sigma_{\text{sub}} \leftarrow \text{RecursivePrithvi}(m, \mathcal{L}, \mathcal{A})$
                \IF{$\sigma_{\text{sub}} = \text{True}$}
                    \STATE Add $\mathcal{T}_{\text{sub}}$ to $\mathcal{T}$ as child
                \ELSE
                    \STATE $\text{all\_solved} \leftarrow \text{False}$
                \ENDIF
            \ENDFOR
            \IF{$\text{all\_solved} = \text{True}$}
                \STATE $\sigma \leftarrow \text{True}$
                \STATE \textbf{break} \COMMENT{Complete pathway found}
            \ENDIF
        \ELSE
            \STATE \COMMENT{Single molecule precursor}
            \STATE $\mathcal{T}_{\text{sub}}, \sigma \leftarrow \text{RecursivePrithvi}(p, \mathcal{L}, \mathcal{A})$
            \STATE Add $\mathcal{T}_{\text{sub}}$ to $\mathcal{T}$ as child
            \IF{$\sigma = \text{True}$}
                \STATE \textbf{break} \COMMENT{Pathway solved}
            \ENDIF
        \ENDIF
    \ENDFOR
\ENDIF

\RETURN $\mathcal{T}, \sigma$
\end{algorithmic}
\end{algorithm}

\begin{algorithm}
    \caption{ASK\_LLM: Interface for querying the LLM for single-step retrosynthesis}
    \label{alg:llm-call}
    \textbf{Input}: $m$: target molecule (SMILES string);
    $L$: an LLM instance;
    $k$: number of suggestions to request; \\ 
    \textbf{Output}: $proposed\_steps, explanations, confidences$: list of suggested steps, associated explanations, confidence scores;
    \begin{algorithmic}[1]
        \STATE Define prompt template for single-step retrosynthesis (e.g., "Given the molecule [SMILES], propose k possible single-step retrosynthetic disconnections. For each, list the precursor SMILES strings and the reaction type.").
        \STATE Format prompt with input molecule $m$ and $k$.
        \STATE $response$ = $L(\text{prompt})$; \COMMENT{Send prompt to LLM API/model}
        \STATE $proposed\_steps, explanations, confidences$ = $\text{parse\_llm\_response}(response)$; \COMMENT{Extract structured data}
        \RETURN $proposed\_steps, explanations, confidences$
    \end{algorithmic}
\end{algorithm}

The \texttt{ASK\_LLM} function (Algorithm \ref{alg:llm-call}) encapsulates the interaction with the LLM. It involves careful prompt engineering to instruct the LLM to provide single-step retrosynthetic disconnections for the input molecule $m$. The prompt requests $k$ suggestions, asks for precursors in SMILES format and a brief justification. The raw text output from the LLM is then parsed to extract the proposed precursor molecules and any associated metadata. Effective prompting is key to eliciting useful and correctly formatted responses from the LLM.

\section{Pipeline}
\label{app:pipeline}

We show the LLM Pipeline that is used in an algorithmic format

\begin{algorithm}[H]
\caption{LLM-based Retrosynthesis Pipeline}
\label{alg:llm_pipeline}
\begin{algorithmic}[1]
\REQUIRE Target molecule $M$, LLM model $\mathcal{L}$, stability flag $S$, hallucination check flag $H$
\ENSURE Retrosynthesis pathways $\mathcal{P}$, explanations $\mathcal{E}$, confidence scores $\mathcal{C}$

\STATE Initialize $\mathcal{P} \leftarrow \emptyset$, $\mathcal{E} \leftarrow \emptyset$, $\mathcal{C} \leftarrow \emptyset$
\STATE Set $\text{run} \leftarrow 0$, $\text{max\_run} \leftarrow 1.5$ if $S$ or $H$ is true, else $0.6$

\WHILE{$\mathcal{P} = \emptyset$ \AND $\text{run} < \text{max\_run}$}
    \STATE Select current model $\mathcal{L}_{\text{curr}}$ based on run number
    
    \STATE $\text{response} \leftarrow \text{CallLLM}(M, \mathcal{L}_{\text{curr}}, \text{temperature}=\text{run})$; // {Call LLM for retrosynthesis prediction}
    
    \STATE $\text{split\_response} \leftarrow \text{SplitResponse}(\text{response}, \mathcal{L}_{\text{curr}})$; // {Parse LLM response}
    
    \STATE $\text{molecules}, \text{explanations}, \text{confidence} \leftarrow \text{ValidateJSON}(\text{split\_response})$; // {Extract structured data}
    
    \STATE $\mathcal{P}, \mathcal{E}, \mathcal{C} \leftarrow \text{ValidityCheck}(M, \text{molecules}, \text{explanations}, \text{confidence})$; // {Chemical validity check}
    
    \IF{$S$ is true \AND $\mathcal{P} \neq \emptyset$}
        \STATE $\mathcal{P} \leftarrow \text{StabilityChecker}(\mathcal{P})$; //{Stability verification}
    \ENDIF
    
    \IF{$H$ is true \AND $\mathcal{P} \neq \emptyset$}
        \STATE $\mathcal{P} \leftarrow \text{HallucinationChecker}(M, \mathcal{P})$; //{Hallucination detection}
    \ENDIF
    
    \STATE $\text{run} \leftarrow \text{run} + 0.1$
\ENDWHILE

\RETURN $\mathcal{P}, \mathcal{E}, \mathcal{C}$
\end{algorithmic}
\end{algorithm}
\section{Customizability}
\label{app:customization_parameters}
Our implementation allows the end-user to customize several aspects of the search process, enhancing flexibility and practical applicability:
\begin{enumerate}
    \item \textbf{Stock files}: Users specify available starting materials. This defines the termination condition for the recursive search and ensures pathway feasibility based on available chemicals.
    \item \textbf{Expansion policy (for Tool T)}: If $T$ uses MCTS, users can select different policies (e.g., template-based, neural network guided) to guide its search.
    \item \textbf{Filter model (for Tool T)}: Users can employ models within $T$ to quickly filter out unpromising reaction steps based on predicted yield or feasibility scores.
    \item \textbf{Set of starting materials}: Explicitly defines the chemical inventory (same as stock files).
    \item \textbf{Bad reactions/reagents}: Users can specify reaction types (e.g., based on SMARTS patterns) or specific reagents to avoid, reflecting safety concerns or process constraints.
    \item \textbf{Min/Max number of steps}: Constrains the length of the desired pathways.
    \item \textbf{Min/Max number of pathways}: Controls the number of distinct solutions the system attempts to find.
    \item \textbf{Min yield \%age}: Sets a threshold for estimated yield per step, if yield prediction is incorporated into $T$ or the check stages.
\end{enumerate}

\subsection{Open-Source Release}
As part of this work, we are open-sourcing DeepRetro at \href{https://github.com/deepforestsci/DeepRetro}{https://github.com/deepforestsci/DeepRetro} . To ensure transparency and reproducibility, we have publicly released the prompts, model configurations, and evaluation metrics. The datasets used for benchmarking (the 250 subset of the USPTO-50k test set and USPTO-190) are uploaded at \href{https://github.com/deepforestsci/DeepRetro/tree/main/data}{https://github.com/deepforestsci/DeepRetro/tree/main/data}


\section{DeepRetro GUI}
\label{app:gui}
\begin{figure}[!ht]
    \centering
    \includegraphics[width=1\linewidth]{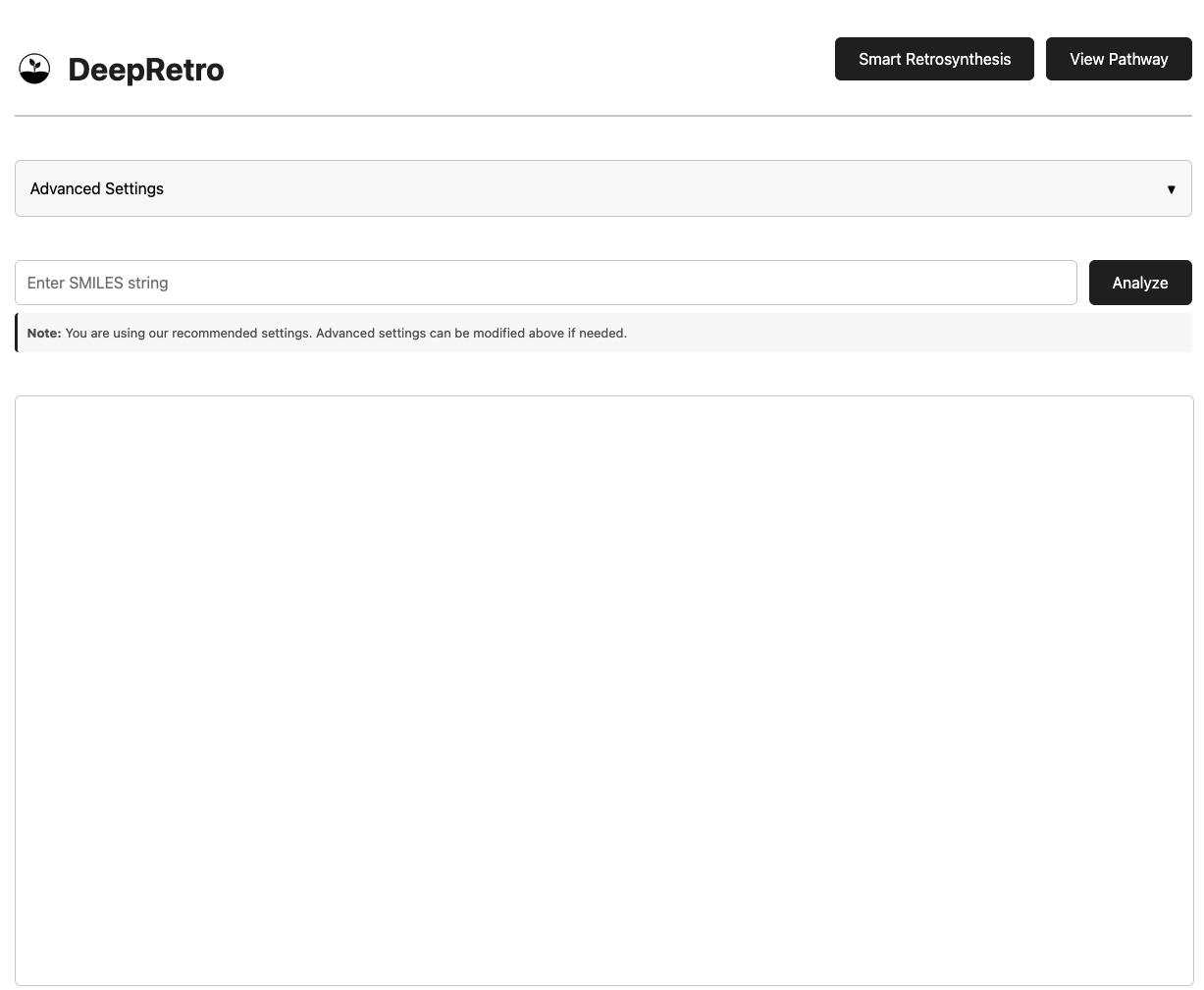}
    \caption{The DeepRetro Landing Page. The graphical interface allows you to set custom settings, run and view the Smart Retrosynthesis Pathway in the dedicated viewer. The user also has the option to use the View Pathway tab which allows them to view a previously run pathway by uploading the relevant JSON pathway file.}
    \label{fig:DR_gui_1}
\end{figure}

\begin{figure}[!ht]
    \centering
    \includegraphics[width=1\linewidth]{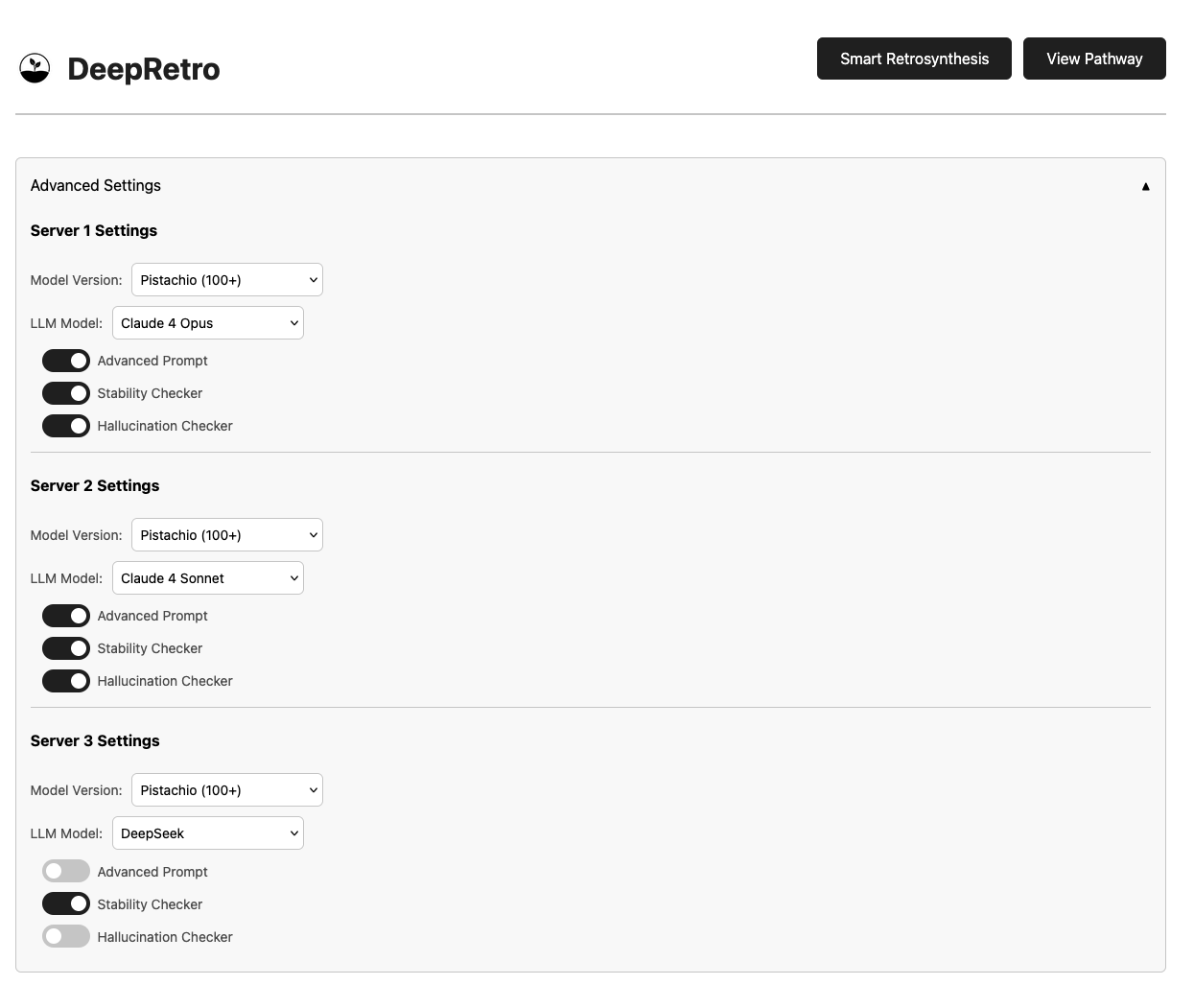}
    \caption{DeepRetro Configuration Selection options. The user has options to select the backend model, LLM model, whether to use advanced prompt, stability checker and hallucination checker or not. These configurations have to be selected for every server allowing granular control to the user.}
    \label{fig:DR_gui_2}
\end{figure}

\begin{figure}[!ht]
    \centering
    \includegraphics[width=1\linewidth]{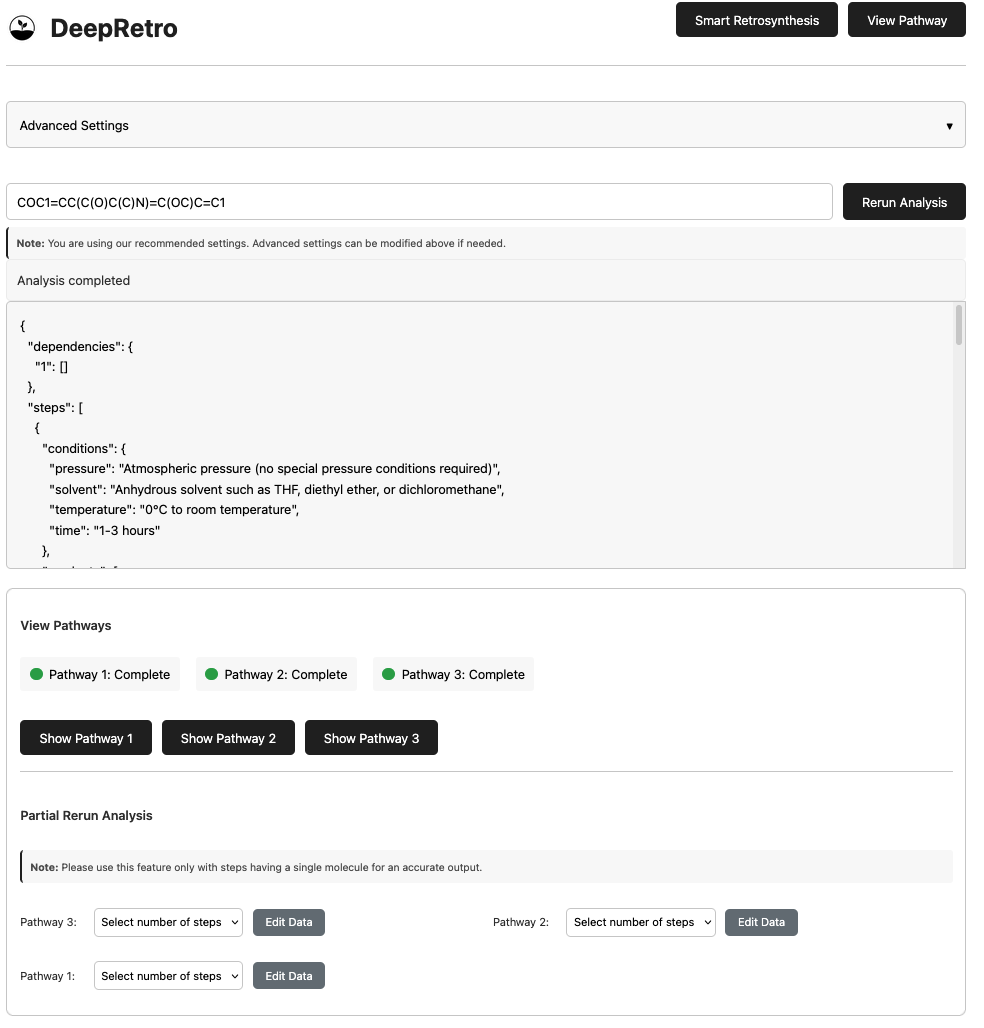}
    \caption{DeepRetro Human-in-the-loop editor. The user can kick off a partial run of the retrosynthesis pathway by selecting the step number post which the pathway would be regenerated, keeping the previous steps as is. This allows the user finer control over the pathway.}
    \label{fig:DR_gui_3}
\end{figure}

\begin{figure}[!ht]
    \centering
    \includegraphics[width=1\linewidth]{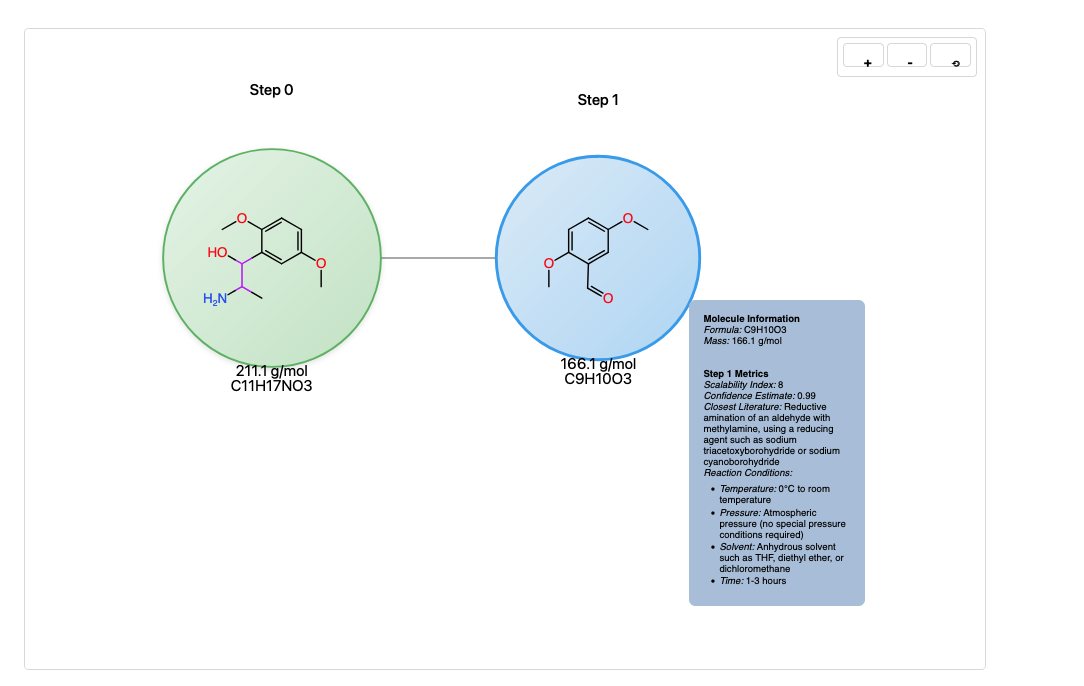}
    \caption{DeepRetro Retrosynthesis pathway viewer with metadata. The user is able to see the entire pathway generated starting from Step 0 being the target molecule. Hovering on any molecule would show the user the reaction metadata.}
    \label{fig:DR_gui_4}
\end{figure}

Figures \ref{fig:DR_gui_1},\ref{fig:DR_gui_2},\ref{fig:DR_gui_3} and \ref{fig:DR_gui_4} showcase the GUI that was built for chemists to easily interface with the DeepRetro backend. These images showcase the landing page, functions of different tabs, granular advanced settings, the Human-in-the-loop editor and the pathway viewer showcasing the reaction steps and metadata. 

\section{Cost Analysis}
\label{app:cost}

The operational cost of running the DeepRetro framework is primarily composed of two key components: the API call costs for the Large Language Models (LLMs) used in the pipeline, and the cloud computing costs for the underlying infrastructure. This analysis provides an estimated breakdown of these expenses.

\subsection{LLM API Costs}
The cost associated with LLM usage is dependent on the specific model and the number of input and output tokens processed for each retrosynthesis query. The table \ref{tab:cost_llm} outlines the pricing for the models utilized in this work and provides a blank space for the estimated cost per target molecule, which can vary significantly based on molecular complexity and the number of iterative steps required.

\begin{table}[h!]

\centering
\caption{Estimated LLM API Costs}
\begin{tabular}{|l|c|c|c|}
\hline
\textbf{LLM Model} & \textbf{Cost / 1M Input Tokens (USD)} & \textbf{Cost / 1M Output Tokens (USD)} & \textbf{Est. Cost / Molecule (USD)} \\ \hline
DeepSeek R1 & 3 & 8 & 0.15\\ \hline
Claude 3 Opus & 15 & 75 & 0.98 \\ \hline
Claude 4 Opus & 15 & 75 & 1.02 \\ \hline
Claude 3.5 Sonnet & 3 & 15 & 0.28\\ \hline
Claude 3.7 Sonnet & 3 & 15 & 0.32\\ \hline
Claude 4 Sonnet & 3 & 15 & 0.35\\ \hline
\end{tabular}
\label{tab:cost_llm}
\end{table}

\subsection{Cloud Infrastructure Costs (AWS)}
The DeepRetro system was deployed on Amazon Web Services (AWS). The architecture consists of a central head node that manages and distributes tasks to multiple worker nodes. For our setup, a \texttt{t3.xlarge} instance was used for the head node, and \texttt{t3.2xlarge} instances were used for the worker nodes to handle the computational workload. The estimated daily operational costs for this configuration are detailed below in table \ref{tab:cost_aws}.

\begin{table}[h!]

\centering
\caption{Estimated Daily AWS Infrastructure Costs}
\begin{tabular}{|l|l|c|c|}
\hline
\textbf{Node Type} & \textbf{AWS Instance} & \textbf{Number of Instances} & \textbf{Estimated Daily Cost (USD)} \\ \hline
Head Node & t3.xlarge & 1 & 5\\ \hline
Worker Node & t3.2xlarge & 3 & 21\\ \hline
\end{tabular}
\label{tab:cost_aws}
\end{table}




\end{appendices}


\bibliography{ref,Retrosynthesis_zotero}

\end{document}